%% file: review_final_twocol.tex
\documentclass[rmp,aps,twocolumn,floatfix,showpacs]{revtex4}

\usepackage{amsmath}
\usepackage{amssymb}
\usepackage{graphicx}
\usepackage{subfigure}
\usepackage{bm}
\usepackage{mathrsfs}
\usepackage{times}
\usepackage[sort&compress]{natbib}

\newcommand{\ppi}{\mathbf{{\Pi}}}
\newcommand{\mno}{\mathnormal{\Omega}}
\newcommand{\mng}{\mathnormal{\Gamma}}
\newcommand{\msl}{\mathscr{L}}
\newcommand{\msf}{\mathscr{F}}
\newcommand{\msg}{\mathscr{G}}

\begin{document}

\unitlength = 1mm

\title{Nonlinear collective effects in photon--photon \\ and 
  photon--plasma interactions}    

\author{Mattias Marklund}  
\affiliation{Department of Physics, Ume{\aa}
  University, SE--901 87 Ume{\aa}, Sweden}
\altaffiliation[Also at: ]{Institut f\"ur Theoretische Physik IV, 
  Ruhr-Universit\"at Bochum, D--44780 Bochum,
  Germany}
\altaffiliation{Centre for Fundamental Physics, Rutherford Appleton 
  Laboratory, Chilton, Didcot, Oxon, OX11 0QX, U.K.} 
  
\author{Padma K.\ Shukla}
\affiliation{Department of Physics, Ume{\aa}
  University, SE--901 87 Ume{\aa}, Sweden}
\altaffiliation[Also at: ]{Institut f\"ur Theoretische Physik IV, 
  Ruhr-Universit\"at Bochum, D--44780 Bochum,
  Germany}
\altaffiliation{Centre for Fundamental Physics, Rutherford Appleton 
  Laboratory, Chilton, Didcot, Oxon, OX11 0QX, U.K.} 
\date{\small Accepted version, submitted Feb.\ 3, 2006, to appear in Rev.\ Mod.\ Phys.\ \textbf{78} (2006)}

\begin{abstract}
We consider strong-field effects in
 laboratory and astrophysical plasmas and high intensity laser and
 cavity systems, related to quantum electrodynamical (QED) photon--photon
 scattering. Current state-of-the-art laser facilities are close to
 reaching energy scales at which laboratory astrophysics will become
 possible. In such high energy density laboratory astrophysical systems, quantum electrodynamics will
 play a crucial role in the dynamics of plasmas and indeed the 
 vacuum itself. Developments such as the free electron laser may also give a means for exploring remote violent
 events such as supernovae in a laboratory environment. At the same
 time, superconducting cavities have steadily increased their quality
 factors, and quantum non-demolition measurements are capable of
 retrieving information from systems consisting of a few
 photons. Thus, not only will QED effects such as elastic
 photon--photon scattering be important in laboratory experiments, it
 may also be directly measurable in cavity experiments.           
 Here we describe the implications of collective interactions
 between photons and photon-plasma systems. We give an overview
 of strong field vacuum effects, as
 formulated through the Heisenberg--Euler Lagrangian. Based on the
 dispersion relation for a single test photon travelling in 
 a slowly varying background electromagnetic field,
 a set of equations describing the nonlinear propagation of an electromagnetic 
 pulse on a radiation-plasma is derived. The stability of the
 governing equations is discussed, and it
 is shown using numerical methods that electromagnetic pulses may collapse and split into 
 pulse trains, as well as be trapped in a relativistic electron hole. 
 Effects, such as the generation of novel
 electromagnetic modes, introduced by QED in pair plasmas is
 described. Applications to laser-plasma systems and astrophysical
 environments are discussed.
\end{abstract}

\maketitle

\tableofcontents


\section{Introduction}

Nonlinear effects, in which a given phenomenon affects its own evolution and dynamics, 
are prominent components in a large variety of physical, chemical, and 
biological systems. The examples range from optical and nerve fibers, autocatalytic
chemical reactions, to ocean waves \cite{Scott}. The field of hydrodynamics
has been especially important for the development of nonlinear physics, both
concerning analytical and computational tools, since there the nonlinear effects
can play a major role in systems with important applications, e.g.\ meteorology. 
The subject of plasma physics is a natural generalization of the field of
hydrodynamics, since it builds on the fluid or kinetic equations, while adding the 
electromagnetic interaction.
The plasma state of matter is prominent 
in large regions of the Universe, such as our closest star, the Sun, accretion discs, and even interstellar clouds. 
It has since long also been noted
within the field of plasma physics that both nonlinear effects and collective interactions
can give rise to important new physical effects, such as the ponderomotive force 
concept and Landau damping \cite{Hasegawa}. 
The low-frequency ponderomotive force, which  arises due to
nonlinear couplings between high-frequency electromagnetic
fields, plays a central role in the physics of laser-plasma
interactions. This force in an unmagnetized plasma is
expressed as the gradient of the electromagnetic field
intensity, which pushes electrons locally and thereby
creating a hugee space charge electric fields and the plasma density
cavitaties. Due to the radiation ponderomotive force, one
has the possibility of many interesting nonlinear pheneomena
in plasmas, e.g. the generation of intense wakefields, stimulated
scattering of electromagnetic  waves off plasmons and phonons,
localization of electromagnetic fields, etc.\ \cite{Eliezer}.
The momentum and energy transfer from the laser field to the plasma
particles can be harnessed in, e.g. inertial confinement fusion \cite{Eliezer}.
Moreover, the intense electromagnetic radiation generated in 
state-of-the-art lasers can be used to model certain astrophysical plasma
conditions in a laboratory environment \cite{Remington}. Questions 
of astrophysical interest that can be approached within the field 
of high energy density laboratory astrophysics range from
the equations of state of planetary interiors to supernova shock 
formation (see \citet{HEDLA} for an overview). 
In the next generation laser-plasma systems 
the influence of 
quantum electrodynamics will become important, and 
fundamental questions related to the nonlinearity of the quantum 
vacuum can be approached in laboratory systems 
\cite{mou05}.

Currently, lasers are capable of reaching intensities of
$10^{21}-10^{22}$ W/cm$^2$ \cite{Bahk-etal,mou98,taj02,taj03,mou05}. 
At such high field strengths, the quiver
velocity of the electrons is highly relativistic, and the 
radiation pressure, manifesting itself as a ponderomotive force term
in the evolution equations for the plasma, gives rise to local electron
expulsion. Moreover, at these intensities, the nonlinear relativistic dynamics of the laser-plasma system gives
rise to a number of other interesting phenomena as well, such as
soliton formation and pulse collapse \cite{Shukla-etal}. 
The latter could be of interest when using laser-plasma systems to generate 
electromagnetic field intensities approaching the Schwinger intensity limit
\cite{cai04,Bulanov-etal,Shukla-Eliasson-Marklund1,bob,bob2,%
Shukla-Marklund-Eliasson,mou05}. 

\begin{figure}[ht]
  \includegraphics[width=.9\columnwidth]{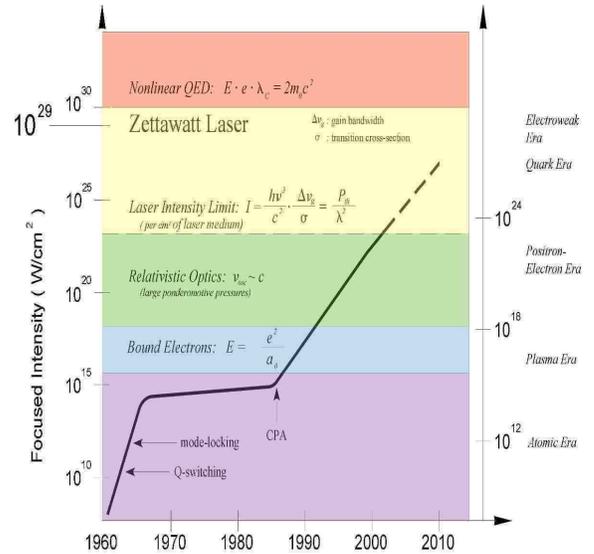}
  \caption{The evolution of laser intensity \citep[reprinted with permission from][]{taj02}.}
\label{fig:laserevol}
\end{figure}

The event of future ultra-short (in the femto-second range)
intense ($10^{23}-10^{25}$ W/cm$^{2}$) lasers \cite{mou98,taj02,taj03,mou05} 
could generate new physics within
the next few years (see Fig. \ref{fig:laserevol}). This is based on the development of
chirped pulse amplification, and the evolution of laser power is
predicted to continue evolving for quite some time \cite{mou98,mou05}. 
The X-ray free electron lasers (XFEL) under construction at SLAC 
\cite{SLAC} and DESY \cite{DESY} will 
be a major source of experimental data not achievable with today's systems,
ranging from molecular properties \cite{pat02} to astrophysical conditions
\cite{Chen}, such as supernova shocks \cite{Woolsey-etal}. 
The intensity at the XFEL focus is expected to reach intensities
making the quantum vacuum directly accessible for observations 
\citep[][see Fig. \ref{fig:brilliance}]{Ringwald,Ringwald1,Ringwald2}.      
Moreover,  
combined effects of laser pulse collapse and ponderomotive force
electron expulsion would be able to create plasma channels in which
ultra-high intensity field strengths are reached \cite{Yu-etal}, such
that the  nonlinear vacuum effect of elastic photon--photon scattering 
could become important \cite{Shen-Yu,Shen-etal}. 

\begin{figure}[ht]
  \includegraphics[width=.7\columnwidth]{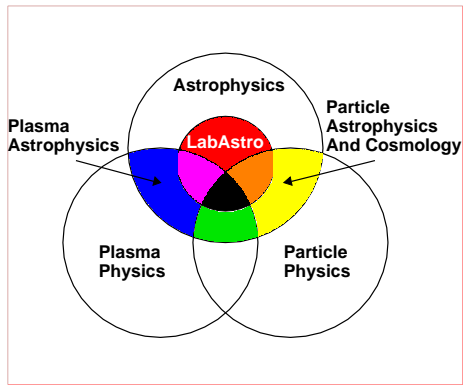}
  \caption{The connection between different high-energy regions in physics  
  and experiments \citep[reprinted with permission from][]{Chen}.}
\label{fig:blobs}
\end{figure}

A majority of studies 
have not taken into account the influence of electron--positron
pair creation or elastic photon--photon scattering  
on the dynamics of laser-plasma systems (there are however important exceptions,
see e.g. \citet{Bulanov-etal2005}). Effects of this kind
will be of the utmost importance when laser compression
schemes approaches the critical field strength
\begin{equation}\label{eq:criticalfield}
  E_{\text{crit}} = \frac{m_ec^2}{e\lambda_e} \sim 10^{18}\,\mathrm{V/m} , 
\end{equation}
as the nonlinearity of the quantum vacuum becomes pronounced.
Here $m_e$ is the electron rest mass, $c$ is the speed of light in vacuum,
$e$ is the magnitude of the electron charge, $\lambda_e = \hbar/m_ec$
is the Compton wavelength, and $\hbar$ is the Planck constant divided by $2\pi$.
Thus, for 
such extreme plasma systems, the concept of photon--photon scattering, 
both elastic and inelastic, has to be
taken into account. 

The interaction of high-intensity laser pulses with   
plasmas has applications to other fields of science, e.g. table-top particle 
accelerators \cite{bob}. Also, 
achieving field strengths capable of producing pair plasmas in the laboratory 
could facilitate a means of producing anti-matter on a more or 
less routine basis, as they currently are at high energy accelerators. 
However, it should be emphasized that the
production of pairs from intense lasers requires that 
severe technical constraints can be overcome, such as phasing of the
two interacting short electromagnetic pulses.  
Even if routine pair production via laser systems is not to be reached 
within the near future, the
related possibility of directly detecting elastic photon--photon
scattering is indeed a fascinating possibility
\cite{Brodin-Marklund-Stenflo,Soljacic-Segev}.  
Furthermore, the creation of multi-dimensional high intensity
electromagnetic pulses, using guiding structures (such as plasma
boundaries), could result in pulse collapse \cite{Brodin-etal,Shukla-Eliasson-Marklund}. Such pulse collapse would give rise to intensities close to the Schwinger critical field (\ref{eq:criticalfield}). Thus, the combination of laser-plasma
interactions and QED effects, such as pair production and
photon--photon scattering could spark new
methods for producing conditions reminiscent of astrophysical
environments in future experiments (see Fig. \ref{fig:blobs}). In fact, some of the 
most pertinent in today's fundamental physics research , such as the question
of dark matter (e.g. through the effects of light pseudoscalar fields, such as the axion 
field, on QED interactions and light propagation, see \citet{Bernard,Dupays-etal,Bradley-etal}), 
cosmic accelerators (such as through laboratory plasma wakefield accelerator 
tests \cite{Chen}), and possible new high-density states of 
matter \cite{Remington}, are related to the high energy events for which laboratory astrophysics
would yield valuable insight. Thus, it is of 
interest 
to study such high energy scenarios,
e.g.\ photon--photons scattering in the context of 
laser-plasma systems \cite{Bulanov1}, as these are, in the near future, likely 
to yield the right conditions for such events to take place.

\subsection{Nonlinear quantum electrodynamics}

In classical electrodynamics, as described by Maxwell's equations,
photons are indifferent to each other as long as there is no material
medium present. This is not so in quantum electrodynamics (QED). 
Due to the interaction of photons with virtual electron--positron
pairs, QED offers the possibility of
photon--photon scattering \cite{Heisenberg-Euler,Schwinger}. 
This is commonly expressed through the effective
field theory approach represented by the 
Heisenberg--Euler Lagrangian \cite{Heisenberg-Euler,Weisskopf,Greiner,Grib-etal,Fradkin-etal,Schwinger}, neglecting dispersive effects. This
Lagrangian [see Eq.\ (\ref{eq:lagrangian1}) below] and its
generalizations \cite{Valluri-etal,Dunne,Dittrich-Gies}, 
gives nonlinear corrections to Maxwell's vacuum equations,
similar to the self-interaction terms encountered in nonlinear optics
due to the presence of a Kerr medium \citep{Agrawal,Bloembergen},
and higher order corrections can easily be incorporated by taking into
     account higher vertex order diagrams. 
Since the lowest order effective
     self-interaction term is proportional to the fine structure
     constant squared, the field strengths need to
     reach appreciable values until these effects becomes important,
     see Eq. (\ref{eq:criticalfield}) \cite{Greiner,Grib-etal,Fradkin-etal}.
The corrections 
give rise to both single particle effects, such as closed photon paths
\cite{Novello-etal}, vacuum birefringence \cite{Heyl-Hernquist}, 
photon splitting \cite{Adler} and lensing effect in strong magnetic 
fields (see, e.g. \citet{Harding,DeLorenci-etal}), as well as collective effects, 
like the self-focusing of beams \cite{Soljacic-Segev} or the formation
of light bullets \cite{Brodin-etal}. Recently, it has also been shown,
using analytical means, that these effects give rise to collapsing structure 
in radiation gases \cite{Marklund-Brodin-Stenflo}, results that have been 
extended and confirmed by numerical simulations \cite{Shukla-Eliasson}. 
Possible detection
     techniques, as well as physical implications, of the effects of
     photon--photon scattering have attracted a rather constant
     interest since first discussed (e.g.       
     \citet{Erber,Tsai1,Tsai2,Greiner,Grib-etal,Fradkin-etal,Bialynicka-Birula,%
     Ding,Kaplan-Ding,Latorre-etal,Dicus-etal}), and the
     concept of self-trapping 
     of photons due to vacuum nonlinearities was discussed 
     independently by \citet{Rozanov93,Rozanov} and 
     \citet{Soljacic-Segev} in the context of the
     nonlinear Schr\"odinger equation. 
     
The above studies assume that the dispersive/diffractive effect of 
vacuum polarization is negligible, and this, of course, puts
constraints on the allowed space and time variations of the
fields \cite{Soljacic-Segev}. In the
context of pair creation, rapidly varying fields have been analyzed,
since when individual photons pass the pair creation energy threshold
$2m_ec^2$, real electron-positron pairs may be created from the
vacuum by a ``down-conversion'' process of photons. Similar processes are
thought to be of importance in the neighborhood of strongly
magnetized stars, where the magnetic field induces
photon splitting \cite{Erber,Adler,Adler-etal,Adler-Shubert,Baring-Harding},  
and may effectively absorb the photons \cite{Heyl-Hernquist,Duncan}. 
It has been suggested that the
     nontrivial 
     refractive index due to photon--photon scattering could induce a
     lensing effect in the neighbourhood of a magnetar \cite{Shaviv}.

The physics of elastic photon--photon scattering has interested researchers
for a long time, and      
     several suggestions for ways to detect this scattering 
     in the laboratory have been made during the last decades
     \cite{Dewar,Alexandrov-etal}, and the recent strong increase in
     available laser intensities have stimulated various schemes
     \cite{Rozanov93,Rozanov,Ding-Kaplan}. It has
     been suggested by \citet{Brodin-Marklund-Stenflo,Brodin-Marklund-Stenflo2} 
     that the effect of photon--photon
     scattering could be detected using fields significantly weaker ($10\,\mathrm{MV/m}$) than state-of-the-art
     laser fields. 
           
     Next, we present some physical systems in which the effects of 
     photon--photon interactions may either be of importance (magnetars), or become 
     important in the near
     future (such as state-of-the-art laser-plasma systems).
     
\subsubsection{Intense field generation}
\paragraph{Electromagnetic cavities}

High performance, 
i.e.\ large electromagnetic fields combined with low dissipative losses, can 
be found in superconducting cavities, which among other things are 
used for particle acceleration \cite{Graber}. 
The waves that can be sustained within such a cavity can have a field-strength  
$E \sim 10\,\mathrm{MV/m}$, i.e. close to the maximum that 
can be tolerated by the walls without field emissions. 
For such cavities, the different high intensity wave modes can act as
pump waves for the quantum vacuum. Through the interaction
between these waves and virtual electron--positron pairs, new modes with 
well-defined frequencies and wavenumbers will be generated. Those satisfying 
the dispersion criteria for the given cavity could then also reflect
within the cavity with very small losses, thus yielding a method for 
detection of the quantum vacuum nonlinearities. For example, 
for a cavity 
resistance $R \sim 1\,{\rm n}\Omega$, corresponding to  
superconducting niobium at a temperature $1.4 \, \mathrm{K}$ and a  
frequency $\omega \sim 2\times 10^{10}\,\mathrm{rad/s}$ of the 
wave mode generated via the nonlinear quantum vacuum,  
one finds that the saturated energy  
flux $P_3$ of the generated mode is of the order of $10^{-6}\,\mathrm{W/m}^{2}$
(see Secs. \ref{sec:cavity} and \ref{sec:cavityexp}) 
\cite{Brodin-Marklund-Stenflo,Eriksson-etal}.   
This energy flux is above the detection level by 
several orders of magnitude. However, one should note the importance of the 
superconducting walls for the output level of the excited mode. For copper 
at room temperature, the cavity resistance increases by a factor 
 $\sim 10^7$ as compared to the above example, and consequently the 
energy flux of the excited mode falls by a factor $\sim 10^{-14}$. In 
the latter case, it is questionable whether the excited signal can be 
detected. The concepts of cavity mode interactions and cavity experiments
will be further discussed in Secs.\ \ref{sec:cavity} and \ref{sec:cavityexp}.

\paragraph{Laser development}

The event of ultra-short (in the femto-second range)
intense ($10^{22}-10^{24}$ W/cm$^{2}$) lasers \cite{mou98,taj02,taj03,mou05} 
holds the promise of generating large amounts of new physics within
the next few years. This promise is based on the development of
chirped pulse amplification, and the increase of laser power is
predicted to continue for quite some time \cite{mou98,mou05}. 
There are two ways for reaching high intensities within laser systems.
The method most common is to shorten the pulse duration ($\lesssim 100\,\mathrm{fs}$),
while keeping the energy content in each pulse rather modest ($\sim 1-10\,\mathrm{J}$). 
Such pulse generation techniques can have high repetition rates, which can be advantageous 
in certain experiments where a large number of shots is needed. The other route
is to increase the pulse contents while keeping the pulse duration of the order
$0.5 - 1\,\mathrm{ps}$. Such systems has the advantage of providing a high signal to
noise ratio for some experiments.
The Nova Petawatt laser at the Lawrence Livermore National 
Laboratory,USA, used this principle, and each pulse, which had a duration
of $\sim 500\,\mathrm{fs}$, had an energy contents of $\sim 500\,\mathrm{J}$.
Similar systems are operating at ILE/Osaka, Japan, \cite{ILE} and the Rutherford
Appleton Laboratory, U.K., \cite{RAL}. 
The OMEGA EP laser under construction at the University of Rochester, USA, 
will also work according to the high energy principle, and have
pulse energies $1-2.6\,\mathrm{kJ}$ with durations $1-10\,\mathrm{ps}$ 
\cite{OMEGA}.
Apart
from being a tool for practical use, such as inertial confinement
fusion and material science, intense laser facilities are now of international interest for 
basic research (such as at, e.g.\ the National 
Ignition Facility at the Lawrence 
Livermore National Laboratory (USA), the Laboratory for Laser Energetics at the University 
of Rochester (USA), the Advanced Photon Research Center (Japan), the
Institute for Laser Engineering at Osaka University (Japan), 
LULI Laboratoire pour l'Utilisation des Lasers Intenses (France),
LIL/Laser M\'egajoule at CEA (France), 
or the Central Laser Facility, Rutherford Appleton Laboratory (UK)), and the increased
laser output power also gives the opportunity, for the first time, to
obtain astrophysical energy scales in a controlled laboratory setting
\cite{Chen,Remington,HEDLA}.

The generation of high (electromagnetic) field
strengths is at the heart of understanding a variety of phenomena,
such as astrophysical shocks and jets, in a laboratory setting, and furthermore
forms the basis for a number of applications, e.g., table-top plasma
accelerators. Thus, schemes and mechanisms for generating such high
fields, other than the direct laser pumping, will be of 
importance to the development of a wide range of scientific areas. 

\paragraph{The free electron laser}

\begin{figure}[ht]
  \includegraphics[width=.7\columnwidth]{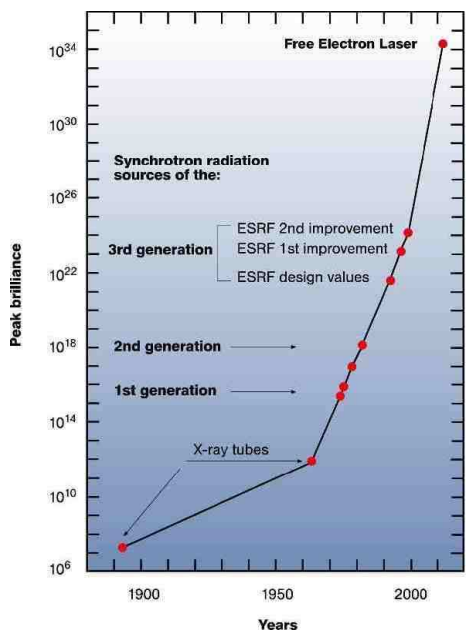}
  \caption{Evolution of peak brilliance, in units of 
  photons/(s\,mrad$^2$\,mm$^2$\,0.1\,\%$\times$bandwidth), of X-ray sources. 
  Here, ESRF stands for the European Synchrotron Radiation Facility in 
  Grenoble. (Source: DESY Hamburg, \texttt{http://xfel.desy.de/})}
\label{fig:brilliance}
\end{figure}

The x-ray free electron laser (XFEL) is an alternative to the current laser
generation techniques, and it has as its base the particle
accelerator, where high energy electrons are generated to obtain
high-frequency radiation \cite{DESY,SLAC}. Within these lasers, a large number of
coherent photons are generated (10 orders of magnitude more than regular
synchrotron sources). The XFEL concept has
a wide variety of interesting applications, among these the
possibility to probe the structure of large molecules, commonly found
within molecular biology systems \cite{pat02}. It is also hoped that
XFEL could form the basis of electron--positron pair creation
\cite{alk01,Ringwald,Roberts-etal}. Both at the TESLA collider at DESY
and LCLS at SLAC, the energy density at the focus (with a spatial 
width $\sim 10^{-10}\,\mathrm{m}$ and on the time-scale $10^{-13}\,\mathrm{s}$) 
of the XFEL
is expected to reach energy densities of $10^{29}\,\mathrm{J/m^3}$ 
\cite{Ringwald}, see Fig. \ref{fig:brilliance}. This
corresponds to electric field strengths of the order $10^{20}\,\mathrm{V/m}$,
i.e. two orders of magnitude \emph{above} the Schwinger critical
field (\ref{eq:criticalfield}), at which pair creation is expected to take place.
The possibility to `fuel' the generation of
electron--positron pairs by nonlinear effects is therefore a very
promising prospect, and as some authors have noted, this essentially
amounts to `boiling the vacuum' \cite{Ringwald1,Ringwald2}. 
Moreover, it is believed that nonlinear QED effects
will be a very important component of the interaction of XFEL
generated radiation with dense media. Therefore, it is of 
interest to achieve an understanding of the
influence of those nonlinear effects within the parameter regime obtainable
by the XFEL.

\subsubsection{Laser-plasma systems}

High laser powers 
generate enormous radiation pressures, and accelerate particles to
relativistic velocities. The fields generated by the particles will, 
therefore, backscatter and create a nonlinear feedback, something which
can give rise to, e.g. laser pulse compression and electron density
cavitation. This may play a significant role in different
proposed schemes of laser self-focusing  
\cite{Bulanov-etal,puk03,Shukla-Eliasson-Marklund1},
completely changing the dynamics of the suggested methods and altering
the final results in nontrivial ways. This serves as an important example of 
nonlinear effects in the evolution of laser-plasma systems.

The interaction of high power lasers, reaching intensities of 
$10^{20} - 10^{22}\,\mathrm{W/cm^2}$, with plasmas has long been 
the backbone in different schemes for table-top particle 
accelerators \cite{bob}, and is also essential for the concept
of inertial confinement fusion \cite{Eliezer}. The ponderomotive
force generated in high power laser-plasma systems due
to laser intensity gradients may give rise to 
a plethora of phenomena, such as laser pulse self-focusing and filamentation,
soliton formation, parametric instabilities, and magnetic field generation 
(see \citet{Eliezer} and references therein).  

Laser-plasma systems have been suggested as sources
of high intensity radiation. Due to the laser ponderomotive force
plasma electrons will be pushed out of the path of the laser pulse,
trapping and compressing the laser pulse, such that further electrons are pushed 
out. The plasma can sustain high field strengths, and the pulse compression
can therefore reach appreciable intensity values \cite{mou05}. In fact, 
using a Langmuir wave as a plasma `mirror' for the laser pulse,
pulse intensities could reach, and even surpass,
the Schwinger critical field (\ref{eq:criticalfield}) \cite{Bulanov-etal,mou05}.
Thus, the interaction between intense lasers and plasmas is an 
intriguing tool for generating pulse intensities above the laser limit
\cite{mou98}.

\subsubsection{Astrophysical and cosmological environments}

Astrophysical environments and events display truly enormous energy releases. 
Supernova explosions, black hole accretion,
magnetar and pulsar systems are a few examples of such
extreme situations. Moreover, the energy scales in the early universe are
equally immense, or even greater, and our understanding of the origin
of the universe is hampered by the fact that the energy density scales are so
far from anything that can be generated in a laboratory, except perhaps in relativistic 
heavy ion collisions \cite{RHIC}. 
It is therefore not surprising that these
environments can often act as laboratories for phenomena that we 
currently do not have technology to reproduce in earth based laboratories.
Quantum electrodynamical nonlinear vacuum effects have received 
a fair amount of attention in strongly magnetized systems, such as 
pulsars \cite{CurtisMichel,Beskin-etal} and magnetar environments 
\cite{magnetar,Harding}.
The magnetic field strengths of magnetars can reach energy levels comparable, 
or even surpassing, the energy corresponding to the Schwinger critical field 
strength $10^{18}\,\mathrm{V/m}$, thus making the vacuum truly
nonlinear (nonlinear QED effects in the magnetized vacuum are described in a large 
number of  publications, and for a representative but incomplete list, see 
\citet{Erber,Adler,Adler-etal,Adler-Shubert,Tsai1,Tsai2,Mentzel-Berg-Wunner,%
Baier-Milstein-Shaisultanov,Chistyakov-etal} and references therein). 
The effect of a nonlinear vacuum may even be of crucial 
importance for our understanding of these objects \cite{Baring-Harding}.

\section{Effective field theory of photon--photon scattering}

The development of quantum electrodynamics was the result of a long and collective 
effort, and paved the way for an understanding of the weak and strong forces 
as well. It has, since its advent, been confirmed to an unprecedented accuracy,
compared to any physical theory. It solved some of the long-standing conceptual 
problems of relativistic quantum theory, as proposed by Dirac and others, and
it furthermore changed the way we look at the elementary interactions between
particles and fields. The theoretical proposal, due to Dirac,  of anti-matter as a 
result of relativistic quantum theory was put on a firm foundation with QED.

The quantization of the vacuum has lead to some remarkable insights and discoveries.
Consider for example the Casimir effect \cite{Casimir1,Casimir2},\footnote{Casimir 
considered 
particle--particle and particle--plate \cite{Casimir1}, and plate--plate systems 
\cite{Casimir2}, since the
problem stemmed from research on colloidal solutions, but is most
clearly represented by the parallel plate example. The Casimir effect 
has since been confirmed by many different experiments (see, e.g., 
\citet{Sukenik-etal,Mostepanenko-Trunov,Lamoreaux,%
Bordag-Mohideen-Mostepanenko,Bressi-etal,Harber-etal} and 
references therein).} 
in which two parallel conducting plates with area $A$ separated by a 
distance $d$. Due to the different boundary conditions between and 
outside the plates, there will be a net attractive force $F \propto A/d^4$ 
between the plates. In a heuristic sense, the vacuum between the plates
is `emptier' than outside, since fewer states are allowed due the finite distance 
between the plates. Related to this is the much debated Scharnhorst effect 
\cite{Scharnhorst,Barton,Barton-Scharnhorst,Scharnhorst2}, at which the phase 
(and group) speed exceeds the speed of light $c$ in vacuum. As will be demonstrated 
later, the opposite occurs in the electromagnetic vacuum, i.e.\ the phase and group 
velocities decrease due to the electromagnetic influence on the quantum vacuum.   

In conjunction with any description of photon--photon
scattering, it should also be mentioned the large amount of literature and interest in
finite temperature effective field theory effects. Thermal effects
generalize the classical results of Schwinger in the weak field limit 
\cite{Heisenberg-Euler,Weisskopf,Schwinger}. It was pioneered by
\citet{Dittrich} who investigated the thermal effects in combination with an 
external magnetic field, and later a comprehensive study using the real time 
formalism in the case of an general electromagnetic field background was 
performed by \citet{Elmfors-Skagerstam} (see also \citet{Gies}). The dispersion
relation, including dispersive effects, was discussed by \citet{Gies2}, and it
was later shown that in a thermal vacuum, in contrast to the non-thermal 
one, two-loop corrections will dominate over the one-loop effects
\cite{Gies3}. 
However, treating all these studies (we have by no means exhausted the list 
of papers in this short expos\'e) in detail is outside the scope of the present 
paper, and since we moreover are interested in the problem of collective effects,
the treatment of thermal effects, although of interest, is left for a future review.
  
\subsection{The concept of elastic scattering among photons}

\begin{figure}
\includegraphics[width=0.5\columnwidth]{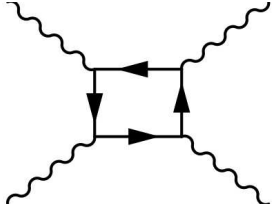}
\caption{The Feynman box diagram for the lowest order
  photon--photon scattering at the one-loop level.}
  \label{fig:weak}
\end{figure}

Photon--photon scattering is a non-classical effect arising in
quantum electrodynamics (QED) due to virtual electron--positron
pairs in vacuum, see Fig. \ref{fig:weak}. In the low energy limit, 
i.e. $\hbar\omega \ll m_ec^2$
the magnitude of photon--photon can be descibed in terms of the 
differential cross-section \cite{Berestetskii-etal}
\begin{equation}\label{eq:diffcross}
  \frac{d\sigma_{\gamma\gamma}}{d\Omega} = \frac{139\alpha^2r_e^2}{32400\pi^2}%
  (3 + \cos^2\theta)\left(\frac{\hbar\omega}{m_ec^2}\right)^6 ,
\end{equation}
where $\omega$ is the photon frequency in the center-of-mass
system, $r_e$ is the classical electron radius, and $\alpha = e^2/4\pi\epsilon_0\hbar c 
\approx 1/137$ is the fine structure constant. 
Integrating (\ref{eq:diffcross}) gives the total cross-section
\begin{equation}
  \sigma_{\gamma\gamma} = \frac{973\alpha^2r_e^2}{10125}%
  \left(\frac{\hbar\omega}{m_ec^2}\right)^6 \approx
  0.7\times10^{-65}\left(\frac{\hbar\omega}{1\,\mathrm{eV}}\right)^6
  \,\mathrm{cm}^2.
\end{equation}
We note that the cross-section decreases very fast with decreasing 
photon energy. In the high energy limit, the cross-section on the 
other hand goes like $\omega^{-2}$. The cross-section
reaches a maximum of $\sigma_{\gamma\gamma} \approx 2\times 10^{-30}\,\mathrm{cm}^2$ 
for photon energies $\hbar\omega \sim m_ec^2$ \cite{Berestetskii-etal},
indeed a very small number.

Instead of a microscopic description, the interactions of photons
may be described by an effective field theory. Formulated in
terms of such an effective
field theory, using the Heisenberg--Euler Lagrangian (valid in the long wavelength and
weak field limit, see Eq.\ (\ref{eq:constraint1}))
\cite{Heisenberg-Euler,Schwinger}, this results
in nonlinear corrections to Maxwell's vacuum equations, which to
lowest order in the fine structure constant are 
cubic in the electromagnetic (EM) field. These correction
takes the same form as in nonlinear optics, where the material
properties of, e.g. optical fibres, gives rise to cubic nonlinear
terms in Maxwell's equations, so called Kerr nonlinearities
\cite{Agrawal,Kivshar-Agrawal}. Since the effective
self-interaction term is proportional to the fine structure
constant squared, this means that the field strengths under most 
circumstances need to
reach values close to the critical field (\ref{eq:criticalfield}) until these effects becomes 
important
\cite{Greiner,Grib-etal,Fradkin-etal}. With this at hand, we now continue to focus on the 
concept of photon--photon scattering. 

\subsection{Weak field limit}

We first derive the general dispersion
relation in the low photon energy and weak field limit, assuming 
(see Eq.\ (\ref{eq:criticalfield}))
\begin{equation}
  \omega \ll \omega_e \equiv m_ec^2/\hbar \,\,\text{ and }\,\, |\mathbf{E}| \ll
  E_{\text{crit}} ,
  \label{eq:constraint1} 
\end{equation}
where $E_{\text{crit}}$ is given by (\ref{eq:criticalfield}), and $\omega_e \approx 
8\times 10^{20}\,\mathrm{rad/s}$ is the Compton frequency. 
When these constraints are valid,
pair-creation, both due to single photons and collective effects, will
be unimportant, and the effective Lagrangian may therefore be treated
solely in terms of its real part. It should be remembered that
the second of these constraints comes from the pair creation
probability of Schwinger \cite{Schwinger}, which was derived for a
pure electric field and may therefore not strictly be applied to the
case of a radiation gas. Thus, this investigation goes far beyond 
the description of the thermodynamics of nonlinearly interacting
incoherent photons with a photon gas in a plasma environment \cite{Nodar}.

Photon-photon scattering
is a second order effect (in terms of the fine structure constant $\alpha$), 
and for constant or weakly
varying fields it can be formulated in standard notation using 
the Euler--Heisenberg Lagrangian density 
\cite{Heisenberg-Euler,Schwinger} 
\begin{equation}  \label{eq:lagrangian1}
  {\mathscr{L}}  = \mathscr{L}_0 + \mathscr{L}_c
  = \epsilon_0{\mathscr{F}} + \epsilon_0^2\kappa(4{\mathscr{F}}^2 +  
    7{\mathscr{G}}^2) \ , 
\end{equation} 
where 
\begin{equation} \label{eq:parameter}
  \kappa \equiv \frac{2\alpha^2\hbar^3}{45m_e^4c^5} 
  = \frac{\alpha}{90\pi}\frac{1}{\epsilon_0E_{\text{crit}}^2} 
  \approx \frac{1}{3\times10^{29}\,\mathrm{J/m^3}}.
\end{equation} 
Moreover, the field invariants are defined in terms of the 
field tensor $F_{ab}$ according to 
\begin{equation}
   \msf \equiv \tfrac{1}{4}F_{ab}F^{ab} = 
  \tfrac{1}{2}(c^2\mathbf{B}^2 - \mathbf{E}^2) , \, 
  \msg \equiv \tfrac{1}{4}F_{ab}\widehat{F}^{ab} = -c\mathbf{E}\cdot\mathbf{B} ,
  \label{eq:invariants}
\end{equation} 
$\widehat{F}^{ab} = \epsilon^{abcd}F_{cd}/2$, and ${\mathscr{F}}^2$ and
${\mathscr{G}}^2$ are the lowest order QED corrections.  
We note that ${\mathscr{F}} = {\mathscr{G}} = 0$ in the 
limit of parallel propagating waves. 
The latter terms in (\ref{eq:lagrangian1})
represent the effects of vacuum 
polarization and magnetization, and the QED corrected Maxwell's vacuum 
equations take the classical form, using  
\begin{equation}
  {\bf D} = \epsilon_0{\bf E} + {\bf P} \ , \quad  
  {\bf H} = \frac{1}{\mu_0}{\bf B} - {\bf M} \ ,
  \label{eq:constituent}  
\end{equation}
where ${\bf P}$ and ${\bf M}$ are of third order in the field amplitudes  
${\bf E}$ and ${\bf B}$ (see Eqs. (\ref{eq:polarization}) and (\ref{eq:magnetization}) below), and $\mu_0 = 1/c^2\epsilon_0$.
Furthermore, they contain terms 
${\mathscr{F}}$ and ${\mathscr{G}}$ such that
${\bf P} = {\bf M} = 0$ in the limit of parallel propagating waves.  
It is therefore necessary to use nonparallel waves in 
order to obtain an effect from these QED corrections.

From the constituent relations (\ref{eq:constituent}) we can deduce 
the general wave equations for ${\bf E}$ and ${\bf B}$ 
according to   
  \begin{equation}
    \frac{1}{c^2}\frac{\partial^2{\bf E}}{\partial t^2}  
      - \nabla^2{\bf E} =  
      -\mu_0\left[ \frac{\partial^2{\bf P}}{\partial t^2}   
      + c^2\nabla(\nabla\cdot{\bf P})  
      + \frac{\partial}{\partial t}(\nabla\times{\bf M)} \right]    
    \ , \label{WaveE}
  \end{equation}
and 
  \begin{equation}
    \frac{1}{c^2}\frac{\partial^2{\bf B}}{\partial t^2}  
      - \nabla^2{\bf B} = 
      \mu_0\left[ \nabla\times(\nabla\times{\bf M})  
      + \frac{\partial}{\partial t}(\nabla\times{\bf P)} \right]    
      \ . \label{WaveB} 
  \end{equation}
Furthermore, the effective polarization and magnetization, 
appearing in (\ref{eq:constituent}), in vacuum due to 
photon-photon scattering induced by the exchange of virtual 
electron-positron pairs can be obtained from the Lagrangian (\ref{eq:lagrangian1}) 
and are given by (see, e.g., \citet{Soljacic-Segev})
\begin{equation}
    {\bf P} = 2\kappa\epsilon_0^2\left[ 2(E^2 - c^2B^2){\bf E}  
      + 7c^2({\bf E\cdot B}){\bf B} \right]  \ ,
      \label{eq:polarization}
\end{equation}
and
\begin{equation}
    {\bf M} = 2\kappa\epsilon_0^2c^2\left[ -2(E^2 - c^2B^2){\bf B}  
      + 7({\bf E\cdot B}){\bf E} \right] \ .
      \label{eq:magnetization} 
\end{equation}
Equations (\ref{WaveE})--(\ref{eq:magnetization}) offer the starting point for the study of
a weakly nonlinear electromagnetic vacuum in terms
of the classical field strength vectors $\mathbf{E}$ and $\mathbf{B}$.

The corrections in the Lagrangian (\ref{eq:lagrangian1}) is the power series
expansion in the field strengths of the full one-loop correction given 
by \cite{Heisenberg-Euler,Weisskopf,Schwinger}
\begin{eqnarray}
&&\!\!\!\!\!\!\!\!\!
 \msl_{c} = -\frac{\alpha}{2\pi}\epsilon_0E_{\text{crit}}^2\int_0^{\mathrm{i}\infty}
  \frac{dz}{z^3}\mathrm{e}^{-z}\times
\label{eq:lagrangian2} \\ &&\!\!\!\!\!\!\!\!\!
  \bigg[
  z^2\frac{ab}{E_{\text{crit}}^2}\,\coth\left(\frac{a}{E_{\text{crit}}}z\right)\,%
  \cot\left(\frac{b}{E_{\text{crit}}}z\right)
  - \frac{z^2}{3}\frac{(a^2 - b^2)}{E_{\text{crit}}^2} - 1 \bigg] ,
\nonumber
\end{eqnarray}
where 
\begin{equation}
  a = \left[(\msf^2 + \msg^2)^{1/2} + \msf\right]^{1/2} , \,
  b = \left[(\msf^2 + \msg^2)^{1/2} - \msf \right]^{1/2} .
\label{eq:ab}
\end{equation}  
Thus, $\msf = (a^2 -
b^2)/2$ and $|\msg| = ab$. The Lagrangian correction (\ref{eq:lagrangian2})
is the starting point of the effective field theory analysis of a strongly nonlinear 
quantum vacuum, such as used in some studies of photon splitting  
(see \citet{Dittrich-Gies2,Dittrich-Gies} and references therein), and defining 
some of the properties of a strongly nonlinear 
gas of photons \cite{Marklund-Shukla-Eliasson}. 

\subsection{The dispersion function}\label{sec:dispersionfunction}

One may find the dispersion relation of photons in an arbitrary 
constant, or weakly varying, electromagnetic background 
\cite{Bialynicka-Birula,DeLorenci-etal,Thoma}. 
Starting from the Lagrangian (\ref{eq:lagrangian1}), 
introducing the four-potential $A^b$ such that
$F_{ab} = \partial_aA_b - \partial_bA_a$, the Maxwell 
equations resulting from the variation with respect to the four-potential
becomes
\begin{equation}
  \partial_aF^{ab} = 
   2\epsilon_0\kappa\partial_a\left[ (F_{cd}F^{cd})F^{ab} +
   \tfrac{7}{4}(F_{cd}\widehat{F}^{cd})\widehat{F}^{ab} \right] ,
\label{eq:exact-evol1}
\end{equation}
where we adopt the convention 
$(-1,1,1,1)$ for the metric $\eta_{ab}$, used for raising and lowering
four-indices $a, b, \ldots = 0, 1, 2, 3$.
Next, assuming that $F_{ab}
= f_{ab} + \phi_{ab}$, where $f_{ab}$ denotes the varying 
background field,
and $\phi_{ab}$ ($\ll{f_{ab}}$) is a weak field propagating on this
background, we find that the background satisfies $\partial_af^{ab} = 0$, 
and  
$\partial_a\widehat{F}^{ab} = 0$ is identically satisfied due to 
the definition of $F_{ab}$ in terms of the four-potential $A^b$.  

Linearizing Eq.\ (\ref{eq:exact-evol1}) with respect to $\phi$, and Fourier decomposing 
perturbations according to 
$
  \phi_{ab}(x) = (k_a\epsilon_b - k_b\epsilon_a)\exp(ik\cdot x) + \text{c.c.},
$ 
where $\epsilon_a$ is the polarisation vector, $k\cdot x \equiv
k_ax^a$ and c.c.\ denotes the complex conjugate,
we obtain  the following algebraic set of equations for the
polarization vector 
\begin{equation}
  M^0_{ab} \epsilon^b = \big[k^2g_{ab} - k_ak_b
  - \kappa\epsilon_0\left( 8a_a a_b + 14\hat{a}_a\hat{a}_b\right) \big]
  \epsilon^b = 0, 
  \label{eq:algebraic}
\end{equation}
where $a_b \equiv f_{bc}k^c$ and $\hat{a}_b \equiv
\widehat{f}_{bc}k^c$ has the properties $k^ba_b = k^b\hat{a}_b 0$. From this it follows that $M^0_{ab}k^b = 0$, and that the polarization may
therefore, without loss of generality, be taken to obey $\epsilon_b
k^b = 0$, corresponding to the Lorentz gauge. 
In order to simplify the analysis, it is assumed that the background 
is slowly varying in spacetime compared to the perturbation. 
Using $a_b$ and $\hat{a}_b$ as the polarization eigenvectors 
gives two equations 
\begin{equation}\label{eq:equations}
  k^2 = 8\kappa\epsilon_0f_{ab}f^{ac}k^bk_c ,
\,\text{ and }\, 
  k^2 = 14\kappa\epsilon_0f_{ab}f^{ac}k^bk_c ,
\end{equation}
from Eq.\ (\ref{eq:algebraic}) for the polarization $a^a$ and $\hat{a}^a$, respectively.  
Since we can decompose $f_{ab} = u_aE_b
-u_bE_a + \epsilon_{abc}B^c$ for an observer with four-velocity $u^a$, 
 we have
\begin{equation}
  a^2 = -(\mathbf{k}\cdot\mathbf{E})^2 - c^2(\mathbf{k}\cdot\mathbf{B})^2 +
  \omega^2\mathbf{E}^2 + c^2\mathbf{k}^2\mathbf{B}^2 -
  2c\omega\mathbf{k}\cdot(\mathbf{E}\times\mathbf{B}) 
\end{equation}
and $ \hat{a}^2 \approx a^2$. 
Equations (\ref{eq:equations}) can be written in the clear 
and compact form
\cite{Bialynicka-Birula} 
\begin{equation}
  \omega \approx c|\mathbf{k}|(1 -
  \tfrac{1}{2}\lambda|\mathbf{Q}|^2 ) ,  
\label{eq:HE-disprel}
\end{equation}
where $\lambda$ is $8\kappa$ or $14\kappa$,
respectively, depending on the polarization state of the photon, and 
\begin{eqnarray}
  &&\!\!\!\!\!\!\!\! |\mathbf{Q}|^2 \equiv \epsilon_0|\hat{\mathbf{k}}\times\mathbf{E} +
   c\hat{\mathbf{k}}\times(\hat{\mathbf{k}}\times\mathbf{B})|^2 
\label{eq:Q2}
\\ &&\!\!\!\!\!\!\!\!
   =
   \epsilon_0\Big[ \mathbf{E}^2 + c^2\mathbf{B}^2 -
   (\hat{\mathbf{k}}\cdot\mathbf{E})^2 - c^2(\hat{\mathbf{k}}\cdot\mathbf{B})^2
   - 2c\hat{\mathbf{k}}\cdot(\mathbf{E}\times\mathbf{B}) \Big] . 
\nonumber
\end{eqnarray}
Here, the hat denotes the unit vector. 
The expression (\ref{eq:HE-disprel}) is valid for arbitrary, slowly varying background fields.
It is straightforward to show that expression (\ref{eq:Q2}) 
vanishes in the case of a self-interacting plane wave field.
The two different possible polarization directions can be given 
in a similar manner (see \citet{Bialynicka-Birula}).

\subsection{Corrections due to rapidly varying fields}

As we saw in the previous section, it is possible to derive
a dispersion relation for photons moving in a given background 
field. However, the field variations were neglected, and in order
to take them into account, a modified weak field Lagrangian 
must be used.

It is well-known that the weak field theory of photon--photon
scattering can be formulated using the effective Lagrangian density
  $\msl = \msl_0 + \msl_{\text{HE}} + \msl_D$,
where $\msl_0$ is the classical free field Lagrangian $\msl_{\text{HE}}$ 
is the Heisenberg--Euler correction as given in Eq.\ (\ref{eq:lagrangian1}),
The derivative corrections are given by \cite{Mamaev-etal}
\begin{equation}
  \msl_D = \sigma\epsilon_0\left[ (\partial_aF^{ab})(\partial_cF^{c}\!_b)
  - F_{ab}\square F^{ab}\right] ,
\end{equation}
where $\square = \partial_a\partial^a$, and 
$\sigma = (2/15)\alpha c^2/\omega_e^2 
\approx 1.4\times10^{-28}\,\mathrm{m}^2$. 
As we have seen in the dispersion relation (\ref{eq:HE-disprel}), 
the parameter $\kappa$ gives the nonlinear coupling. Here, we find that  
the parameter $\sigma$ gives the dispersive
effects in the polarized vacuum.  
Physically, setting the parameter $\sigma \neq 0$ corresponds to 
to taking correction due to rapidly varying perturbation into account. 
Since the Compton frequency is $\sim 10^{20} \, \mathrm{rad/s}$, we 
see that  $\omega/\omega_e \ll 1$ in most applications. Thus, the dispersive term is 
normally a small correction. 

In the previous section, the requirement (\ref{eq:constraint1}) was assumed 
to be satisfied. Here, even though we include effects of the 
rapidly varying fields, we require that there is no electron--positron pair 
creation, neither by single
photons nor by collective effects, i.e. the conditions
(\ref{eq:constraint1}) should still hold. Furthermore, 
the dispersive/diffractive effects must be small, otherwise the limit of
weak fields would  imply unphysical branches in the dispersion
relation \cite{Rozanov}.

As in the previous section, we set $F_{ab} = \partial_aA_b - \partial_bA_a$, 
and obtain the field equations from the Euler--Lagrange equations 
$\partial_b[\partial\msl/\partial F_{ab}] = 0$. Thus, we have 
\cite{Rozanov,Shukla-Marklund-Tskhakaya-Eliasson}
\begin{equation}
  \tfrac{1}{2}(1 + 2\sigma\square)\partial_aF^{ab} = 
   \epsilon_0\kappa\partial_a\left[ (F_{cd}F^{cd})F^{ab} +
   \tfrac{7}{4}(F_{cd}\widehat{F}^{cd})\widehat{F}^{ab} \right] .
\label{eq:exact-evol}
\end{equation}
Equation (\ref{eq:exact-evol}) describes the nonlinear evolution of
the electromagnetic field through the nonlinear dispersive vacuum. We
note that when $\sigma, \kappa \rightarrow 0$, we obtain 
the classical Maxwell's equations, as we should.  

Repeating the procedure leading up to Eq.\ (\ref{eq:algebraic}), we
find the corresponding expression in the case of a dispersive
vacuum  
\begin{eqnarray}
  M_{ab} \epsilon^b = \left[ M^0_{ab} - 2\sigma k^2(k^2g_{ab} - k_ak_b) \right]
  \epsilon^b = 0, 
  \label{eq:algebraic2}
\end{eqnarray}
where $M^0_{ab}$ is given by Eq.\ (\ref{eq:algebraic}). 
With $a_b$ and $\hat{a}_b$ as the principal polarization
direction, Eq.\ (\ref{eq:algebraic2}) yields  
\begin{subequations}
\begin{equation}
  (1 - 2\sigma k^2)k^2 = 8\kappa\epsilon_0f_{ab}f^{ac}k^bk_c ,
\end{equation}
and 
\begin{equation}
  (1 - 2\sigma k^2)k^2 = 14\kappa\epsilon_0f_{ab}f^{ac}k^bk_c ,
\end{equation}
\label{eq:equations2}
\end{subequations}
for the two different polarizations $a_b$ and $\hat{a}_b$, respectively. 
When $\sigma = 0$, Eqs.\ (\ref{eq:equations2}) reduces to Eqs.\ (\ref{eq:equations}), which 
yield the dispersion 
relation (\ref{eq:HE-disprel}). With $\sigma \neq 0$,
we may use Eq.\ (\ref{eq:HE-disprel}) in the dispersive term
of Eqs.\ (\ref{eq:equations2}). Thus, the final
dispersion relation is of the form \cite{Rozanov,Shukla-Marklund-Tskhakaya-Eliasson}
\begin{equation}
  \omega \approx c|\mathbf{k}|\Big[ 1 - \tfrac{1}{2}\lambda|\mathbf{Q}|^2
  \left( 1 + 2\sigma\lambda|\mathbf{Q}|^2|\mathbf{k}|^2 \right) \Big] ,
\label{eq:gen-disprel}
\end{equation}
where $|\mathbf{Q}|^2$ is given by Eq.\ (\ref{eq:Q2}). 
Thus, we see that the effect of the dispersive parameter is, as expected, to make 
$\omega$ a nonlinear function of $\mathbf{k}$. 

\subsection{Special cases of weak field dispersion}

\subsubsection{Magnetized background}\label{sec:weakmagnetic}

In the case of a background magnetic field $\mathbf{B}_0$, 
the dispersion relation (\ref{eq:HE-disprel}) becomes \cite{Erber,Adler,Adler-etal,%
Adler-Shubert,Dittrich-Gies2}
\begin{equation}
  \omega \approx c|\mathbf{k}|\left( 1 
    - \tfrac{1}{2}\lambda\epsilon_0c^2|\mathbf{B}_0|^2%
    \sin^2\theta \right) ,
    \label{eq:magnetized}
\end{equation}
where $\theta$ is the angle between the background magnetic field 
$\mathbf{B}_0$ and the wavevector $\mathbf{k}$. Thus, photons propagating
parallel to the background magnetic field will not experience any 
refractive effects, while a maximum refraction is obtained for
perpendicular propagation. This dispersion relation will be relevant
for photon propagation in pulsar magnetospheres and in magnetar environments,
where for example the vacuum becomes birefringent \cite{Heyl-Hernquist,Tsai-Erber}.
This in turn may affect the optical depth of neutron star thermal emission,
and thereby also influence the interpretation of pulsar observations 
\cite{Lodenqual-etal,Ventura,Heyl-Shaviv,Heyl-etal}.

For a magnetar with surface field strength $|\mathbf{B}_0| = 10^{11}\,\mathrm{T}$
\cite{magnetar}, 
the field has the energy density $\epsilon_0c^2|\mathbf{B}_0|^2 \approx 
8\times 10^{27}\,\mathrm{J/m^3}$. 
Since $\lambda \sim 10\kappa \approx 1/(3\times 10^{28}\,\mathrm{J/m^3})$ 
(see Eqs.\ (\ref{eq:parameter}) and (\ref{eq:HE-disprel})), we find from 
(\ref{eq:magnetized}) that the phase velocity $v = \omega/|\mathbf{k}|$
satisfies $v/c \approx 1 - 0.13\sin^2\theta$. However, 
the magnetar field strength does not qualify as a weak field, and one may
question if it is appropriate to use Eq.\ (\ref{eq:magnetized}) in this case 
(see Sec.\ \ref{sec:strongmagnetic}).

\subsubsection{Random photons in a magnetic field}

From the previous example, we saw that the effect of a magnetic field
on the thermal or random distribution of photons could be observationally important.  
Thus, for a random ensemble of photons 
in a strong magnetic field we have a direction independent dispersion 
relation
\begin{equation}
  \omega \approx c|\mathbf{k}|\left( 1 
    - \tfrac{1}{3}\lambda\epsilon_0c^2|\mathbf{B}_0|^2 \right) .
\end{equation}
We note that the value of the effective action charge $\lambda$ 
still depends on the polarization of the thermal photons.
As the crust of magnetars are subject to enormous stresses due to 
the immense field strengths ($\sim 10^{10} - 10^{11}\,\mathrm{T}$ \cite{magnetar}), 
it will 
suffer from crust quakes, at which bursts of low-frequency random photons
are released \cite{Kondratyev}. In such a scenario, the above dispersion 
relation may be of relevance.

\subsubsection{Random photons in a plane wave field}

Analogously, we may treat the case of incoherent photons on an
intense plane wave background $\mathbf{E}_p$. Then, in the equilibrium state of the
radiation gas, the propagation directions of the photons in the gas
are random and the EM pulse is a superposition
of uni-directional plane waves such that $\mathbf{B}_p \hat{\mathbf{k}}_p\times\mathbf{E}_p/c$.
Thus, we obtain the direction independent dispersion 
relation
\begin{equation}
  \omega \approx c|\mathbf{k}|\left( 1 
    - \tfrac{2}{3}\lambda\epsilon_0|\mathbf{E}_p|^2 \right) .
    \label{eq:planewave}
\end{equation}
Equation (\ref{eq:planewave}) is the proper dispersion relation to use in some laser-plasma 
interaction applications, where a large number of incoherent photons
are produced \cite{bob,bob2,cai04}.

\subsubsection{Radiation gas background}

Consider a single photon transversing a dense 
background radiation gas with energy density $\mathscr{E}$. Then the 
dispersion relation can be written as \cite{Marklund-Brodin-Stenflo,Dittrich-Gies}
\begin{equation}
  \omega \approx c|\mathbf{k}|\left( 1 
    - \tfrac{2}{3}\lambda\mathscr{E} \right) .
    \label{eq:weakgas}
\end{equation}
As one considers higher redshifts $z$, the cosmic microwave background will 
increase in energy density, since $\mathscr{E}(z) = (1 + z)^4\mathscr{E}_0$ 
(we note that radiation decouples from matter at a redshift $\sim 10^3$ \cite{Peacock}), 
where
$\mathscr{E}_0 = a T_0^4$ denotes the current energy density,
and $a = 8\pi^5k_B^4/15h^3c^3 \approx 7.6\times 10^{-16}\,\mathrm{J/m^3K^4}$ 
is the radiation constant. Thus, Eq.\ (\ref{eq:weakgas}) gives
  $v/c = 1 - (2/3)\lambda a(1 + z)^4T_0^4 $,
for the phase velocity $v$. Using $T_0 = 2.7\,\mathrm{K}$, a correction of 
10\,\% to the phase velocity in vacuum is obtained for 
a redshift $z_c \sim 10^{10}$, i.e. roughly at the time for neutrino--matter
decoupling \cite{Peacock}.
 
Including the dispersive correction, as given by Eq.\ (\ref{eq:gen-disprel}),
the dispersion relation for a background of incoherent photons takes the form
\begin{equation}
  \omega \approx c|\mathbf{k}|\left[ 1 
    - \tfrac{2}{3}\lambda\mathscr{E}\left( 1 + 
    \tfrac{8}{3}\sigma\lambda\mathscr{E}|\mathbf{k}|^2\right) \right] 
    \label{eq:weakgas-disp}
\end{equation}
i.e. high-frequency pulses may suffer spectral dilution
when propagating through a radiation gas. 

\subsubsection{Other field configurations}

Similar dispersion relations can be found for other background 
field configurations, e.g. plane wave backgrounds or 
partially coherent electromagnetic fields, as is relevant in ultra-high
intensity laser applications. 
Of special interest for detection purposes 
is the configuration of photon propagation perpendicular to a
collection of constant electric and magnetic fields
\cite{Bakalov-etal,Rikken-Rizzo1,Rikken-Rizzo2} 

We should also note that in all the cases above the 
group and phase velocities of the test photons are
subluminal, as expected, since we have excited the 
quantum vacuum by using electromagnetic fields,
analogous to a normal dispersive material medium.
This can be contrasted with the Scharnhorst effect \cite{Scharnhorst,Barton,%
Barton-Scharnhorst,Scharnhorst2}, for which 
we obtain superluminal phase and group velocities between
two conducting plates. This can be traced back to the
Casimir effect \cite{Casimir1,Casimir2}, where the quantization between
two conducting plates allows fewer states than for field
with boundary conditions at infinity. Thus, in this sense, the vacuum
between the plates is `emptier' than outside, giving rise to
superluminal velocity.

\subsection{Ultra-intense fields}

The dispersion relations treated so far have used the weak field expansion 
of the general Heisenberg--Euler correction (\ref{eq:lagrangian2}). However,
both from an application point of view and due to theoretical issues, the 
inclusion of fully nonlinear vacuum effects deserve attention 
(see also \citet{Dittrich-Gies} for a thorough discussion
of the strong magnetic field case).

Within astrophysical 
and cosmological settings, such as neutron stars and magnetars
\cite{magnetar},  
the strong field conditions can be met. Even in
laboratory environments, such conditions could be encountered in
future high energy laser configurations. While today's lasers can produce $10^{21}-10^{22}$
W/cm$^2$ \cite{mou98}, it is expected that the next generation laser-plasma
systems could reach $10^{25}$ W/cm$^2$ 
\cite{cai04,bob,bob2}, where field
strengths close to the Schwinger critical value could be reached
\cite{Bulanov-etal}. Thus, nonlinear effects
introduced by photon--photon scattering will be significant, and the
weak field approximation no longer holds. In terms of Feynman diagrams
the discussion to follow will consider the full one-loop correction
\begin{equation}
\includegraphics[width=0.05\columnwidth]{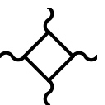}\, {}^{\displaystyle{+}} \, 
\includegraphics[width=0.05\columnwidth]{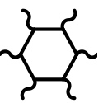}\, {}^{\displaystyle{+}} \,
\includegraphics[width=0.05\columnwidth]{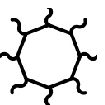}\, {}^{\displaystyle{+}} \,
\includegraphics[width=0.05\columnwidth]{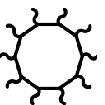}\, {}^{\displaystyle{+ \ldots}}
\end{equation}

The general dispersion relation for a test photon in a vacuum dressed by 
a strong electromagnetic field is given by \cite{Dittrich-Gies2}
\begin{eqnarray}
 && \!\!\!\!\!\!\!\!\!\!\!\!\!\!\!\!
  \left[1 + \tfrac{1}{2}{\lambda}\epsilon_0(\mathbf{E}^2 + c^2\mathbf{B}^2)\right]\frac{v^2}{c^2} -
  2{\lambda}\epsilon_0c\hat{\mathbf{k}}\cdot(\mathbf{E}\times\mathbf{B})\frac{v}{c}
    \nonumber \\ && \!\!\!\!\!\!\!\!\!\!\!\!\!\!\!\!
  + {\lambda}\epsilon_0\left[ \tfrac{1}{2}(\mathbf{E}^2 + c^2\mathbf{B}^2) -
  (\hat{\mathbf{k}}\cdot\mathbf{E})^2 - 
  c^2(\hat{\mathbf{k}}\cdot\mathbf{B})^2\right]  = 1  ,
\label{eq:disprel-full}
\end{eqnarray}
where $v = \omega/|\mathbf{k}| \equiv c/n$ is the photon phase velocity, and 
$n$ is the refractive index. The effective action charge ${\lambda}$ is no longer a 
constant, but instead defined through
\cite{Dittrich-Gies2} 
\begin{equation}
  {\lambda} = \frac{1}{\epsilon_0}\frac{(\partial^2_{\msf} +
  \partial^2_{\msg})\msl}{-2\partial_{\msf}\msl +
  \msf(\partial^2_{\msf} + \partial^2_{\msg})\msl -
  2(\msf\partial^2_{\msf} + \msg\partial^2_{\msf\msg})\msl} .
\label{eq:Q}
\end{equation}
In many cases, the part of the denominator of (\ref{eq:Q}) stemming
from the nonlinear QED correction (\ref{eq:lagrangian2}) can be neglected,
since it will be much smaller than the remaining terms, even for 
fields $\gg E_{\text{crit}}$.

\subsubsection{Pure magnetic field}\label{sec:strongmagnetic}

The case of pure magnetic fields 
enables a simplification of the evaluation
of the Lagrangian (\ref{eq:lagrangian2}). The refractive index for the
strongly magnetized vacuum can be determined in terms of special functions. 
For a pure magnetic field $\mathbf{B}_0$, we immediately obtain using 
Eq.\ (\ref{eq:ab}) that $a = c|\mathbf{B}_0|$ and $b = 0$, respectively.
\citet{Dittrich-Gies2} (see also \citet{Dittrich-Gies}), starting from the Lagrangian 
(\ref{eq:lagrangian2}), devised the general expression (\ref{eq:Q}) for the effective action
charge $\lambda$ (as also can be found in the work of, e.g. \citet{Bialynicka-Birula}). 
 
Using $a = c|\mathbf{B}_0|$ and $b = 0$ and the approximation 
$\lambda \approx (\partial^2_{\msf} + \partial^2_{\msg})\msl/2\epsilon_0$ 
(see Eq.\ (\ref{eq:Q})), the effective action charge takes the form
\begin{eqnarray}
  \lambda = \frac{\alpha}{2\pi\epsilon_0c^2|\mathbf{B}_0|^2}\Bigg[ 
  \left( 2x^2 - \tfrac{1}{3}\right)\psi(1 + x)  - x - 3x^2 
\nonumber \\ 
  - 4x\ln\Gamma(x)
  + 2x\ln 2\pi + \frac{1}{6} + 4\zeta^{\prime}(-1, 4x) + \frac{1}{6x} 
  \Bigg]  , 
\label{eq:exact}
\end{eqnarray} 
where $x = E_{\text{crit}}/2c|\mathbf{B}_0|$, $\psi$ is the logarithmic derivative of the 
$\Gamma$ function, and $\zeta^{\prime}$ is the derivative of the Hurwitz zeta 
function with respect to the first index. Thus, the refractive index, given in 
Eq. (\ref{eq:disprel-full}), becomes
\begin{equation}
  n^{-2} \approx 1 - \lambda\epsilon_0c^2|\mathbf{B}_0|^2\sin^2\theta
  \geq 0, 
\end{equation}
where higher order terms in the fine structure constant $\alpha$ has been neglected,
and $\theta$ is the angle between the background magnetic field $\mathbf{B}_0$ and
the wavevector $\mathbf{k}$. 

The refractive effects of a super-strong magnetic field is of interest in neutron stars 
and in magnetar environments, since they generate extreme conditions in terms
of the field strength. Comparing with the case presented in Sec.\ 
\ref{sec:weakmagnetic} we see that given the magnetic field strength 
$|\mathbf{B}_0| \sim 10^{11}\,\mathrm{T}$, we have $\lambda\epsilon_0c^2|\mathbf{B}_0|^2 
\approx 15\alpha/\pi \approx 0.03$. Thus, the effect of the magnetic field 
on the refractive index is weaker than predicted by the lowest order calculation.

\subsubsection{Crossed field background}

For a crossed field configuration, i.e.    
$|\mathbf{E}| = c|\mathbf{B}|$ and $\mathbf{E}\cdot\mathbf{B} = 0$, it immediately 
follows that  $a = b = 0$. For a
test photon belonging to an ensemble of random photons (such as in a 
photon gas), we obtain \cite{Marklund-Shukla-Eliasson} 
\begin{equation}
 n = \left(\frac{1 + {\lambda}\epsilon_0|\mathbf{E}|^2}%
{1 - \tfrac{1}{3}{\lambda}\epsilon_0|\mathbf{E}|^2} \right)^{1/2} ,  
\label{eq:refractive1}
\end{equation}
from Eq. (\ref{eq:disprel-full}), 
where the effective action charge in (\ref{eq:Q}) is a constant due to the random 
properties of the test photons, 
  ${\lambda}^{-1}  = 
    ({45}/{22})({4\pi}/{\alpha})\epsilon_0{E}_{\text{crit}}^2 $.
Note that this is the same charge as the 
geometrical average of the coefficient obtained from the polarization 
tensor in the weak field limit \cite{Bialynicka-Birula}, 
i.e. an average over polarization states. The refractive 
index diverges as $\epsilon_0|\mathbf{E}|^2 \rightarrow 3{\lambda}^{-1}$. As these field 
strengths are reached, it is not correct that the test radiation gas is in 
thermodynamical equilibrium, and the assumptions behind the derivation of
(\ref{eq:refractive1}) are no longer valid.

\subsubsection{Incoherent radiation background}

We may characterize single photons in terms of plane electromagnetic waves. Thus, 
for an electromagnetic wave moving in an isotropic and homogeneous medium
with the refractive index $n$ we have $|\mathbf{B}| = n|\mathbf{E}|/c$, 
$\msg = - c\mathbf{E}\cdot\mathbf{B}$ = 0, and 
$\msf = \tfrac{1}{2}(n^{2} - 1)|\mathbf{E}|^2 \geq 0$. Here we have assumed
that $n > 1$, which implies $a = [(n^2 - 1)\mathscr{E}/\epsilon_0]^{1/2}
\neq 0$ and $b = 0$, while for superluminal velocities, we have 
$a = 0$ and $b \neq 0$, which allows for
spontaneous pair production \cite{Schwinger}. 
The $n > 1$ assumption is consistent with elastic photon--photon scattering. 
The vacuum is now treated fully nonlinearly so we have to 
take into account the backreaction of the random photons onto themselves, 
an interaction mediated by the refractive index.
The effective action charge (\ref{eq:Q}) will therefore depend on both the field 
strength and the refractive index. Since the refractive index itself  
depends on the effective action charge ${\lambda}$, it will be nonlinearly determined 
via (\ref{eq:disprel-full}). 
For incoherent photons the Poynting flux in the gas rest frame vanishes, and we may uniquely 
characterize the gas by its energy density $\mathscr{E}$. 
From (\ref{eq:disprel-full}) we then obtain \cite{Marklund-Shukla-Eliasson}
\begin{equation}
  \frac{1}{n^2} = \frac{1 - \tfrac{2}{3}{\lambda} \mathscr{E} + \sqrt{1 -
  2{\lambda}\mathscr{E} +
  \tfrac{1}{9}({\lambda}\mathscr{E})^2}}{2 + 
  {\lambda}\mathscr{E}} ,
\label{eq:refractive2} 
\end{equation}
while the effective action charge takes the form 
\begin{equation}
  {\lambda} = \frac{\alpha}{4\pi\epsilon_0
  a^2}\,\frac{F(a/E_{\text{crit}})}{2 +  
  (\alpha/8\pi)[ F(a/E_{\text{crit}}) +
  G(a/E_{\text{crit}}) ]} . 
  \label{eq:Qgas}
\end{equation}
Here 
\begin{eqnarray}
  &&
  F(a/E_{\text{crit}}) = 
  \frac{4\pi}{\alpha\epsilon_0}a^2\lim_{b\rightarrow 0}\,(\partial^2_{\msf} + \partial^2_{\msg})\msl 
\label{eq:FG} \\  && 
   =  \frac{1}{4\pi}
  \int_0^{\mathrm{i}\infty} \frac{dz}{z}\,\mathrm{e}^{-E_{\text{crit}}
  z/a} \left( \frac{1 - z\coth z}{\sinh^2 z} + \frac{1}{3}z\coth z 
  \right) ,
\nonumber 
\end{eqnarray}
and 
\begin{equation}
 G(a/E_{\text{crit}}) = \frac{8\pi}{\alpha\epsilon_0}
 \lim_{b\rightarrow 0}\,[ -2\partial_{\msf}\msl_c
  -a^2\partial^2_{\msf}\msl_c ]  
   .
\end{equation} 
Note that the latter function only gives a small correction to the effective action charge, 
and can in most cases safely be neglected. Moreover, the function $F$ may be expressed
in terms of special functions, see Eq. (\ref{eq:exact}) \cite{Dittrich-Gies}.

\begin{figure}
  \includegraphics[width=.7\columnwidth]{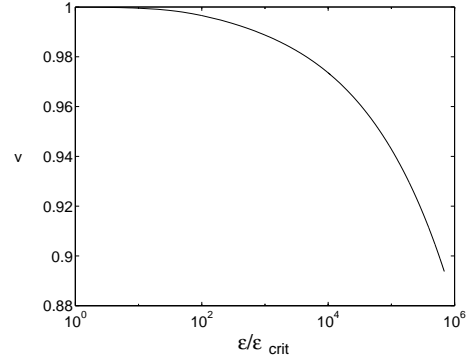}
  \caption{The phase velocity $v$ in units of $c$, as given by Eqs.\
    (\ref{eq:refractive2}) and (\ref{eq:Qgas}), plotted as a function of the
    normalised energy density $\mathscr{E}/\mathscr{E}_{\text{crit}}$
(reprinted with permission from \citet{Marklund-Shukla-Eliasson}).}
\label{fig:velocity}
\end{figure}

The weak field limit of (\ref{eq:refractive2}) and (\ref{eq:Qgas}) takes the form 
(\ref{eq:weakgas}) 
\cite{Marklund-Brodin-Stenflo,Dittrich-Gies2,Bialynicka-Birula}. On
the other hand, in the ultra-strong field limit 
($ \mathscr{E} / \epsilon_0{E}_{\text{crit}}^2 
\rightarrow \infty$), we obtain the asymptotic constant phase velocity 
\cite{Marklund-Shukla-Eliasson}
\begin{equation}
v_{\infty} = c/\sqrt{5} \approx 0.45 c ,
\label{eq:constant}
\end{equation}
valid in the low frequency approximation. 
Thus, for very high radiation densities, we expect the phase and group velocities to 
be approximately half that of the speed of light in vacuum.  
In this limit, the one-loop radiation gas
essentially evolves by free streaming, and it is therefore likely
that  higher-order loop corrections are important \cite{Ritus}. 
The phase velocity as a function of intensity is depicted in Fig. \ref{fig:velocity}.

Under most circumstances, the contribution proportional to $\alpha$ in 
the denominator of Eq.\ (\ref{eq:Qgas}) is small and may be
neglected. When $\mathscr{E} \geq \epsilon_0{E}_{\text{crit}}^2$, 
we obtain $F \approx a/3E_{\text{crit}}$ from Eq.\ (\ref{eq:FG}), 
and we have  \cite{Marklund-Shukla-Eliasson}
\begin{equation}
  {\lambda} 
  \approx \frac{\alpha}{8\pi \epsilon_0a^2}F(a/E_{\text{crit}}) 
  \approx \frac{\alpha}{24\pi \epsilon_0^{1/2}E_{\text{crit}}} \frac{1}{\sqrt{n^2 -
  1}} \frac{1}{\sqrt{\mathscr{E}}} \ .
\label{eq:approxlambda}
\end{equation} 
This expression for the effective action charge can be used in conjunction 
with (\ref{eq:refractive2}) to analyse the propagation of single photons
in a radiation gas where the energy density is in the intermediate range.

\section{Nonlinear collective photon interactions}

In the preceding section, we presented the dispersion relations 
for single photons interacting nonlinearly with the vacuum, for 
a number of special cases. The single-photon picture contains
many interesting physical phenomena, such as photon splitting and
the birefringence of the quantum vacuum. However, in many
applications, collective effects among photons may be dominant \cite{Mendonca}. 

\subsection{Coherent field interactions}

The formulation of the interaction between coherent electromagnetic 
waves and possible background field configurations is a starting 
point for the discussion concerning possible detection techniques
of elastic photon--photon collisions. Experiments for detecting 
elastic photon--photon scattering are important tests of QED, and 
constitute a new type of  
tests of, e.g. Lorentz invariance in extensions of the Standard Model 
such as supersymmetric field theories 
\cite{Colloday-Kostelecky,Jackiw-Kostelecky,Lipa-etal,Nibbelink-Pospelov}.

The Maxwell equations that results from the weak field 
Heisenberg--Euler corrected electromagnetic Lagrangian 
(\ref{eq:lagrangian1}) are
\begin{subequations}
\begin{eqnarray}
  \nabla\cdot\mathbf{E} &=& (\rho - \nabla\cdot\mathbf{P})/\epsilon_0 , \\
  \nabla\cdot\mathbf{B} &=& 0 , \\
  \frac{\partial\mathbf{B}}{\partial t} + \nabla\times\mathbf{E} &=& 0 ,\\
  \frac{1}{c^2}\frac{\partial\mathbf{E}}{\partial t} - \nabla\times\mathbf{B} 
  &=& -\mu_0\left( \mathbf{j}  + \frac{\partial\mathbf{P}}{\partial t} 
    + \nabla\times\mathbf{M} \right)\!\! , 
\end{eqnarray}
\label{eq:maxwell}
\end{subequations}
where the vacuum polarization and magnetization are given by the expressions 
(\ref{eq:polarization}) and (\ref{eq:magnetization}), respectively, and $\rho$ and
$\mathbf{j}$ are the charge and current densities, respectively. From these, it is 
straightforward to derive the wave equations (\ref{WaveE}) and 
(\ref{WaveB}). These may in turn be used to derive the dispersion
function for the appropriate wave field on a given background.

As noted above, the interaction between waves in parallel propagation does
not yield any interaction due to the vacuum dispersion function $D_{\text{vac}} 
= \omega^2 - |\mathbf{k}|^2c^2$ appearing as an overall factor in the wave
equations (\ref{WaveE}) and (\ref{WaveB}). Thus, 
dispersive effects need to be introduced in the wave propagation. 
This can be done in a multitude of ways, such as crossing 
light beams \cite{Soljacic-Segev}, photon propagation on a constant 
coherent field background \cite{Rozanov93,Rozanov,Ding,Ding-Kaplan,%
Kaplan-Ding,Bakalov-etal,Rikken-Rizzo1,Rikken-Rizzo2},\footnote{%
  Here it should be noted that the paper by \citet{Bakalov-etal} is 
  a progress report to one of the few actual experimental setups 
  within photon--photon scattering, and their detection techniques 
  are based on the work by \citet{Iacopini-Zavattini} and \citet{Bakalov-etal2}.} 
  cavity fields \cite{Brodin-Marklund-Stenflo,%
Brodin-Marklund-Stenflo2}, waveguide propagation \cite{Brodin-etal,%
Shukla-Eliasson-Marklund,Shen-Yu,Shen-etal}, plasma interactions 
\cite{Shen-Yu,Shen-etal,Stenflo-etal,Marklund-Shukla-Stenflo-Brodin-Servin,%
Marklund-Shukla-Brodin-StenfloC}, and interaction between coherent and  
incoherent photons \cite{Marklund-Brodin-Stenflo,Shukla-Eliasson,%
Shukla-Marklund-Tskhakaya-Eliasson,r1,Marklund-Shukla-Brodin-Stenflo,%
Marklund-Shukla-Brodin-Stenflo2,Shukla-Marklund-Brodin-Stenflo}.

\subsubsection{Nonlinear vacuum magneto-optics}

We have seen that the propagation of photons on a magnetized background can 
be expressed according to the dispersion relation (\ref{eq:magnetized}).
We may also start from the constituent relations (\ref{eq:constituent}) 
together with the expressions (\ref{eq:polarization}) and (\ref{eq:magnetization}) 
for the polarization and magnetization, respectively. This was first done by 
\citet{Klein-Nigam1,Klein-Nigam2}, and later for arbitrary intensities by 
\citet{Heyl-Hernquist}.
Denote the slowly varying background 
magnetic field by $\mathbf{B}_0$, and the perturbation 
fields by $\mathbf{E}$ and $\mathbf{B}$. Then, we find that  
$D_i = \epsilon_{ij}E_j$ and $H_i = \mu_{ij}B_j$, where the quantum 
vacuum electric permittivity and magnetic permeability  are 
given by
\begin{eqnarray}
  && \epsilon_{ij} = \epsilon_0\left[ \delta_{ij} 
  + 4\kappa\mathscr{B}_0\left(-\delta_{ij} + \tfrac{7}{2}b_ib_j  \right) \right] , \\
  && \mu_{ij} = \mu_0\left[ \delta_{ij} 
  + 4\kappa\mathscr{B}_0\left( \delta_{ij} + 2b_ib_j \right)\right] ,
\end{eqnarray}
respectively. Here, we have introduced the background magnetic field energy density
$\mathscr{B}_0 = |\mathbf{B}_0|^2/\mu_0$ and the background magnetic field direction
$\mathbf{b} = \mathbf{B}_0/|\mathbf{B}_0|$. Thus, the permittivity and permeability 
are diagonal when using the magneto-optical axis as eigen-direction. Denoting this 
direction by $z$, we have $\epsilon_{xx} = \epsilon_{yy} 
= \epsilon_0(1 - 4\kappa\mathscr{B}_0)$ and $\epsilon_{zz} 
= \epsilon_0(1 + 10\kappa\mathscr{B}_0)$, while $\mu_{xx} = \mu_{yy}
= \mu_0(1 + 4\kappa\mathscr{B}_0)$ and $\mu_{zz} 
= \mu_0(1 + 12\kappa\mathscr{B}_0)$.

For an electromagnetic wave propagating 
perpendicular to $\mathbf{B}_0$, there are essentially two different 
polarization states, and we may write $\epsilon_{ij} = \epsilon\delta_{ij}$, 
$\mu_{ij} = \mu\delta_{ij}$ . When $\mathbf{E} \perp \mathbf{B}_0$,  
we have $\epsilon = \epsilon_{\perp}$, $\mu = \mu_{\perp}$ according to
\begin{equation}
  \epsilon_{\perp} = \epsilon_0(1 - 4\kappa\mathscr{B}_0) , \quad\text{ and }\quad
  \mu_{\perp} = \mu_0(1 + 12\kappa\mathscr{B}_0) , 
\end{equation} 
while if $\mathbf{E} \| \mathbf{B}_0$, we find $\epsilon = \epsilon_{\|}$, 
$\mu = \mu_{\|}$, where
\begin{equation}
  \epsilon_{\|} = \epsilon_0(1 + 10\kappa\mathscr{B}_0) , \quad \text{ and }\quad 
  \mu_{\|} = \mu_0(1 + 4\kappa\mathscr{B}_0) .
\end{equation} 
Thus, we see that for strong magnetic fields, there is a significant difference
in the behavior of the two polarization modes. This has been exploited in various
scenarios (e.g. \citet{Bakalov-etal2,Bakalov-etal,Ding,Ding-Kaplan,Kaplan-Ding,%
Rikken-Rizzo1,Rikken-Rizzo2,Heyl-Hernquist}). The procedure is straightforward to 
perform for other uni-directional background field configuration.

\subsubsection{Nonlinear self-interactions}

Based on these results, the self-action of an electromagnetic pulse on a given background 
can also be considered. Taking into account the lowest order cubic 
nonlinear terms of the (complex) pulse amplitude $E$, and employing the slowly varying 
envelope approximation \cite{Hasegawa,Kivshar-Agrawal}, 
one can derive a nonlinear Schr\"odinger equation (NLSE) for $E$.
Letting $E = \int E_k\exp[i(\mathbf{k} - \mathbf{k}_0)\cdot\mathbf{r} 
- i(\omega - \omega_0)t]\,d\mathbf{k}$, and expanding the frequency 
around the background values (denoted by $0$), we obtain
\begin{eqnarray}
  && \!\!\!\!\!\!\!\!
  \omega \approx \omega_0 
  + \left.\frac{\partial\omega}{\partial k_i}\right|_0(k_i - k_{0i}) 
  + \frac{1}{2}\left.\frac{\partial^2\omega}{\partial k_i\partial k_j}\right|_0%
    (k_i - k_{0i})(k_j - k_{0j}) 
  \nonumber \\ && \!\!\!\!\!\!\!\!
  + \left.\frac{\partial\omega}{\partial|\mathbf{Q}|^2}\right|_0%
    (|\mathbf{Q}|^2 - |\mathbf{Q}_0|^2) ,  
\end{eqnarray}
where $|\mathbf{Q}|^2$ is given by Eq. (\ref{eq:Q2}). 
Thus, the envelope will satisfy the NLSE
\begin{equation}
  i\left( \frac{\partial}{\partial t} + v_{gi}\nabla_i\right)E 
  + \frac{1}{2}\frac{\partial v_{gi}}{\partial k_{0j}}\nabla_i\nabla_jE 
  + \frac{1}{2}\lambda k_0c I(|\mathbf{Q}|^2)  = 0 ,
  \label{eq:nlse-gen}
\end{equation}
where $\mathbf{v}_g$ is the group velocity and $I(|\mathbf{Q}|^2) = 
\int E_k(|\mathbf{Q}|^2 - |\mathbf{Q}_0|^2)\exp[i(\mathbf{k} - \mathbf{k}_0)\cdot\mathbf{r} 
- i(\omega - \omega_0)t]\,d\mathbf{k}$ is the nonlinear response. 
We note that the term containing $|\mathbf{Q}_0|^2$ represents
a phase shift, and can be removed by a transformation. 
When high-frequency corrections are added, the NLSE will 
attain a second order derivative along the
propagation direction. 
The group velocity $\mathbf{v}_g = \partial\omega/\partial\mathbf{k}$ 
on an arbitrary background can be written as \cite{Bialynicka-Birula} 
  $\mathbf{v}_g = c\,\hat{\mathbf{k}} - ({c\lambda\epsilon_0}/{2})[
  |\mathbf{E}|^2 + c^2|\mathbf{B}|^2 + (\hat{\mathbf{k}}\cdot\mathbf{E})^2
  + c^2(\hat{\mathbf{k}}\cdot\mathbf{B})^2
  ]\hat{\mathbf{k}} 
  -
  {c\lambda\epsilon_0}[(\hat{\mathbf{k}}\cdot\mathbf{E})\,\mathbf{E} 
  + c^2(\hat{\mathbf{k}}\cdot\mathbf{B})\,\mathbf{B} +
  c\,\mathbf{E}\times\mathbf{B}]$  
in the weak field case. We note that $|\mathbf{v}_g| = \omega/k < c$.
We will now discuss some special cases. 

\paragraph{Constant background fields}

For a constant background configuration, the back reaction of the
photon propagation may, in some cases, be neglected. Then, the nonlinear
contribution to the self-interaction of the pulse will come through
a coupling of higher order in the parameter $\lambda$. 

\citet{Rozanov} considered the perpendicular propagation of high
intensity laser pulses on a background $\mathbf{E}_0 = E_0\hat{\mathbf{x}},
\mathbf{B}_0 = B_0\hat{\mathbf{y}}$. By choosing the polarization directions
of the laser pulse in the direction of the background fields, one obtains 
the NLSE\footnote{%
We note that \cite{Rozanov} obtained a the NLSE (\ref{eq:rozanov}) with
the dispersive correction.} 
\begin{equation}
  i\left( \frac{\partial}{\partial t} + v_g\frac{\partial}{\partial z} \right)E
  + \frac{1}{2}v_g^{\prime}\nabla_{\perp}^2E + \xi|E|^2E = 0
  \label{eq:rozanov}
\end{equation}
where $v_g^{\prime} = c(1 - \lambda|\mathbf{Q}_0|^2/2)/k_0$ 
is the group velocity dispersion, $\xi = k_0c\epsilon_0\lambda^3|\mathbf{Q}_0|^4/8$, and
$|\mathbf{Q}_0|^2 = \epsilon_0(E_0 - cB_0)^2$. 
The NLSE (\ref{eq:rozanov}) could be of interest for laboratory applications, 
when studying high intensity laser pulse propagation in given background 
electromagnetic fields. 

\paragraph{Crossing beams}

When the background is given by the source itself, the nonlinear 
self-interaction term will be of first order in $\lambda$, thus requiring 
weaker background conditions. 

\citet{Soljacic-Segev} derived a NLSE for the dynamics of the envelope $A(x)$ of the
interaction region due to crossing laser beams. By symmetry arguments concerning 
the QED corrected Maxwell's equations, they reduce the problem to a 
1D stationary NLSE
\begin{equation}
  \frac{d^2A}{dx^2} + \Gamma A + \tfrac{1}{2}k^2\kappa\epsilon_0 A^3 = 0 ,
\end{equation} 
where $k$ is the wavenumber of the laser beams, $\Gamma$ 
is the eigenvalue of the equation, and the beams are assumed to be polarized 
in the $x$-direction. The lowest order solitary wave solution
is given by \cite{Kivshar-Agrawal}
  $A(x) = A_0\,\text{sech}\left( \sqrt{-2\Gamma}\,z\right)$ 
where we have denoted the eigenvalue $\Gamma = -k^2\kappa\epsilon_0A_0^2/8$. 
This one-dimensional soliton solution is stable, as opposed to 
higher dimensional solitons \cite{Kivshar-Agrawal,r5}.
Furthermore, \citet{Soljacic-Segev} also suggested the possibility of higher 
dimensional soliton formation, e.g., necklace solitons \cite{Soljacic-etal,Soljacic-Segev2}.

\subsubsection{Propagation between conducting planes}%
\label{sec:planes}

Similar to the case of a Casimir vacuum, one of the simplest geometries 
where dispersive effects makes the presence of QED vacuum
nonlinearities apparent, is given by two parallel conducting planes. 
They are the first
example where multi-dimensional photon configurations can self-compress
to reach intensities above the laser limit \cite{mou98,Brodin-etal,Shukla-Eliasson-Marklund}.  

\paragraph{Variational formulation}

Consider the propagation between two parallel conducting planes with 
spacing $x_0$ of one ${\rm TE}_{\ell0}$-mode ($\ell = 1, 2, ...$) given by  
\begin{equation}\label{eq:TEn0-mode}
  \mathbf{A} = {A}\sin \left( \frac{\ell\pi x}{x_{0}}\right)%
  \exp [{\rm i}(kz-\omega t)]\hat{\mathbf{y}} + {\rm c.c.}
\end{equation}
in the radiation gauge ($\phi =0$). 
The \emph{linear} dispersion relation is $\omega^2/c^2 - k^2 - \ell^2\pi^2/x_0^2 = 0$. 
From Maxwell's equations (\ref{eq:maxwell}) 
a nonlinear dispersion relation can be derived  
by inserting the linear expression for the fields and 
separating into orthogonal trigonometric functions. 
The coefficients in the NLSE can be found from the resulting equation. One
may also start from
the Heisenberg--Euler Lagrangian (\ref{eq:lagrangian1}), and
minimize the resulting expression for the action. This may appear 
more elegant and gives the same result \cite{Brodin-etal}. 

We follow \citet{Brodin-etal}. 
Let ${A} = {A}(t,y,z)$ and assume ${A}$ to be
weakly  
modulated so that $|\partial{A}/\partial t| \ll |\omega{A}|$,
$|\partial{A}/\partial z|  \ll  |k{A}|$. 
To lowest order, the nonlinear terms and the  
slow derivatives in $\mathscr{L}$ are omitted. Averaging over the plate spacing, 
$x_0$, shows that this lowest order Lagrangian is identically zero due to the 
dispersion relation. To the next order of
approximation in the Lagrangian, first order slow derivatives
are included. 
After variation of the corresponding action,  
this leads to an equation where the envelope moves
with the group velocity.  
The next order and final approximation includes second order slow
derivatives. After performing the averaging between the plate inter-spacing, 
the final expression for the Lagrangian is 
\begin{eqnarray}
&&\!\!\!\!\!\!\!\!
{\mathscr{L}} 
= {\rm i}\omega \varepsilon _{0}\left( \frac{\partial
{A}}{\partial t}{A}^{\ast }- 
\frac{\partial {A}^{\ast }}{\partial t}{A}\right) -
{\rm i}kc^{2}\varepsilon _{0}\left( \frac{\partial {A}}{\partial 
z}{A}^{\ast }- 
\frac{\partial {A}^{\ast }}{\partial z}{A}\right)
\nonumber \\
 && \qquad 
 + (c^2 - v_g^2)\varepsilon_0\left| \frac{\partial {A}}{\partial z}\right|^2 + 
\frac{3\ell^4c^4\pi^4\epsilon_0^2\kappa}{x_0^4}|{A}|^{4} .
\label{eq:averagelagrangian}
\end{eqnarray}
Variation of the action due to the Lagrangian (\ref{eq:averagelagrangian}) 
with respect to  ${A}^*$ leads to the NLSE
\begin{equation}
  i \left( \frac{\partial }{\partial t} + v_{g}\frac{\partial }{\partial z}\right)A   
+ \frac{c^{2}}{2\omega}\frac{\partial ^{2}A}{\partial y^{2}}+
\frac{v_{g}}{2}^{\prime }\frac{\partial ^{2}A}{\partial z^{2}}
+ L^2|A|^{2}A 
= 0  ,  
\label{eq:nls2} 
\end{equation}
where $v_{g}$ and $v_{g}^{\prime } = \partial v_g/\partial k$ follow from the linear 
dispersion relation, and $L^2 = (3\ell^{4}c^{4}\pi^{4}\kappa\epsilon_0)/(\omega x_{0}^{4})$. 
The nonlinear correction in Eq.\ (\ref{eq:nls2}) is due to the self-interaction
of the $\mathrm{TE}_{\ell0}$-mode (\ref{eq:TEn0-mode}) via the quantum vacuum.
In one space dimension, i.e. $\partial^2 A/
\partial y^2 = 0$, Eq. (\ref{eq:nls2}) reduces to the cubic Schr\"odinger equation which admits
an envelope soliton solution \cite{Kivshar-Agrawal}.

Changing to a system moving with the
group velocity while rescaling the coordinates and the amplitude according
to $
    \tau  = \omega t/2$, $\upsilon =\,\omega y/c$, 
    $\zeta = (\omega/v_{g}^{\prime })^{1/2}(z-v_{g}t)$, and $a  = \sqrt{2}\,LA$,  
and assuming cylindrical symmetry, Eq. (\ref{eq:nls2}) takes the form
\begin{equation}
{\rm i}\frac{\partial a}{\partial \tau }+
\frac{1}{\rho }\frac{\partial }{\partial \rho }
\left( \rho \frac{\partial a}{\partial \rho }\right)
+|a|^{2}a=0 ,  \label{eq:nls3}
\end{equation}
where 
$a=a(t,\rho )$, and $\rho ^{2}=\upsilon ^{2}+\zeta ^{2}$.

Equation (\ref{eq:nls3}) is a 2-dimensional radially symmetric NLSE, to which 
exact solutions are not available. However, an accurate analytical 
approximation of the dynamics of the pulse-like solutions  
of Eq.\ (\ref{eq:nls3}) can be obtained by means of Rayleigh--Ritz optimization 
based on suitably chosen trial 
functions (see e.g.\ \citet{Desaix-Anderson-Lisak,Anderson-Cattani-Lisak} and 
references therein).  
An accurate approximate solution, mimicking the solitary behavior as well 
as capturing the collapse properties of Eq. (\ref{eq:nls3}) is given by 
\cite{Desaix-Anderson-Lisak} 
\begin{equation}\label{eq:approximatesolution}
a_T(\tau, \rho) = 
F(\tau)\,{\rm sech}\!\left[ \frac{\rho}{f(\tau)} 
\right]\exp\left[ {\rm i} b(\tau)\rho^2 \right] ,
\end{equation}
where\footnote{%
The complex amplitude $F(\tau)$ and the phase function $b(\tau)$
can be expressed in terms of the pulse width $f(\tau)$ 
\cite{Desaix-Anderson-Lisak}.  
Here $\gamma = 4(\ln 4 + 1)/(27\zeta(3)) \approx 0.29$, $I(\tau) = 
f^2(\tau)|F(\tau)|^2 = f^2(0)|F(0)|^2 = I_0$, 
and $I_c = (2\ln2 + 1)/(4\ln2 - 1) \approx 1.35$. %
}
$f(\tau) = 
[f^2(0) + {\gamma}\left(1 -
{I_0}/{I_c}\right)\tau^2]^{1/2}$,   
showing the instability of the stationary solution $I_0 = I_c$, either
collapsing to zero width in a  
finite time when $I_0 > I_c$, or diffracting monotonously towards
infinite width when $I_0 < I_c$. 

In the next section, a perturbation analysis will show that the exact equations
produced unstable solutions.

\paragraph{Instability analysis}

Following \citet{Shukla-Eliasson-Marklund}, a rescaling of Eq. (\ref{eq:nls2})
gives the dimensionless 
equation
\begin{equation}
\label{eq:nls4}
\mathrm{i}\left(\frac{\partial}{\partial
  t}+\sqrt{1-\beta^2}\,\frac{\partial}{\partial z}\right)A 
+\frac{1}{2}\frac{\partial^2 A}{\partial y^2}+
\frac{\beta^2}{2}\frac{\partial^2A}{\partial z^2}+|A|^2A=0,
\end{equation}
where $\beta=\ell\pi c/x_0\omega$, and the time is scaled by $\omega^{-1}$,  
the spatial variables by $c/\omega$, and the 
vector potential by $(\kappa\epsilon_0/2\omega^2)^{1/2}$.

Conditions for the 
modulational and filamentation instabilities can be obtained as follows. 
Let $A=[A_0+A_1\exp(i\phi)+A_2\exp(-i\phi)]\exp(-i\omega_0 t)$, where 
$\phi={\bf K}\cdot {\bf r}- {\mno} t$ is a phase, $\omega_0$ 
is a constant frequency and the constants $A_0$, $A_1$ and $A_2$
are the complex amplitudes of the pump wave and the two electromagnetic
sidebands, respectively. The wavevector and the frequency of modulating 
perturbations are denoted by ${\bf K} = \hat {\bf y} K_y +  \hat {\bf z} K_z$
and ${\mno}$, respectively, where $\hat {\bf y}$ and $\hat {\bf z}$ 
are the unit vectors along $x$ and $y$ axes, respectively. 
Following the standard procedure of the modulational/filamentational 
instability \cite{Shukla-etal,dan99}, the nonlinear dispersion relation
becomes  
\begin{eqnarray}
  &&\!\!\!\!\!\!\!
  \left({\mno} -K_z\sqrt{1-\beta^2}\right)^2
\nonumber \\ &&\!\!\!\!
  +\left[ |A_0|^2-\tfrac{1}{4}\left(K_y^2+\beta^2 K_z^2 \right)\right]
  \left(K_y^2+\beta^2K_z^2 \right) = 0.
  \label{eq:Shu-Eli-disprel}
\end{eqnarray}
Letting ${\mno} = K_z\sqrt{1-\beta^2} + i \gamma$ in (\ref{eq:Shu-Eli-disprel}), 
one obtains the modulational
instability growth rate    
\begin{equation}\label{eq:plategrowthrate}
 \gamma=\left[ |A_0|^2-\tfrac{1}{4}(K_y^2+\beta^2 K_z^2)\right]^{1/2}
  (K_y^2+\beta^2K_z^2)^{1/2} .
\end{equation}
The instability grows quadratically with the amplitude $A_0$, 
and attains a maximum value at a critical wavenumber. Values
of $\beta$ different from unity make the instability
region asymmetric with respect to $K_y$ and $K_z$. 
On the other hand, the spatial amplification rate 
${\mng} = iK_z$ of the filamentation instability in the 
quasi-stationary limit ({\it viz}.~$\Omega=0$)  
and for $\beta^2 {\mng}^2 \ll K_y^2$ is 
\begin{equation}
  {\mng} =  \left(|A_0|^2-\tfrac{1}{4}K_y^2\right)^{1/2}\frac{K_y}{\sqrt{1-\beta^2}}. 
\end{equation}

\begin{figure}[ht]
    \includegraphics[width=.7\columnwidth]{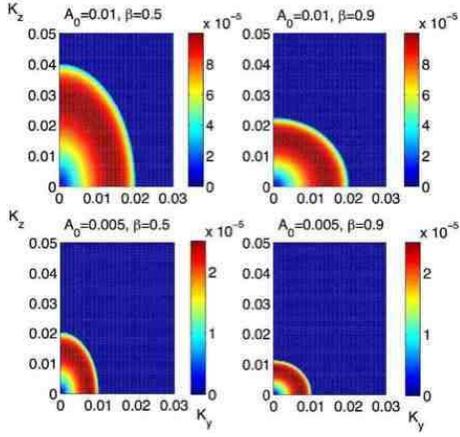}
    \caption{The growth rate $\gamma$ as given by (\ref{eq:plategrowthrate}) 
    of the modulational instability for an 
    initially homogeneous radiation field as a function of the wavenumber
    $(K_y,\,K_z)$, for different values of $\beta$ and pump strength $A_0$. 
(Reprinted from \citet{Shukla-Eliasson-Marklund}, Copyright (2004), with permission from Elsevier.)}
\label{fig:plates1}
\end{figure}

\begin{figure}[ht]
    \includegraphics[width=.7\columnwidth]{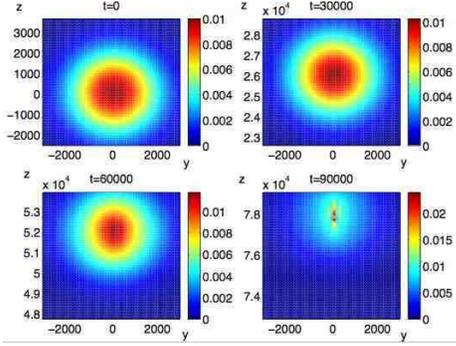}  
    \caption{The time-evolution of $|A|^2$ as given by 
    Eq.\ (\ref{eq:nls4}) for an initially Gaussian shaped electromagnetic  
    pulse. The pulse slowly self-focuses and finally collapses. Here 
    $\beta=0.5$ and a Gaussian pulse  
    $A=10^{-2}\exp[-(y^2+z^2)/(2\times 10^3)^2]$ is used.  
    In the upper right and lower left panels, the pulse self-compresses
    and in the lower right panel the field strength of the pulse has reached 
    a critical limit where Eq.\ (\ref{eq:nls4}) is no longer valid.
    (Reprinted from \citet{Shukla-Eliasson-Marklund}, Copyright (2004), with permission from Elsevier.)}
\label{fig:plates2}
\end{figure}

\citet{Shukla-Eliasson-Marklund} performed a numerical study of Eq. 
(\ref{eq:Shu-Eli-disprel}) showing the instabilities indicated by the approximate
solution (\ref{eq:approximatesolution}), see Fig. \ref{fig:plates1}. 
Using the normalized Eq.\ (\ref{eq:nls4}) an initially Gaussian pulse
was shown to collapse or disperse in accordance to the collapse
criterion in \citet{Brodin-etal}, see Fig. \ref{fig:plates2}. 
The collapse is unbounded in the weakly nonlinear model given by (\ref{eq:nls4}). 
As the collapse pursues, the intensity of the 
pulse will reach values at which the weakly nonlinear theory breaks down
and higher order effects \cite{Bialynicka-Birula,Marklund-Shukla-Eliasson}, 
and possible pair creation processes
\cite{Schwinger}, has to be taken into account. 
For the latter, 
a significant energy dissipation into the electron--positron
plasma will take place.

In practice, the trapping of an electromagnetic pulse could be achieved 
by two highly conduction layers. As an example, we consider the generation of
two-dimensional plasma channels due to the interaction of a plasma
with high--intensity lasers \cite{Shen-etal}. In this case, a vacuum will
be created within the plasma due to the complete evacuation of the electrons 
by the ponderomotive force of intense laser beams. The resulting plasma
waveguides can sustain very high field intensities \cite{bob}, and
with future laser systems \cite{mou98,mou05} the intensities could surpass
even the theoretical laser limit \cite{mou98,she02,puk03}. 
The trapping of intense laser fields could yield the
right conditions for the electromagnetic modes to self-interact
via the nonlinear quantum vacuum, giving rise to pulse evolution according
to Eq.\ (\ref{eq:nls4}).

\subsubsection{Cavity mode interactions}\label{sec:cavity}

Wave-wave interactions can give rise to a host of interesting phenomena, 
well-known in optics and plasma physics \cite{Agrawal,Kivshar-Agrawal,%
Weiland-Wilhelmsson}. 
As cubic nonlinearities act within a cavity environment,
they will produce wave-wave couplings, and given certain resonance 
conditions, a new mode will be generated, that will satisfy 
the cavity dispersion relation. \citet{Brodin-Marklund-Stenflo} showed 
that these new modes could reach detectable levels within state of the art 
cavities, and \citet{Brodin-Marklund-Stenflo2} were able to derive a
NLSE for the self-interaction of a single mode in a rectangular cavity.

Calculations of the three-wave coupling strength between various eigenmodes can be made
including the nonlinear polarization (\ref{eq:polarization}) and magnetization
(\ref{eq:magnetization}), see e.g. \citet{Brodin-Marklund-Stenflo}. 
However, a more convenient and elegant approach, which was pioneered by 
\citet{Brodin-etal} and which gives
the same result, starts directly with the Lagrangian density (\ref{eq:lagrangian1}).

A general procedure for finding the cavity eigenmode coupling and the
saturated amplitudes of the excited mode can be formulated 
\cite{Brodin-Marklund-Stenflo,Brodin-etal,Eriksson-etal,Weiland-Wilhelmsson}:

1. Determine the linear eigenmodes of the cavity in terms of
the vector potential . 

2. Choose resonant eigenmodes fulfilling frequency matching
conditions for modes 1, 2, and 3, such as  
\begin{equation}
  \omega _{3}=2\omega _{1}-\omega _{2}  . \label{frequency-matching}
\end{equation}

3. Assume a slowly varying amplitude of the vector potential eigenmode 
amplitudes and minimize the effective action obtained from the 
Lagrangian (\ref{eq:lagrangian1}) and follow steps 1 and 2.
The lowest order linear terms vanish, since the dispersion
relation of each mode is fulfilled. 

4.  In the absence of dissipation, the mode coupling equations imply steady
growth of mode 3, until the energy of that mode is comparable to that of the
pump modes. 
A damping mechanism, such as finite conductivity of the cavity walls, 
may be inserted on phenomenological grounds. 
This saturates the amplitude at a level depending on the
the mode-coupling growth versus losses.

\paragraph{Rectangular cavities}

For a rectangular prism cavity with dimensions $(x_0, y_0, z_0)$, 
choosing the radiation gauge, the 
pump modes have vector potentials of the form 
\begin{equation}
\mathbf{A}_{j}=A_{j}\sin \left( \frac{\pi x_j}{x_{j0}}\right) \sin \left( \frac{%
\ell_{j}\pi z}{z_{0}}\right) \exp (-\mathrm{i}\omega _{j}t)\hat{\mathbf{y}}+%
\mathrm{c.c}  , \label{vector-rectj}
\end{equation}
where $j = 1,2$, $\ell_{j}=1,2,3...$ are the 
mode numbers the pump waves, and $x_1 = x$, $x_2 = y$, $x_{10} = x_0$, 
and $x_{20} = y_0$. The dispersion relations are 
\begin{equation}
  \omega _{j}^{2} = \frac{\ell_{j}^{2}\pi ^{2}c^{2}}{z_{0}^{2}}
 + \frac{\pi^{2}c^{2}}{x_{j0}^{2}} .
\end{equation} 
The mode excited due to the QED nonlinearities is given by  
\begin{equation}
\mathbf{A}_{3}=A_{3}\sin \left( \frac{\pi y}{y_{0}}\right) \sin \left( \frac{%
\ell_{3}\pi z}{z_{0}}\right) \exp (-\mathrm{i}\omega _{3}t)\widehat{\mathbf{x}}+%
\mathrm{c.c.} , \label{vector-rect3}
\end{equation}
where 
\begin{equation}
  \omega _{3}^{2} = \frac{\ell_{3}^{2}\pi^{2}c^{2}}{z_{0}^{2}} 
  + \frac{\pi^{2}c^{2}}{y_{0}^{2}} .
\end{equation}
 
Following the scheme as given by \citet{Brodin-Marklund-Stenflo,Brodin-etal} 
and \citet{Eriksson-etal},
with the resonance condition $\omega _{3}=2\omega _{1}-\omega _{2}$, one obtains 
the evolution equation for mode 3  
\begin{equation}
\frac{dA_{3}}{dt}=-\frac{\mathrm{i}\varepsilon _{0}\kappa \omega _{3}^{3}}{8}%
K_{\mathrm{rec}}A_{1}^{2}A_{2}^{\ast } , \label{Evolution1}
\end{equation}
where the dimensionless coupling coefficient $K_{\mathrm{rec}}$ is 
\begin{eqnarray}
  &&
  K_{\mathrm{rec}}  = \frac{\pi ^{2}c^{2}}{\omega _{3}^{4}}\Bigg\{(-,-,+)\bigg[%
  \frac{8\pi ^{2}c^{2}}{x_{0}^{2}y_{0}^{2}}+\left( \frac{4}{x_{0}^{2}}+\frac{%
  7\ell_{1}^{2}}{z_{0}^{2}}\right) \omega _{2}\omega _{3}\bigg] 
\nonumber \\
  && + \frac{%
  \ell_{2}\ell_{3}\pi ^{2}c^{2}}{z_{0}^{2}}\left( \frac{7\ell_{1}^{2}}{z_{0}^{2}}-\frac{%
  3}{x_{0}^{2}}\right)  
\nonumber \\ &&
  + \frac{7\omega _{1}\ell_{1}}{z_{0}^{2}}\left( (-,+,-)\omega
  _{2}\ell_{3}(+,-,-)\omega _{3}\ell_{2}\right) \Bigg\}  .\label{Coupling1}
\end{eqnarray}
The different signs in the expression (\ref{Coupling1}) for the coupling strength 
correspond to the mode number matchings 
$2\ell_{1}-\ell_{2}+\ell_{3} = 0$, 
$2\ell_{1}+\ell_{2}-\ell_{3}  = 0$, and   
$2\ell_{1}-\ell_{2}-\ell_{3}  = 0$, 
respectively, that must be fulfilled for nonzero coupling. 
The coupling coefficient for
specific mode numbers and geometries can thus be evaluated. 
If a saturation mechanism is included in (\ref{Evolution1}),
one may solve for the steady-state value of $A_3$. 

\paragraph{Cylindrical cavities}

\begin{figure}[ht]
\subfigure[]{\includegraphics[height=3.5cm,width=6cm]{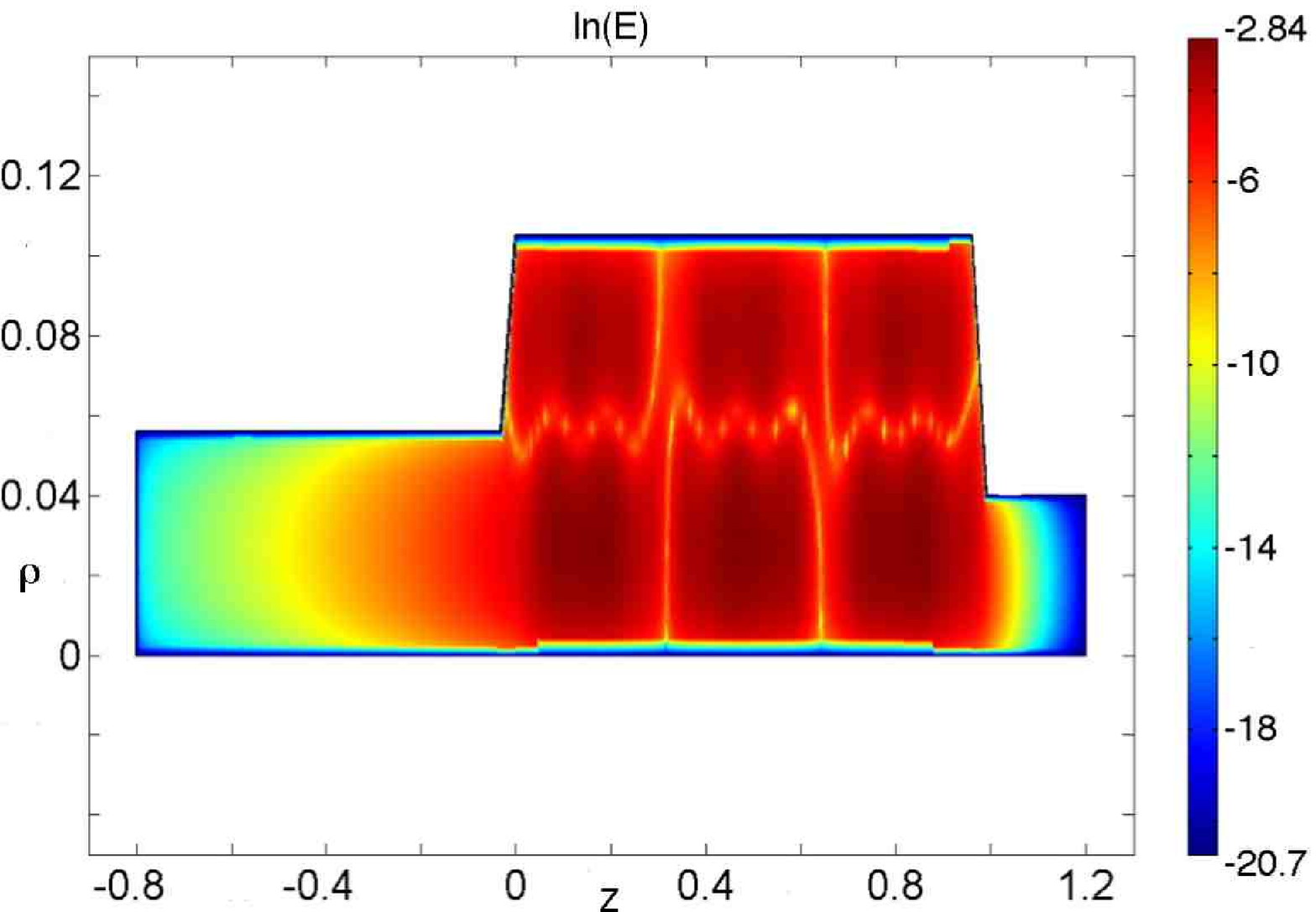}}
\subfigure[]{\includegraphics[height=3.5cm,width=6cm]{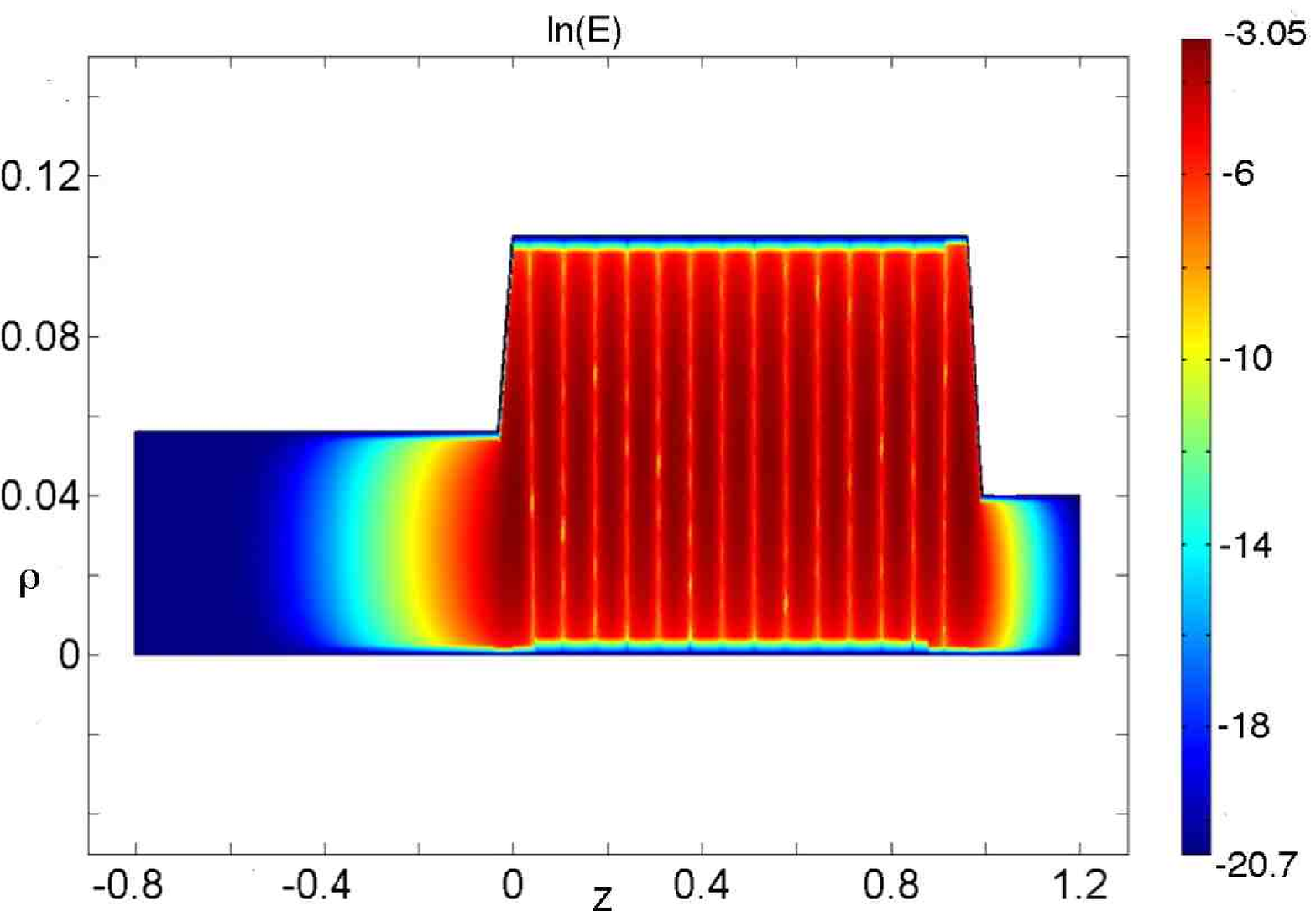}}
\subfigure[]{\includegraphics[height=3.5cm,width=6cm]{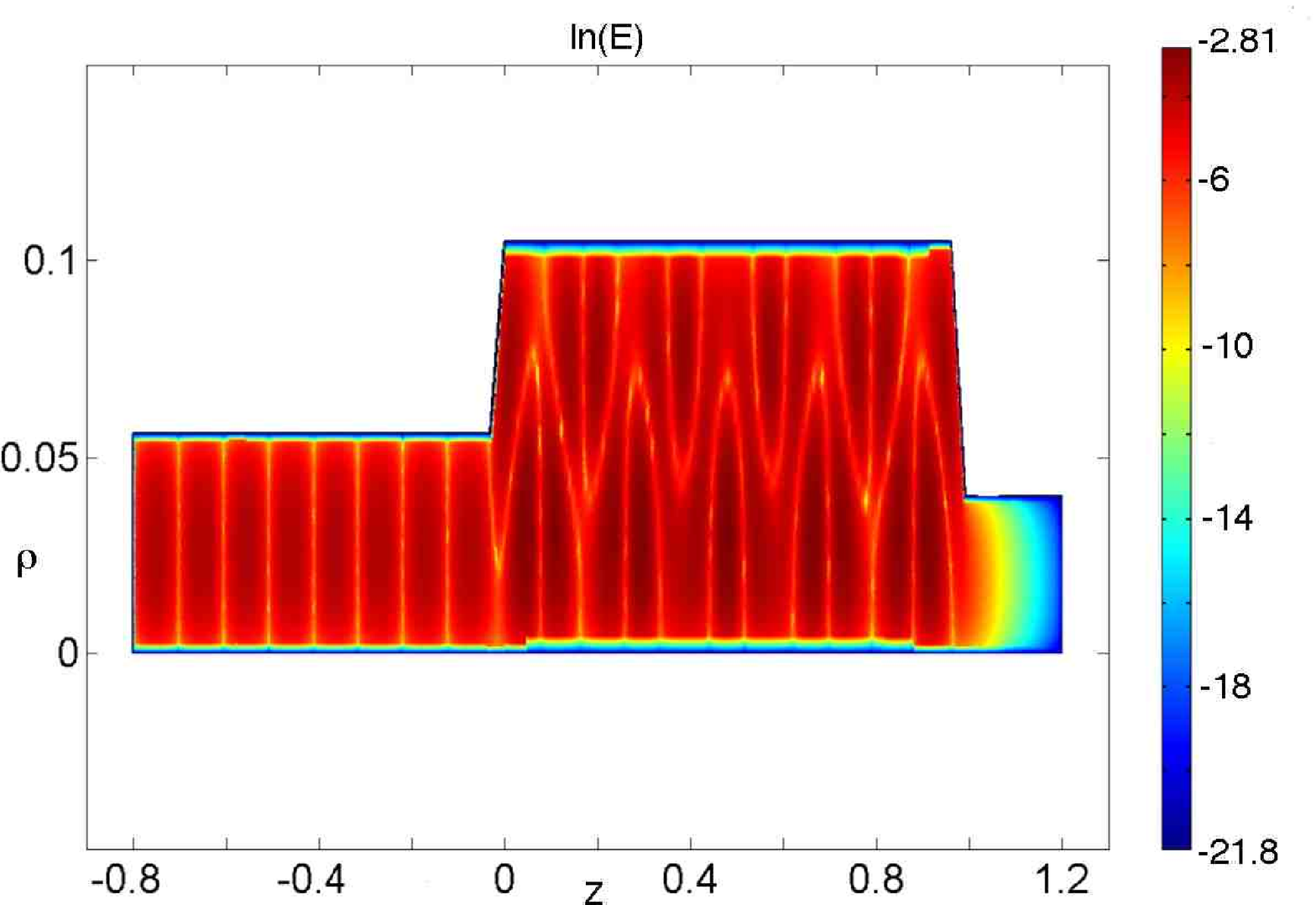}}
\caption{The mode structure for the filtering
geometry (figure adopted from \citet{Eriksson-etal}). 
The small region to the right is the entrance region, the large middle
region is where the interaction takes place, and the region to the left is
the filtering region. Here $\ln|E|$ is plotted using an arbitrary 
normalization. We have the cavity distance $z$ and the cylindrical radius $\rho$
in units of meter. 
(a) Pump mode 1: We see that spatial exponential decay  
has diminished the amplitude by a factor 
$10^{-6}$ in the filtered region to the left. 
(b) Pump mode 2: As for mode 1, exponential decay 
 diminishes the amplitude in the filtering region, here by a factor  
$10^{-8}$. 
(c) Excited mode: The amplitude of the mode excited by the QED
vacuum nonlinearities is almost unaffected when passing into the filtering
region.}
\label{fig:cavity}
\end{figure}

As was shown by \citet{Eriksson-etal}, the efficiency of the mode
conversion can be slightly improved by the choice of a cylindrical 
cavity. The results can be obtained along the lines of the previous 
example. For \textrm{TE}-modes with no
angular dependence, the vector
potential 
\begin{equation}
\mathbf{A}=AJ_{1}(\rho \beta /a)\sin \left( \frac{\ell\pi z}{z_{0}}\right) \exp
(-\mathrm{i}\omega t)\hat{\mathbf{\varphi }}+\mathrm{c.c.}
\label{Vector-cyl}
\end{equation}
gives a complete description of the fields. 
Here $a$ is the cylinder radius,  
$z_{0}$ the length of the cavity, $J_{1}$
the first order Bessel function, $\ell$ is the mode number, 
and $\beta $ one of its zeros. The cylinder
occupies the region $0\leq z\leq z_{0}$ centered around the $z$-axis. We
have here introduced cylindrical coordinates $\rho $ and $z$ as well as the
unit vector $\hat{\mathbf{\varphi }}$ in the azimuthal direction. 
The eigenfrequency is given by   
  $\omega ^{2} = c^{2}[ ({\beta }/{a})^2 
  + ({\ell\pi}/{z_{0}})^2 ] $, 
for all modes $\omega =\omega _{1,2,3}$. From the
matching condition $\omega _{3}=2\omega _{1}-\omega _{2}$ it follows that all
the eigenmodes cannot have the same order of their respective $\beta$,
and one thus introduces $\beta =\beta_{1,2,3}$. Proceeding along the lines
of the previous section, one obtains 
\begin{equation}
\frac{dA_{3}}{dt}=-\frac{\mathrm{i}\varepsilon _{0}\kappa \omega _{3}^{3}}{8}%
K_{\mathrm{cyl}}A_{1}^{2}A_{2}^{\ast }  \label{Evolution2}
\end{equation}
for the mode number matching $\ell_3 = 2\ell_1 + \ell_2$. 
Here the cylindrical coupling coefficient $K_{\mathrm{cyl}}$ is 
defined in terms of integrals of Bessel functions
and can be found in \citet{Eriksson-etal}.
As in the case of a rectangular geometry, the linear growth of $A_3$ as dictated 
by Eq. (\ref{Evolution2}) will be saturated by dissipative mechanisms. The intensity of the 
generated field amplitudes and the pump field amplitudes is shown in Fig. \ref{fig:cavity}.

\subsection{Incoherent field interactions}

Above we have considered the interaction via the nonlinear 
quantum vacuum between coherent electromagnetic waves.
However, in many situations where the Heisenberg--Euler
Lagrangian is important, such as in astrophysical and laser-plasma
applications, there can be intense incoherent electromagnetic fields
present.   
We will below study two scenarios. First, a plane wave pulse
propagating on a vacuum dressed by an intense gas of incoherent 
photons is analysed, and, secondly, the effects on a radiation gas by the 
quantum vacuum excited by an intense 
electromagnetic (EM) pulse is
investigated. These two results are then used in conjunction to obtain
the relevant equations governing the nonlinear interaction between the
pulse and the radiation gas.

\subsubsection{Coherent pulse interaction with incoherent photons}

The dispersion relation (\ref{eq:weakgas-disp}) represents the propagation
of test photons in an intense incoherent radiation background.
Using standard methods for slowly varying envelopes \cite{Hasegawa} in
conjunction with (\ref{eq:weakgas-disp}), we obtain the special case of (\ref{eq:nlse-gen}),  
for a pulse in an intense photon gas background 
\cite{Marklund-Brodin-Stenflo,r1,Shukla-Marklund-Tskhakaya-Eliasson}
(see also \citet{Rozanov} for a similar result using a different strong background field)
\begin{equation}
  i\left(\frac{\partial}{\partial t} + v_g\frac{\partial}{\partial z} 
  \right)E_p + \frac{v_g}{2k_0}\left(\nabla_{\perp}^2 - 
    \beta_z\frac{\partial^2}{\partial z^2} \right)E_p + \mu\, 
  \delta\mathscr{E}\,E_p = 0 .
\label{eq:nlse}
\end{equation}
where $\delta\mathscr{E} = \mathscr{E} - \mathscr{E}_0$ is a radiation gas 
perturbation due to the pulse propagation and $\mathscr{E}_0$ is the unperturbed 
background radiation energy density. 
Here we have adapted the coordinates such that $\mathbf{k}_0 = k_0\hat{\mathbf{z}}$, 
$\mathbf{k}_0$ being the background wavevector of the
pulse, included the high-frequency correction represented by $\beta_z$,
and denoted the (complex) pulse amplitude by $E_p$. 
Moreover, $\mathbf{v}_g = (\partial\omega/\partial\mathbf{k})_0$ 
is the group velocity on the
background, $\beta_z = (32/3)\sigma\lambda^2\mathscr{E}_0^2k_0^2$ is the
vacuum dispersion parameter, $\mu
= (2/3)ck_0\lambda [1 + (16/3)\sigma\lambda\mathscr{E}_0k_0^2]$ is the
nonlinear refraction parameter, $\nabla^2_{\perp} = \nabla^2 -
(\hat{\mathbf{k}}_0\cdot\nabla)^2$, and $\lambda = 8\kappa$ of $14\kappa$ 
depending on the polarization of the pulse (see Sec.\ \ref{sec:dispersionfunction}). 

\subsubsection{Radiation gas response of pulse propagation}

We saw that the effects of a plane wave on incoherent photons could 
be expressed via the dispersion relation (\ref{eq:planewave}).
Following \citet{Marklund-Brodin-Stenflo}, the response of the
radiation gas can be determined using a kinetic theory.  
For a dispersion relation $\omega = ck/n(\mathbf{r}, t)$, where $n$ is the spacetime
dependent refractive index, we have the Hamiltonian ray equations
\begin{equation}\label{eq:groupvelocity} 
  \dot{\mathbf{r}} = \frac{\partial\omega}{\partial\mathbf{k}} = \frac{c}{n}\,\hat{\mathbf{k}}  ,
\text{ and } 
  \dot{\mathbf{k}} = -\nabla\omega = \frac{\omega}{n}\nabla n ,
\end{equation}
where $\dot{\mathbf{r}}$ denotes the group velocity of the photon,
$\dot{\mathbf{k}}$ the force on a photon, and the dot denotes a time
derivative.

The equation for the collective interaction of photons can
then be formulated as~\cite{Mendonca}
\begin{equation}\label{eq:kinetic}
  \frac{\partial f(\mathbf{k}, \mathbf{r}, t)}{\partial t} +
  \nabla\cdot(\dot{\mathbf{r}} f(\mathbf{k}, \mathbf{r}, t)) +
  \frac{\partial}{\partial\mathbf{k}}\cdot(\dot{\mathbf{k}}f(\mathbf{k}, \mathbf{r},
  t)) = 0 ,
\end{equation}
where the distribution function $f(\mathbf{k}, \mathbf{r},t)$ 
has been normalized such that $\int f(\mathbf{k},
\mathbf{r},t)\,d\mathbf{k}$ gives the photon gas 
number density. In what follows, we will neglect the dispersive
effects on the evolution of the radiation gas \footnote{The dispersive
correction, due to the variations in the photon-field, will give rise
to higher order effects in the final fluid equation for the energy
density of the radiation gas. Thus, we may at this stage neglect the
dispersive term from the fluid equations.}. 
Taking the moments of the kinetic equation (\ref{eq:kinetic})
\cite{r1}, we obtain the energy conservation equation
\begin{subequations}
\begin{equation}\label{eq:energy}
  \frac{\partial\mathscr{E}}{\partial t} +
  \nabla\cdot\left( \mathscr{E}\mathbf{u} + \mathbf{q} \right) =  -\frac{\mathscr{E}}{n}\frac{\partial n}{\partial t} ,
\end{equation}
where $\mathscr{E}(\mathbf{r}, t) = \int\hbar\omega f\,d\mathbf{k}$
is the energy density, and $\mathbf{q}(\mathbf{r}, t) = \int\hbar\omega\mathbf{w} f\,d\mathbf{k}$ is the energy 
(or Poynting) flux.
Here we have introduced $\dot{\mathbf{r}} = \mathbf{u} + \mathbf{w}$, where
$\int\mathbf{w} f\,d\mathbf{k} = 0$. Thus
$\mathbf{w}$ represents the random velocity of the photons.
Equation (\ref{eq:energy}) is coupled to the momentum conservation
equation
\begin{eqnarray}\label{eq:momentum}
  \frac{\partial{\ppi}}{\partial t} + \nabla\cdot\Big[
  \mathbf{u}\otimes{\ppi} + \mathbf{\mathsf{P}} \Big] =  \frac{\mathscr{E}}{n}\nabla n ,
\end{eqnarray}
\label{eq:comoving}
\end{subequations}
where ${\ppi} = \int \hbar\mathbf{k} f\,d\mathbf{k}$
is the momentum density and
$\mathbf{\mathsf{P}} = \int \mathbf{w}\otimes(\hbar\mathbf{k}) f\,d\mathbf{k}$
is the pressure tensor. It follows immediately from
the definition of the pressure tensor that the trace 
satisfies $\mathrm{Tr}\,\mathbf{\mathsf{P}} = \int \hbar k\mathbf{w}\cdot\hat{\mathbf{k}} f\,d\mathbf{k}
  = (n/c)\int \hbar\omega\mathbf{w}\cdot\hat{\mathbf{k}} f\,d\mathbf{k}$.
For an observer comoving with the fluid, i.e. a system in
which $(\mathbf{u})_0 = 0$ (the $0$ denoting the comoving system),
Eq.~(\ref{eq:groupvelocity}) shows that
$(\mathbf{w}\cdot\hat{\mathbf{k}})_0 = (n)_0/c$, so that the trace of the
pressure tensor in the comoving system becomes 
$(\mathrm{Tr}\,\mathbf{\mathsf{P}})_0 = (\mathscr{E})_0$.
For an isotropic distribution function, the pressure can be expressed
in terms of the scalar function $P = \mathrm{Tr}\,\mathbf{\mathsf{P}}/3$,
satisfying the equation of state $P = \mathscr{E}/3$. We will
henceforth adopt the comoving frame, in which $\mathbf{u} = 0$, and 
the equation of state $\mathsf{P}_{ij} = P\delta_{ij}   \delta_{ij}\mathscr{E}/3$, in order to achieve closure
of the fluid equations. 

\subsubsection{Quasi-linear theory}

We now assume that the radiation gas is perturbed around the
equilibrium state
$\mathscr{E}_0 =$ constant and ${\ppi}_0 = \mathbf{0}$, letting
$\mathscr{E} = \mathscr{E}_0 + \delta\mathscr{E}$, where
$|\delta\mathscr{E}| \ll \mathscr{E}_0$. Then, using Eqs.\
(\ref{eq:comoving}), we obtain an
acoustic equation  
\begin{equation}
  \left( \frac{\partial^2}{\partial t^2} - \frac{c^2}{3}\nabla^2
  \right)\delta\mathscr{E} =  -\frac{2\lambda\epsilon_0\mathscr{E}_0}{3}\left( 
  \frac{\partial^2}{\partial t^2} + {c^2}\nabla^2 
  \right) |E_p|^2 ,
\label{eq:response}
\end{equation}
to lowest order in $\delta\mathscr{E}$.  
This gives the dynamics of a radiation gas due to the pulse propagation
\cite{r1}. Equations (\ref{eq:nlse}) and
(\ref{eq:response}) were first presented by
\citet{r1}, and generalize the Marklund--Brodin--Stenflo (MBS) equations 
\cite{Marklund-Brodin-Stenflo}, to the case of a dispersive vacuum. 
The MBS equations are different from the Karpman equations
\cite{Karpman,Karpman2}, that govern the dynamics of small amplitude
nonlinearly interacting electromagnetic waves and ion-sound waves driven 
by the radiation pressure in an electron-ion plasma, due to the 
difference in the driving term on the right hand side of Eq. (\ref{eq:response}).

The dispersion-free case admits pulse collapse
\cite{Marklund-Brodin-Stenflo,Shukla-Eliasson}, and similar features
appear within the dispersive case, with the difference that pulse
splitting may occur, resulting in a train of ultra-short pulses.  
If the time response of the radiation background is slow, Eq.\
(\ref{eq:response}) may be integrated to yield $\delta\mathscr{E}
\approx 2\lambda\mathscr{E}_0\epsilon_0|E_p|^2$, and from Eq.\
(\ref{eq:nlse}) we obtain the 
standard equation for analyzing ultra-short intense pulses in normal
dispersive focusing media, see
\citet{Chernev-Petrov,Rothenberg,Zharova-Litvak-Mironov,Gaeta,Kivshar-Agrawal}
and references therein. It is well known that the evolution of a pulse
within this equation displays first self-focusing, then pulse
splitting \cite{Chernev-Petrov,Rothenberg}, and the approximate
description of the solutions can be given as a product of bright and
dark soliton solutions \cite{Hayata-Koshiba}. A modulational
instability can be found as well \cite{Kivshar-Agrawal}.

\subsubsection{Instability analysis}

\paragraph{The two-dimensional case}

As $\beta_z$ goes to zero in Eq.\ (\ref{eq:nlse}), we regain the
MBS equations \cite{Marklund-Brodin-Stenflo}. 
In this case, the dispersion relation for the modulational and filamentational
instabilities of a constant amplitude photon pump ($\omega_0,
{\bf k}_0)$ can be found by linearizing the simplified set of Eqs.\
(\ref{eq:nlse}) and (\ref{eq:response}) around  
the unperturbed state $E_p = E_{0} =$ real constant and $\delta\mathscr{E}
=0$. Following the standard procedure of parametric instability   
analysis \cite{Shukla,Kivshar-Agrawal} (see also next section), we
consider perturbations varying according to
$\exp[ i (\mathbf{K}\cdot\mathbf{r} - {\mno} t)]$. Here ${\mno}$ and 
$\mathbf{K}$ are the frequency and wavevector of the acoustic-like
disturbances. 
Then, Eqs.\ (\ref{eq:nlse}) and (\ref{eq:response}),  
yield the nonlinear dispersion relation \cite{Shukla-Eliasson,%
Shukla-Marklund-Tskhakaya-Eliasson}
\begin{eqnarray}
&&
\left[\left({\mno} - c \hat{\mathbf{k}}_0\cdot\mathbf{K} \right)^2 
- \frac{K_\perp^4c^2}{4 k_0^2}\right](3{\mno}^2 - K^2c^2) 
\nonumber \\ &&
= \frac{4
  K_\perp^2 c^2}{3} 
({\mno}^2 + K^2c^2)\lambda^2 \mathscr{E}_0\epsilon_0 E_{0}^2,
\label{eq:2D-disprel}
\end{eqnarray}
where $K^2 =K_x^2 + K_y^2 + K_z^2 \equiv K_\perp^2 + K_z^2$. 
Defining ${\mno} = c K_z +  i \gamma_{2D}$, $|\gamma_{2D}| \ll
c K_z$, we obtain the approximate modulational instability
growth rate 
\begin{equation} 
  \gamma_{2D} \approx \left\{\frac{c}{2k_0}K_{\perp}^2\Bigg[ \chi_{2D}\,
  \frac{K_{\perp}^2 + 2K_z^2 }{K_{\perp}^2 - 2K_z^2} 
  - \frac{c}{2k_0}K_{\perp}^2  \Bigg] \right\}^{1/2} , 
\label{eq:2Dmod_growth}
\end{equation}   
where $\chi_{2D} = (8/3)ck_0\lambda^2\mathscr{E}_0 \epsilon_0 E_0^2$.

Similarly, in the quasi-stationary limit, ${\mno} =0$, a
filamentational instability may occur. For $K_z \ll K_\perp$, we
obtain 
\begin{equation}
K_z^2 = \frac{K_\perp^4}{4k_0^2} - \frac{4}{3} K_\perp^2 
\lambda^2 \mathscr{E}_0\epsilon_0 E_{0}^2  . 
\end{equation}
Thus, filamentation 
of an intense photon beam on a radiation background takes
place when $\epsilon_0 |E_{p0}|^2 > 3K_\perp^2/(16 k_0^2\lambda^2
\mathscr{E}_0)$, due to elastic photon--photon scattering.

\begin{figure}[ht]
  \includegraphics[width=.7\columnwidth]{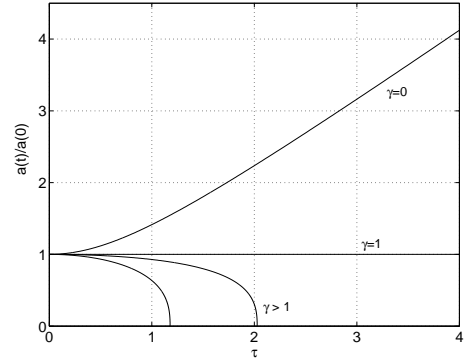}
  \caption{The pulse width as a function of normalized time 
  ($\tau \rightarrow ck_0\tau$) in the two-dimensional case. 
  Note that the solitary solution ($\gamma = 1$) is unstable.}
\label{fig:2D-collapse}
\end{figure}

In the case of a slow acoustic response, we may integrate Eq. 
(\ref{eq:response}) in comoving coordinates, to obtain a relation
between the radiation gas perturbation and the pulse intensity. 
By inserting the relation into Eq. (\ref{eq:nlse}) (with $\beta_z = 0$), 
the equation
\begin{equation}
  i\frac{\partial E_p}{\partial\tau} +
  \frac{c}{2k_0}\nabla_{\perp}^2E_p
  + \frac{4}{3}\lambda^2ck_0\varepsilon_0\mathscr{E}_0|E_p|^2E_p = 0 ,
\label{eq:transformednlse}
\end{equation}
is obtained. The collapse properties of Eq. (\ref{eq:transformednlse}) can be 
obtained by approximate analytical means. 
Starting from an two-dimensional approximately Gaussian pulse 
$
  E_p = A(\tau)\,\text{sech}(r_{\perp}/a(\tau))\exp[ib(\tau)r_{\perp}^2] , 
$ 
where 
$r_{\perp}^2 = x^2 + y^2$, an approximate solution can be found 
\cite{Desaix-Anderson-Lisak}. The relation $|A|/|A_0| = a_0/a$, where the 0
demotes the initial value. Moreover, the parameter 
$\gamma = (4/3)\lambda^2\epsilon_0\mathscr{E}_0k_0^2|A_0|^2a_0^2(I_1/I_2)$,
where 
  $I_1 = \int_0^{\infty} x\,\text{sech}^4(x) \,dx = (4\ln2 -1)/6$ and
  $I_2 = \int_0^{\infty} x^3\,\text{sech}^2(x) \,dx = 9\zeta(3)/8$,
and $\zeta$ is the Riemann zeta function, characterizes the critical behaviour 
of the solution in terms of the initial data. The collapse criteria can be seen 
in Fig. \ref{fig:2D-collapse}.

\begin{figure}[ht]
  \includegraphics[width=.8\columnwidth]{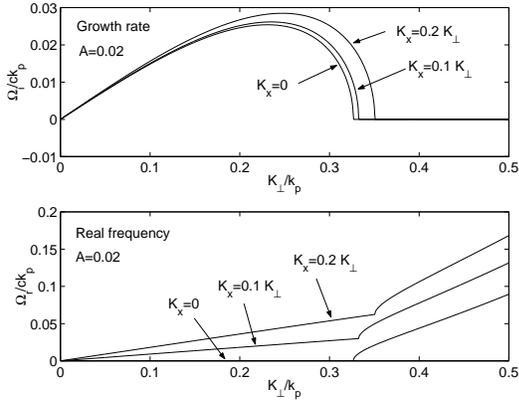}
  \caption{The growth rate and real part of the frequency, respectively,
  as a function the orthogonal wave number for different values of 
  the parallel wave number (see Eq.\ (\ref{eq:2Dmod_growth})). 
  The dimensionless pump strength is $A = \lambda^2\mathscr{E}_0 \epsilon_0E_0^2  = 0.02$
(reprinted with permission from \citet{Shukla-Eliasson}).}
\label{fig:2D-1}
\end{figure}

\begin{figure}[ht]
  \includegraphics[width=.8\columnwidth]{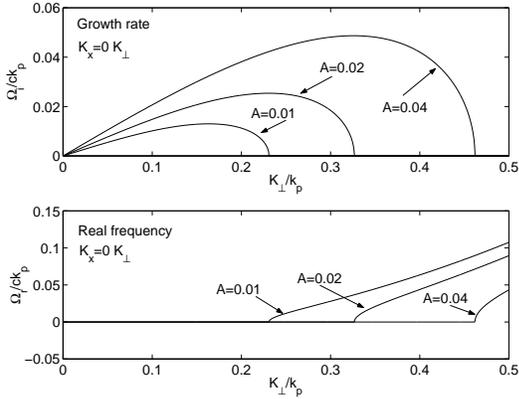}
 \caption{The growth rate and the real part of the frequency, respectively,
  as a function the orthogonal wave number for different dimensionless 
  pump strengths $A = \lambda^2\mathscr{E}_0 \epsilon_0E_0^2$ 
(see Eq.\ (\ref{eq:2Dmod_growth}))
(reprinted with permission from \citet{Shukla-Eliasson}).}
\label{fig:2D-2}
\end{figure}

\begin{figure}[ht]
  \includegraphics[width=.7\columnwidth]{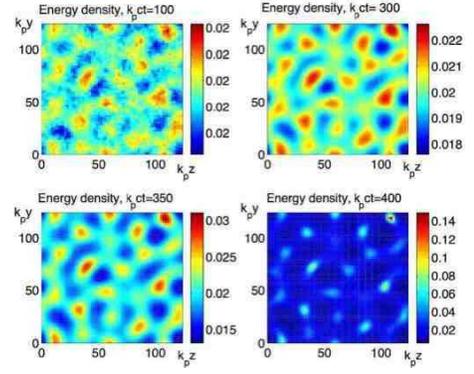}
  \caption{The two-dimensional evolution, as given by Eqs.\ (\ref{eq:nlse}) and (\ref{eq:response})
with $\beta_z = 0$, 
  of the dimensionless pulse energy density $\lambda^2\mathscr{E}_0\epsilon_0|E_p|^2$  
  for initially random perturbations on a radiation gas background. The color bars give
  the dimensionless pulse energy density  
(reprinted with permission from \citet{Shukla-Eliasson}).}
\label{fig:2D-3}
\end{figure}

Exact results regarding the two-dimensional modulational and
filamentational instabilities was found by 
\citet{Shukla-Eliasson}, where numerical solutions
of Eq.\ (\ref{eq:2D-disprel}) were presented, see Figs. \ref{fig:2D-1} and \ref{fig:2D-2}.
The growth of random seeds on a radiation gas background was also investigated
and is shown in Fig. \ref{fig:2D-3}. 

\paragraph{The three-dimensional case}

We will now show the presence of modulational and filamentational 
unstable modes for the three-dimensional case, as given by Eqs.\
(\ref{eq:nlse}) and (\ref{eq:response}).  
Fourier analysing Eq.\ (\ref{eq:response}) according to
$\delta\mathscr{E}$
and $|E_p|^2 \propto \exp[ i (\mathbf{K}\cdot\mathbf{r} - {\mno} t)]$, we
obtain 
\begin{equation}
  \delta\mathscr{E} = \frac{2}{3}\lambda\mathscr{E}_0W\epsilon_0|E_p|^2 , 
\end{equation}
where $W = (\mno^2 + c^2K^2)/(-\mno^2 + c^2K^2/3)$, with $\mno$ and $\mathbf{K}$ 
the frequency and wavevector, respectively, of the Fourier
component. Next, following \citet{Shukla} (see also
\citet{Kivshar-Agrawal}), we let $E_p = (E_0 +
E_1)\exp( i \delta t)$, where $\delta$ is the nonlinear phase
shift and $E_0$ ($\gg |E_1|$) is a real constant. 
To zeroth order in $E_1$ we have the nonlinear phase shift $\delta = -\kappa E_0^2$. 
We let $E_1 = d_1\,\exp[ i (\mathbf{K}\cdot\mathbf{r} - {\mno} t)] + d_2\,
\exp[- i (\mathbf{K}\cdot\mathbf{r} - {\mno} t)]$, with $d_1$ and $d_2$
real constants. Linearizing Eq.\ (\ref{eq:nlse}) with respect to
$E_1$, we obtain a coupled system of equations for $d_1$ and
$d_2$. Eliminating $d_1$ and $d_2$, we obtain the dispersion relation
\cite{Shukla-Marklund-Tskhakaya-Eliasson}
\begin{eqnarray}
  && ({\mno} - K_z v_g)^2 =  \frac{v_g}{2k_0}(K_{\perp}^2 - \beta_z
  K_z^2)
\nonumber \\&&\quad \times
  \left[ \frac{v_g}{2k_0}(K_{\perp}^2 - \beta_z K_z^2) - 
  \frac{1}{3}\chi W \right]  , 
\label{eq:disprel}
\end{eqnarray}
where $\chi = 4\mu\lambda\mathscr{E}_0\epsilon_0E_0^2$. 
Remembering that $W$ depends on
the perturbation frequency and wavevector, we see that the solution to
Eq.\ (\ref{eq:disprel}) in terms of $\mno$ is nontrivial.

Letting ${\mno} = K_z v_g +  i \gamma_m$ in (\ref{eq:disprel}), $|\gamma_m| \ll
K_z v_g$, we obtain the approximate modulational instability
growth rate 
\begin{eqnarray} 
  &&
  \gamma_m \approx \Bigg\{\frac{v_g}{2k_0}(K_{\perp}^2 -
  \beta_z K_z^2)\Bigg[ \chi\, \frac{K_{\perp}^2 +
  (1 + v_g^2/c^2)K_z^2 }{K_{\perp}^2 + (1 - 3v_g^2/c^2)K_z^2}
\nonumber \\ && \quad 
  - \frac{v_g}{2k_0}(K_{\perp}^2 -
  \beta_zK_z^2)  \Bigg] \Bigg\}^{1/2} .
\label{eq:mod_growth}
\end{eqnarray}   
Thus, when $v_g \approx c$, we see that, unlike the
standard modulational 
instability, we have larger growth rate for smaller length scale, with the
occurring asymptotically for $K_{\perp} \sqrt{2}\,K_z$, where the
approximate expression (\ref{eq:mod_growth}) diverges.

If the perturbations are stationary, Eq.\ (\ref{eq:disprel}) yields
\begin{equation}
  K_z \approx \pm \frac{1}{v_g} \left\{ \frac{v_g}{2k_0}K_{\perp}^2\left[
  \frac{v_g}{2k_0}K_{\perp}^2 - \chi \right] \right\}^{1/2}
\end{equation} 
when $\beta_zK_z^2 \ll K_{\perp}^2$, and we see that a filamentation
instability will occur for $\chi > v_g K_{\perp}^2/2k_0$. 

\begin{figure}[ht]
\includegraphics[width=0.7\columnwidth]{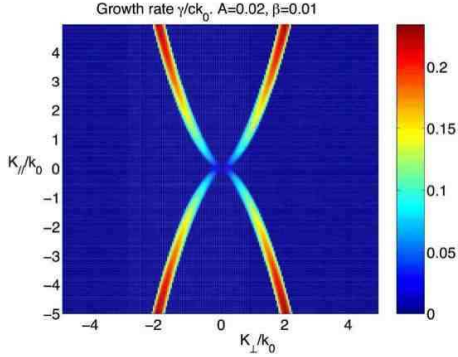}
\caption{The normalized growth rate $\gamma/(ck_0)$,
as given through the dispersion relation (\ref{eq:disprel}), as a
  function of $K_{\perp}/k_0$ and $K_z/k_0$. We note that due to
  cylindrical symmetry, the area of non-zero growth rate is really a
  cone-like structure. 
Here $A = \lambda^2\mathscr{E}_0\epsilon_0E_0^2$ 
(reprinted with permission from \citet{Shukla-Marklund-Tskhakaya-Eliasson}. Copyright 2004, American Institute of Physics).}
\label{fig:mod}
\end{figure}

\begin{figure}[ht]
\subfigure{\includegraphics[width=0.7\columnwidth]{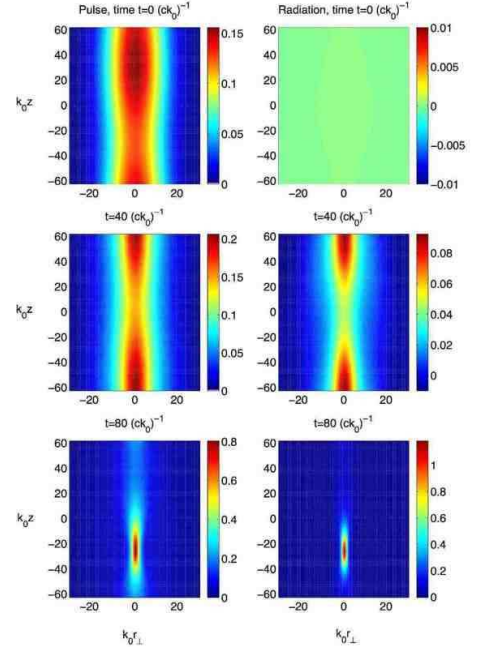}}
\caption{The three-dimensional evolution of a normalized modulated beam energy density 
  $\lambda^2\mathscr{E}_0\epsilon_0|E_p|^2$ (left panels) and the normalized 
  radiation gas energy density $\lambda\,\delta\mathscr{E}$ (right panels). The color gives
  the intensity. The beam collapses after a
  finite time, as does the inhomogeneity in the radiation gas 
(reprinted with permission from \citet{Shukla-Marklund-Tskhakaya-Eliasson}. Copyright 2004, American Institute of Physics).} 
\label{fig:evol1}
\end{figure}

\begin{figure}[ht]
\includegraphics[width=0.7\columnwidth]{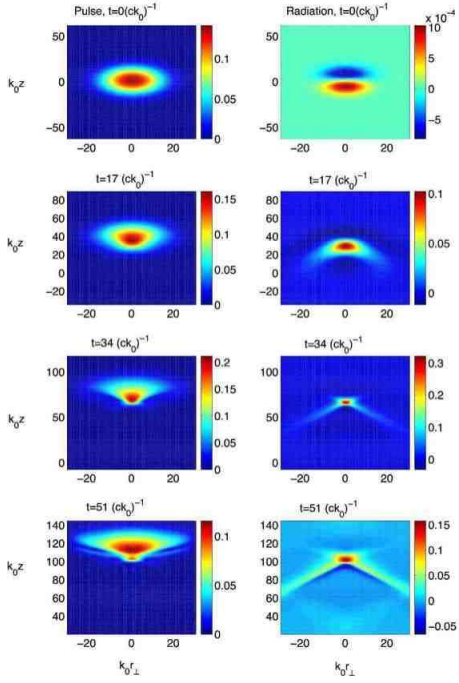}
\caption{The evolution of a normalized spherical pulse energy density $\lambda^2\mathscr{E}_0\epsilon_0|E_p|^2$ 
  (left panels) and the normalized radiation gas energy 
  density $\lambda\,\delta\mathscr{E}$ (right panels). The pulse first undergoes compression, and
  later splitting, while the radiation gas develops a high intensity
  region, from which photon wedges (similar to a Mach cone) are radiated
(reprinted with permission from \citet{Shukla-Marklund-Tskhakaya-Eliasson}. Copyright 2004, American Institute of Physics).} 
\label{fig:evol2}
\end{figure}

Solving (\ref{eq:disprel}) for the growth rate, one obtains the instability regions shown 
in Fig. \ref{fig:mod}.
We note
that the results found by \citet{Shukla-Marklund-Tskhakaya-Eliasson} concerning the modulational 
instability are similar
to the ones obtained by \citet{Karpman-Washimi}, where it was
found that the largest growth rates are due to parametric
instabilities for wave vectors oblique to the pulse propagation direction. 
\citet{Shukla-Marklund-Tskhakaya-Eliasson} performed a numerical simulation of the
three-dimensional system of equations (\ref{eq:nlse}) and (\ref{eq:response}). 
In Fig. \ref{fig:evol1} the collapse of the initially weakly modulated beam
$E_p = (\lambda^2\mathscr{E}_0\epsilon_0)^{-1/2}A [ 1 + 0.1\sin(40\pi k_0z)]\exp(-r_{\perp}^2/2a^2)$
can be seen. Using $A = \sqrt{0.02}$ and $a = 10k_0^{-1}$, the beam interact with the initially 
homogenous radiation gas background. The collapse is seen by the decrease in the beam width
$r_{\perp} = \sqrt{x^2 + y^2}$ and the increase in the beam energy density.
In Fig.\ \ref{fig:evol2} the evolution of an initially spherical pulse 
$E_p = (\lambda^2\mathscr{E}_0\epsilon_0)^{-1/2}A\exp(-r_{\perp}/2a^2)$ is seen.
As the pulse propagated through the initially homogeneous radiation gas, collapse
ensues, as can be seen by the decrease in the pulse width $r = \sqrt{x^2 + y^2 + z^2}$
and the increase in the pulse energy density. The values of $A$ and $a$ are the same as in the 
beam case. We note that the pulse in the last panel
undergoes splitting, and the radiation gas response develops a photon wedge, analogous to
a Mach cone, through which energy is radiated. 

\subsubsection{Pulse collapse and photonic wedges}

The approximate analytical results in conjunction with the numerical
simulations presented above, shows that 
collapse, filamentation and pulse splitting as generic features of the
system (\ref{eq:nlse}) and (\ref{eq:response}). On the other hand,
both these methods of investigation have shortcomings, and it is
therefore of great interest to obtain more analytical prediction valid
for a wider range of parameters. In this section we will show that
under very general conditions, pulse split and collapse are 
essential features of the photon pulse--acoustic system (\ref{eq:nlse}) and
(\ref{eq:response}). 

As we have seen, the dynamics of an intense short photon pulse and the
radiation background response is described by the system of equations 
(\ref{eq:nlse}) and (\ref{eq:response}).

Following \citet{Shukla-Marklund-Tskhakaya-Eliasson}, 
we re-normalize the system of equations
(\ref{eq:nlse}) and (\ref{eq:response}), by introducing the new
variables
  $E  = \left({4k_{0}\mu\lambda\epsilon_0\mathscr{E}_0}/{v_g}\right)^{1/2} 
  E_p(t, r_{\perp}, z - v_g t)$, 
  $\tau = {v_g t}/{k_0}$, and  
  $\delta\mathscr{E} = (v_g/2k_0\mu)\left( |E|^2 - \mathscr{N} \right)$.
It is implied that the wave packet has the cylindrical symmetry and moves
along the $z$-axis. From Eqs.\ (\ref{eq:nlse}) and (\ref{eq:response})
we then obtain
\begin{subequations}
\begin{equation}
2i\frac{\partial E}{\partial \tau }+\nabla^2_{\perp }E_{{}}-\beta _{z}\frac{%
\partial ^{2}}{\partial z^{2}}E_{{}}+\left| E\right| ^{2}E+\mathscr{N}E_{{}}=0,
\label{five}
\end{equation}
and 
\begin{equation}
  \left( \frac{\partial^2}{\partial\tau^2} - \frac{k_0^2c^2}{3v_g^2}
  \nabla^2 \right) \mathscr{N} = \frac{4}{3}\frac{\partial^2}{\partial\tau^2}
  \left| E(t, r_{\perp }, z - v_g t)\right|^2 ,
\label{six}
\end{equation}
\end{subequations}
respectively.

Representing $\delta\mathscr{E}$ in terms of $|E|^2$ and $\mathscr{N}$, the 
part of the perturbation of the radiation gas that is concentrated in the
region of localization of the wave packet and takes part in its
compression, has been isolated. The second part of the representation of
$\delta\mathscr{E}$ corresponds
to the interaction of the pulse with the radiation field $\mathscr{N}$. This
interaction is described by the last term in Eq.\
(\ref{five}). 
As seen by Eq.\ (\ref{six}) [or Eq.\ (\ref{eq:response})], 
the velocity $v_g \approx c$ of the source (the pulse)
exceeds the velocity $c/\sqrt{3}$ of the radiation waves, i.e.
$v_{g}>c/\sqrt{3}$, and it is therefore 
expected that the radiation field $n$ will be localized behind the
pulse. 
Hence, it is assumed that the pulse and the radiation field are localized in
different volumes, and in the region with a possible overlap the relation 
$\mathscr{N} \ll \left| E\right| ^{2}$ 
is satisfied (see also the end of the next section). This inequality allows
us to neglect the last term in Eq.\ (\ref{five}), which means neglecting 
the back reaction of the radiation on the pulse. In this approximation,
the pulse field $E$ drives the radiation field $\mathscr{N}$ through 
Eq.\ (\ref{six}), while the pulse propagation is unaffected by the radiation
field.

From Eq.\ (\ref{five}) we have the conservation of the ''field mass'' parameter
of the pulse  
\begin{subequations}
\begin{equation}
N=\int |E|^2\,d\mathbf{r}_{\perp }dz,  \label{eighteen}
\end{equation}
and the Hamiltonian 
\begin{equation}
H = \int \mathscr{H}\,d\mathbf{r}_{\perp }dz 
 \label{nineteen}
\end{equation}
\end{subequations}
where the Hamiltonian density is given by 
  $\mathscr{H} = |\mathbf{\nabla }_{\perp }E|^2 - \beta_z|\partial_z E|^2 - |E|^4/2$.

Some information on the spatio-temporal behaviour of the initially localized
wave-packet can be found by following the evolution of the characteristic
widths of the packet in the transversal and longitudinal
directions. These widths are defined by 
\begin{equation}
R^{2}(\tau )=\frac{1}{N}\int r_{\perp}^{2}|E|^2\,d\mathbf{r}_{\perp
}dz, \label{twentyone}
\end{equation}
and 
\begin{equation}
Z^{2}(\tau ) = \frac{1}{N}\int z^{2}|E|^2\,d\mathbf{r}_{\perp }dz. 
\label{twentatwo}
\end{equation}
Straightforward calculations give the following evolution equations 
\begin{equation}
\frac{\partial ^{2}R^{2}}{\partial \tau ^{2}} =
\frac{2}{N}\int \left( \left| \mathbf{\nabla }_{\perp }E\right| ^{2}-\frac{%
1}{2}\left| E\right| ^{4}\right)\, d\mathbf{r}_{\perp }dz,
\label{twentythree}
\end{equation}
and 
\begin{equation}
\frac{\partial ^{2}Z^{2}}{\partial \tau ^{2}}=
\frac{2}{N}\beta _{z}\int \left( \beta_z \left| \frac{\partial E}{\partial z}\right|^2 
+ \frac{1}{4}\left| E\right| ^{4}\right)\,
d\mathbf{r}_{\perp }dz > 0 
\label{twentyfour}
\end{equation}
for the widths (\ref{twentyone}) and (\ref{twentatwo}), respectively.

In the two-dimensional (2D) case, the process of self-compression of the
wave-packet can be clearly seen \cite{r2,r3}. A necessary condition for
collapse in this case consists in the predominance of the ''field mass'' $N$ of the
given state over some critical value $N_{c}$, i.e. $N > N_{c},$ where
$N_{c}$ is the ''field mass'' calculated for the stationary ground state \cite{r3}. 
The ground state is described by the (positive) radially-symmetric solution $\Psi
= E(r,\tau )\exp (-i\lambda \tau )$ of the equation 
\begin{equation}
\nabla _{\perp }^2\Psi -\lambda \Psi +\Psi^{3}=0,  \label{twentyfive}
\end{equation}
derived from Eq.\ (\ref{five}) and computed with $\lambda =1.$ In the
2D case, the right-hand side of the expression for the mean square 
transverse radius (\ref{twentythree}) is $2H/N$, i.e. a combination of
the conserved quantities. Thus, integrating Eq.\ (\ref{twentythree})
twice, we obtain   
\begin{equation}
R^{2}(\tau )=R^{2}(0)+C\tau +(H/N)\tau ^{2},  \label{twentysix}
\end{equation}
where the constants $R^{2}(0)$ and $C$ are defined by the initial
conditions. Hence, a sufficient condition for 2D self-focusing,
ultimately leading to complete collapse in a finite time, is $H < 0$.  
Because this is independent of the value of $C$, there will always exists a
finite time $t_{0}$ for which the transverse radius vanishes. In the
three-dimensional (3D) case and for $\beta_z < 0$ (corresponding to
anomalous dispersion) the equality (\ref{twentysix}) must be replaced
by the inequality  
\begin{equation}
R^{2}(\tau )<C_{0}+C_{1}\tau +(H/N)\tau ^{2},  \label{twentyseven}
\end{equation}
and the sufficient condition for the collapse of the wave-packet is again
$H < 0$ \cite{r3}. 

Equation (\ref{five}) with $\beta _{z} > 0$ and $\mathscr{N} = 0$ corresponds to the normal
dispersion region of the wave-packet. In \citet{r4,r5} and \citet{r6} the
features of the pulse self-focusing, described by the solution
to Eq.\ (\ref{five}), have been  
investigated in detail. Their conclusions, 
can be applied directly to the numerical results presented here 
concerning the break-up of the wave-packet. \citet{r4} showed that the
characteristic length $Z^{2}(\tau )$ of the wave-packet localization along
the direction of propagation is bounded from below by a positive constant.
The relation (\ref{twentyfour}) indicates an
asymptotic longitudinal spreading. Thus, a 3D (global) collapse cannot
take place in media with normal dispersion, since the necessary
conditions for the pulse self-focusing to occur is ($%
\partial R^{2}/\partial \tau ) < 0$ and ($\partial Z^{2}/\partial \tau
) > 0$, which explicitly excludes the 3D case. The self-focusing in the
transverse direction is accompanied by a longitudinal spreading and in the
following by the splitting of the pulse. This does not, however, 
immediately exclude partial (local) collapse scenarios. 

The process of pulse splitting close to the time of self-focusing $%
t\rightarrow t_{0}$ is described by \citet{r5,r6} employing a
\textit{ quasi-self-similar} analysis. This approach uses, besides the
description of the solution of Eq. (\ref{five}) (neglecting $\mathscr{N}$),  
the time-dependent characteristic lengths in the longitudinal $l_{z}(t)$
and transversal $l_{\perp }(t)$ directions.
\citet{r5} and \citet{r6} found that the
transverse scale $l_{\perp }(t)$ exhibits a changing behaviour
as one passes some critical point $z^{\ast }(t)$ on the $z$-axis.
Inside the localized region $z<z^{\ast }(t)$, the transverse width collapse
with the rate  
$
l_{\perp }(t) \sim (t_0 - t)^{1/2}/[\ln\{ \ln \left[ 1/(t_0 - t)\right]\}]^{1/2},  
$
while in the complementary (delocalizing) domain, $z>z^{\ast }(t)$,
the transverse pulse width spreads out with the rate  
$
l_{\perp }(t) \sim \sqrt{t\ln\{ \ln[ 1/(t_{0}-t)]\}}. 
$
Hence, the time derivative of $l_{\perp }(t)$ changes sign around the
point $z^{\ast }(t)$. Meanwhile, the self-similar longitudinal scale
$l_{z}(t) $ increases slowly in time. \citet{r5} and \citet{r6} found 
that near the self-focusing time $t\rightarrow t_{0}$, $l_{z}(t)\thicksim t$, 
which implies a linear increase of $z^{\ast }(t)$ in time such that 
$z^{\ast }(t_{0})$ $(\thicksim \sqrt{\beta _{z}}l_{z}(t_{0})).$
The presence of the coefficient $\sqrt{\beta _{z}}$ leads to the decrease of
the distance of the critical point $z^{\ast }(t_{0})$ from the origin for
small $\beta _{z}.$ The scale $l_{\perp }(t)$ remains strictly positive at
times $t\leqslant t_{0}.$ Consequently, the transversal scale reaches a
minimum value at a finite distance $z^{\ast }(t_{0}).$

Since the wave-packet is assumed to be cylindrically symmetric and also
symmetric relative to the origin $z=0,$ the total field distribution 
during self-focusing must exhibit two maxima located at $z=\pm z^{\ast
}(t_{0}),$ respectively. The wave-packet has therefore been split into two
identical smaller cells, symmetrically placed on each side of the origin $%
z=0.$

Hence, a wave-packet, propagating in media with anisotropic dispersion,
will be spread out along the direction of the negative dispersion, and
split up into smaller cells. 
These analytic results have been confirmed by numerical
solutions \cite{r6,r8}. The duration $\Delta t$ of self-focusing,
accompanied by the pulse splitting, can be estimated using Eq.\ (\ref{twentythree}) as 
\begin{equation}
\Delta t\thickapprox \frac{k_{0}}{v_{g}}R(0)\sqrt{\frac{N}{H}} .
\label{twenty.1}
\end{equation}
In the 
first part of our numerical solution --- where the splitting of the
wave-packet takes place --- 
coincides with these results. The small coefficient $\beta _{z}$, which
determines the negative dispersion, changes the distance (from the origin)
along the $z$-axis, at which the field is localized after
splitting. 

\citet{r5} and \citet{r6} have shown that this splitting process can be
continued (multi-splitting process) if the newly formed cells will possess a
transverse energy higher than the self-focusing threshold $N_{c}$. In our
case, the wave-packet splits into two cells only, as the energy localized in
each new cell is below the threshold $N_{c}$. Furthermore, 
the wave-packet also loses energy to the radiation gas during the
splitting process. The analysis of the formation of these photonic wedges, or
the radiation Mach cone, can be analyzed in accordance with the presentation in 
\citet{Shukla-Marklund-Tskhakaya-Eliasson}, such that e.g. the energy loss
from the pulse can be estimated.

In conjunction with pulse collapse in a radiation gas it should be mentioned that
if the field invariant ${\bf E}^2 - c^2{\bf B}^2 > 0$ pair creation will occur as the
pulse intensity increases, and the loss of energy through the photonic wedges will
be negligible in comparison to the energy radiated into Fermionic degrees of freedom. 
This will give rise to a rich and complex dynamical interplay 
between the pulse photons, the radiation gas, and the pair plasma \cite{Bulanov-etal2005}.
For the case of interaction between a pulse and a pure radiation gas, 
we have $c|{\bf B}| > |{\bf E}|$, due to zero dispersion, and pair creation would not occur.
However, since we will in practice always have some ionized particles present, pair creation 
is likely to be the result of pulse collapse due to weak dispersive plasma effects.

\subsubsection{The strong field case}

Our knowledge of the nonlinear refractive properties of the radiation
gas gives a means for investigating the effects of higher-order
nonlinear corrections to the standard first order Heisenberg--Euler
Lagrangian, and to probe the significance of higher-order effects for
photonic collapse 
\cite{Marklund-Brodin-Stenflo,Shukla-Eliasson,r1}. 
The dynamics of coherent photons, travelling through an intense
radiation gas, may be analysed as above, following 
\citet{Marklund-Brodin-Stenflo}. We obtain a
nonlinear Schr\"odinger equation for the slowly varying
pulse envelope $E_p$ according to \cite{Kivshar-Agrawal}   
\begin{equation}
  i\left(\frac{\partial}{\partial t} + \mathbf{v}_0\cdot\mathbf{\nabla}
  \right)E_p + \frac{v_0}{2k_0}\nabla_{\perp}^2 E_p + 
  \omega_0 \frac{n_{\mathrm{nl}}(\delta\mathscr{E})}{n_0} E_p = 0 ,
\label{eq:nonlin-nlse}
\end{equation}
where the subscript $0$ denotes the equilibrium background state, 
$\nabla_{\perp}^2 = \nabla^2 -(\hat{\mathbf{k}}_0\cdot\mathbf{\nabla})^2$, 
$\delta\mathscr{E} = \mathscr{E} - \mathscr{E}_0$ is a
perturbation,  
$v_0 = v(\mathscr{E}_0)$, $n_0 = n(\mathscr{E}_0)$, 
$n_{\mathrm{nl}}(\delta\mathscr{E}) = \sum_{m = 1}^{\infty}
n_0^{(m)}\delta\mathscr{E}^m/ m!$, $n_0^{(m)} = d^m n_0/d
\mathscr{E}_0^m$, and the refractive index $n$ is given through Eqs.
(\ref{eq:refractive2}) and (\ref{eq:Qgas}). 

For a dispersion relation $\omega = |\mathbf{k}|c/n(\mathbf{r}, t)$, the motion 
of a single photon may be described by the Hamiltonian ray equations 
(\ref{eq:groupvelocity}) \cite{Mendonca}.
Since $n = n(\mathscr{E})$ and $d n/d \mathscr{E} > 0$ always hold (see 
\citet{Marklund-Shukla-Eliasson}), a denser 
region of the radiation gas will exercise an attractive force on the photon
\cite{Partovi}, thus creating lensing effects. The single particle dynamics 
thus supports that photonic self-compression is an inherent property
of the one-loop radiation gas, but we note that as the density of a region
increases, the phase velocity approaches a constant value, given by  
(\ref{eq:constant}), i.e. $\mathbf{\nabla} \ln n \rightarrow 0$.

Following \citet{Marklund-Brodin-Stenflo}, 
the response of the
radiation gas to a plane wave pulse may be formulated in terms of an
acoustic wave equation, generalizing Eq. (\ref{eq:response}) to 
the strong field case, according to \cite{Marklund-Shukla-Eliasson}
\begin{equation}
  \left(  \frac{\partial^2}{\partial t^2} - \frac{v_0^2}{3}\nabla^2
  \right)\delta\mathscr{E} = - \frac{\mathscr{E}_0}{n_0} \left(
  \frac{\partial^2}{\partial t^2} + v_0^2 \nabla^2
  \right)n_{\mathrm{nl}}(|E_p|^2 ) .
\label{eq:acoustic}
\end{equation}
If the time response of
$\delta\mathscr{E}$ is slow, Eq.\ (\ref{eq:acoustic}) gives
\begin{equation}
  \delta\mathscr{E} 
  \approx \frac{3\mathscr{E}_0n_0'}{n_0} \left( 
   1 + \frac{n_0''}{2 n_0'}|E_p|^2 \right) |E_p|^2 .
\label{eq:deltaE}
\end{equation}
Using (\ref{eq:deltaE}) and the expression for
$n_{\mathrm{nl}}(\delta\mathscr{E})$, we can write Eq.\ (\ref{eq:nonlin-nlse}) as 
\begin{eqnarray}
  && \!\!\!\!\!\!
  i\left( \frac{\partial}{\partial t} + \mathbf{v}_0\cdot\mathbf{\nabla}
  \right)E_p + 
  \frac{v_0}{2k_0}\nabla_{\perp}^2 E_p 
\nonumber \\ && \!\!\!\!\!\! \quad 
  +
  \omega_0\left(\frac{3\mathscr{E}_0 n_0'}{n_0}\right)^2\!\!\!
  \left(
   1 + \frac{n_0''}{2 n_0'}|E_p|^2\right)|E_p|^2 E_p = 0 .
\label{eq:nlse-strong}
\end{eqnarray}
When ${n_0''}|E_p|^2/{2 n_0'} \ll 1$, we have a self-focusing
nonlinearity in Eq.\ (\ref{eq:nlse-strong}), but as $|E_p|$ grows the
character of the nonlinear coefficient changes. The coefficient is
positive when $|E_p|^2 < E_{\text{sat}}^2 \equiv |2n_0'/n_0''|$, but since
$n_0'' < 0$ for all $\mathscr{E}_0$ \cite{Marklund-Shukla-Eliasson}, 
the sign changes as the pulse amplitudes grow
above the saturation field strength $E_{\text{sat}}$, making the nonlinearity
defocusing and arresting the collapse. 
The numerical value of this turning point is
dependent on the background parameter $\mathscr{E}_0$. For low
intensity radiation gases, $n_0'' \approx 0$, and Eq.\
(\ref{eq:nlse-strong}) always displays self-focusing, i.e. the field
strengths can reach values above the Schwinger field. When
$\mathscr{E}_0$ roughly reaches the critical value
$\mathscr{E}_{\text{crit}}$, the weak field 
approximation breaks down, and Eqs.\
(\ref{eq:refractive2}) and (\ref{eq:approxlambda}) can be used to
derive an expression for $E_{\text{sat}}$. 
As an example displaying the general character of the intense
background case, consider $\mathscr{E}_0 = \mathscr{E}_{\text{crit}}\times
10^{2}$. We find that $E_{\text{sat}} \approx 2\times10^{17}\,\mathrm{V/cm}\, >
E_{\text{crit}}$, i.e. the pulse saturates above the Schwinger
critical field. Thus, both the weak and moderately strong intensity
cases, the latter desribed here by Eq.\ (\ref{eq:nlse-strong}), display 
self-compression above the Schwinger critical field. 

This analysis can be generalized to take into account the 
statistical spread in the coherent pulse, giving rise to a damping
of the instabilities \cite{Marklund}. 

\subsubsection{Other field configurations}

Above, we have seen that the propagation of an electromagnetic 
pulse through a radiation field gives rise to instabilities, wave collapse,
self-focusing, and pulse splitting. These concepts can be carried
over to the case of multiple beams or pulses propagating through
a radiation gas, including pulse incoherence. \citet{Marklund-Shukla-Brodin-Stenflo}
showed that when several pulses are present, they can exchange energy via 
a background radiation gas and instabilities can occur, even if their propagation 
is parallel.  

\citet{Shukla-Marklund-Brodin-Stenflo} considered an incoherent non-thermal high-frequency spectrum of
photons. As will be shown, this spectrum can interact with
low-frequency acoustic-like perturbations. The high-frequency part is
treated by means of a wave kinetic description, whereas the
low-frequency part is 
described by an acoustic wave equation with a driver 
\cite{Marklund-Brodin-Stenflo} which follows from a radiation fluid
description.  The high-frequency photons drive
low-frequency acoustic perturbations according to
\cite{Marklund-Brodin-Stenflo} 
\begin{eqnarray}
  &&\!\!\!\!\!
  \left( \frac{\partial^2}{\partial t^2} - \frac{c^2}{3}\nabla^2
  \right)\mathscr{E} 
\nonumber \\&& \!\!\!
  =  -\frac{2\lambda\mathscr{E}_0}{3}\left(
  \frac{\partial^2}{\partial t^2} + c^2\nabla^2 \right) \int \hbar\omega
  f(\mathbf{k}, \mathbf{r}, t)\, d^3k ,
\label{eq:wave}
\end{eqnarray}
where the constant $\mathscr{E}_0$ is the \emph{background} radiation
fluid energy density and $f$ is the high frequency photon distribution function. 
This hybrid description, where the high-frequency 
part is treated kinetically, and the low-frequency part is described within 
a fluid theory, applies when the mean-free path between 
photon-photon collisions is shorter than the wavelengths of the low-frequency 
perturbations.  We note that the specific intensity $I_k = \hbar\omega
f/\epsilon_0$ satisfies Eq.\ (\ref{eq:kinetic}), and is normalized
such that $\langle|E|^2\rangle = \int\,I_k\,d^3k$, where $E$ is the 
high-frequency electric field strength. These equations  
resemble the photon--electron system in the paper by 
\citet{Shukla-Stenflo-PoP}, where the interaction between randomly phased
photons and sound waves in an electron--positron plasma has been
investigated.   

Next, we consider a small low-frequency long wavelength perturbation of
a homogeneous  
background spectrum, i.e. $f = f_{0} + f_{1}\exp [i(Kz -
  {\mno}t)]$, $|f_{1}| 
\ll f_{0}$ and $ \mathscr{E} = \mathscr{E}_1\exp[i(Kz - {\mno}t)]$ and 
linearize our equations. Thus, we obtain
the nonlinear dispersion relation
\begin{equation}
  1 = -\frac{\mu K}{3}\frac{{\mno}^2 +
  K^2c^2}{{\mno}^2 - K^2c^2/3}\int\frac{k^2}{{\mno}
  -Kc\hat{\mathbf{k}}\cdot\hat{\mathbf{z}}}\hat{\mathbf{z}}\cdot\frac{\partial
  f_{0}}{\partial\mathbf{k}} \,d^3k ,
\label{eq:kinetic-dispersion}
\end{equation}
where  $\mu = \frac{4}{3}\lambda^2c^2\hbar\mathscr{E}_0$. 

(a) For a mono-energetic high
frequency background, we have $f_{0} = n_0\delta(\mathbf{k} - \mathbf{k}_0)$.
The nonlinear dispersion relation (\ref{eq:kinetic-dispersion})
then reduces to 
\begin{eqnarray}
  && 
  ({{\mno}^2 - K^2c^2/3}){({\mno} -
  Kc\cos\theta_0)^2} 
\label{eq:disprel-mono}
  \\ && 
  = \frac{\mu n_0k_0 K}{3}
  ({{\mno}^2 +
  K^2c^2})
  [{Kc + (2{\mno} -3Kc\cos\theta_0)\cos\theta_0}] ,
\nonumber
\end{eqnarray} 
where we have introduced $\cos\theta_0 \equiv
\hat{\mathbf{k}}_0\cdot\hat{\mathbf{z}}$. This mono-energetic background has a
transverse instability when $\theta_0 = \pi/2$, with the
growth rate 
\begin{equation}
  {\mng} =  \frac{Kc}{\sqrt{6}}\left[ \sqrt{ \left(\frac{v_T}{c}\right)^4
  + 14\left(\frac{v_T}{c}\right)^2 + 1} - \left(\frac{v_T}{c}\right)^2 - 1
  \right]^{1/2} ,
\label{eq:growth1}
\end{equation}
where ${\mng} \equiv -i{\mno}$, and $v_T \equiv
(\mu n_0k_0c)^{1/2}$ is a characteristic speed of the
system. The expression in the 
square bracket is positive definite. 

\begin{figure}[ht]
  \includegraphics[width=.7\columnwidth]{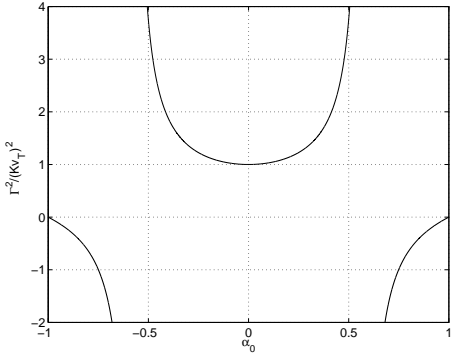}
\caption{$(\mng/Kv_T)^2$, according to Eq.\ (\ref{eq:growth2}),
  plotted as a function of
  $\alpha_0 = \cos\theta_0$ in the mono-energetic case. (Reprinted from \citet{Shukla-Marklund-Brodin-Stenflo}, Copyright (2004), with permission from
Elsevier.)}
\label{fig2}
\end{figure}

In fact, when $\mathscr{E}_{0}\hbar ck_{0}n_{0}/E_{\text{crit}}^{4} 
\ll 1$, a condition which is satisfied due to (\ref{eq:constraint1}), we 
have $v_T \ll c$. Using the   
expression (\ref{eq:disprel-mono}), we then have two
branches. The branch corresponding to $\mno \approx
Kc/\sqrt{3}$ is always stable for small $v_T$, while for the branch
corresponding to $\mno \approx Kc\cos\theta_0$ we obtain the
growth rate  
\begin{equation}
  \mng = Kv_T\sqrt{\frac{1 - \cos\theta_0}{1 - 3\cos\theta_0}} ,
\label{eq:growth2}
\end{equation} 
which is consistent with (\ref{eq:growth1}) in the limit
$\theta_0 \rightarrow \pi/2$. In Fig.\
\ref{fig2}, the behavior of the growth rate 
(\ref{eq:growth2}) is depicted.

(b) The high-frequency photons have generally a 
spread in momentum space. For simplicity, we 
here choose the background intensity distribution as a
shifted Gaussian, i.e. 
\begin{equation}\label{eq:gaussian-distribution}
  I_{k0} = \frac{\mathscr{I}_0}{\pi^{3/2}k_W^3}\exp\left[
  - \frac{(\mathbf{k} - \mathbf{k}_0)^2}{k_W^2}\right] ,
\end{equation}
where $\mathscr{I}_0 = \langle|E_0|^2\rangle$ 
is the (constant) background intensity and $k_W$
is the width of the distribution around $\mathbf{k}_0$. 
Assuming that the deviation of $\mathbf{k}_0$ from the $\hat{\mathbf{z}}$-axis
is small, and that $\delta \equiv k_0/k_W \ll 1$, we can integrate Eq.\
(\ref{eq:kinetic-dispersion}) with (\ref{eq:gaussian-distribution}), keeping terms linear in $\delta$, to
obtain   
\begin{eqnarray}
  1 &\approx& -\pi b^2\frac{\eta^2 + 1}{\eta^2 - 1/3}\Bigg[ 
  \frac{3\sqrt{\pi}}{2} + 8\delta\eta\cos\theta_0
\\ && \!\!\!\!\!\!\!\!\!\!\!\!\!\!\!\! \!\!\!\!\!\!\!\!
  +
  \left(\delta\cos\theta_0 - \frac{3\sqrt{\pi}}{4}\eta
  -4\delta\eta\cos\theta_0 \right)(2\,\text{arctanh}\,\eta - i\pi)  
  \Bigg] ,
\nonumber 
\end{eqnarray}
for $0 < \eta < 1$. Here $\eta \equiv {\mno}/Kc$ and 
$b^2 = (4/9\pi^{3/2})\lambda^2\epsilon_0\mathscr{E}_0
\mathscr{I}_0\exp(-k_0^2/k_W^2)$. 
Thus, we see that the non-zero
width of the distribution complicates the characteristic behavior of
the dispersion relation considerably. It is clear 
that the width will reduce of the growth rate 
compared to the mono-energetic case. 
  
We may also look at the case when the time-dependence is weak, i.e.
$\partial^2\mathscr{E}/\partial t^2 \ll c^2\nabla^2\mathscr{E}$,
such that Eq.\ (\ref{eq:wave}) yields 
  $\mathscr{E} = 2\lambda\mathscr{E}_0\int\hbar\omega_k f\,d^3k $.
Upon using this relation, we find that 
  ${\nabla\omega}  = -\mu k\nabla\int\,k' f' \,d^3k'$.  
Hence, Eq.\ (\ref{eq:kinetic}) becomes
\begin{equation}
  \frac{\partial f}{\partial t} + 
  \mathbf{v}_g\cdot\frac{\partial f}{\partial\mathbf{r}} +
  \mu k
  \left(\frac{\partial}{\partial\mathbf{r}}\int k'f' \,d^3k' \right)%
  \cdot\frac{\partial f}{\partial\mathbf{k}} = 0 .
\label{eq:self}
\end{equation}
A similar equation may be derived for the specific intensity
$I_k$. 
Equation (\ref{eq:self}) describes the evolution of high-frequency photons
on a slowly varying background radiation fluid, and it may be used to
analyze the long term behavior of amplitude modulated intense short
incoherent laser pulses. The results in this section can be generalized to 
several partially coherent pulses \cite{Marklund-Shukla-Brodin-Stenflo}.

\subsection{Effects due to plasmas}

Plasma channels are closely connected to both the plasma dispersion 
and the propagation of wave modes in waveguides. 
\citet{Shen-etal} and \citet{Shen-Yu} first suggested
the use of plasma cavitation as a means of fostering conditions
in which nonlinear quantum vacuum effects, such as photon--photon
scattering, could take place.
As a high intensity electromagnetic pulse propagates through a 
plasma, the interaction with the plasma may completely evacuate
regions giving conditions similar to the that of Secs. \ref{sec:planes}
and \ref{sec:cavity}.

Moreover, although in many cases the presence of a plasma
will swamp the effects due to photon--photon scattering, it can under certain 
circumstances provide a means for the propagation of non-classical 
plasma modes. Due to the nontrivial dispersion of electromagnetic
waves in plasmas, there will be a net effect due to photon--photon 
scattering, such that low-frequency modes will be generated. In general,
the effects of a nonlinear quantum vacuum is expected to become
pronounced for next generation lasers \cite{Bulanov-etal2,Bulanov-etal3,mou05}

Also, for future applications, the combination of photon--photon 
scattering induced pulse compression in conjunction with 
pair creation \cite{Nitta-etal} could provide interesting insights both 
into fundamental properties of the quantum vacuum as well as
into the prospects of creating high power electromagnetic sources.
The effects of plasmas within the environment of a quantum
vacuum therefore deserve further investigations. 

\subsubsection{Plasma cavitation and plasma channels}

If the power of the laser pulse propagating through the plasma 
surpasses the critical value $P_{\text{crit}} = 17(\omega/\omega_p)^2 \, 
\mathrm{GW}$, where $\omega$ is the laser frequency and $\omega_p$ 
the electron plasma frequency, there may be complete expulsion of plasma
particles from the high intensity region \cite{Max-etal}, thus
forming a wave-guide \cite{Shen-Yu}. In such wave-guides, the effects
of photon--photon scattering could be of importance \cite{Shen-etal}

The nonlinear interactions of plasmas with high intensity lasers is of
great current interest (see, e.g.  \citet{gol99,she02,Bulanov-etal,%
Shukla-Eliasson-Marklund,cai04,Tajima-Taniuti}, and \citet{Pukhov,mou05}
for recent reviews).  In the context of doing fundamental physics 
and mimicking astrophysical events in laboratory environments, the
evolution of laser intensities has received a lot of attention. Examples of 
experimental suggestions are axion detection (as a dark matter candidate)
\cite{Bernard,Dupays-etal,Bradley-etal}; pair production (see, e.g. \citet{Ringwald,Ringwald1,%
Ringwald2}, and \citet{Bamber-etal,bur97,Meyerhofer} for a discussion
of the detection of pair production from real photons); laboratory calibration
of observations, relativistic jets, analogue general relativistic event horizon 
experiments (such as Hawking and Unruh radiation \cite{Hawking,Unruh}),
and probing the quantum spacetime properties \cite{Chen,Chen-Tajima}.
The possibility of reaching extreme power levels 
with such setups is one of the promising aspects of laser-plasma 
systems \cite{bob}, and also holds the potential of overcoming the laser 
intensity limit $\sim 10^{25}$ W/cm$^{2}$ \cite{mou98}. As the field strength 
approaches the critical Schwinger field $E_{\text{crit}} \sim 10^{16}$ V/cm
\cite{Schwinger}, there is possibility of photon--photon scattering, 
even within a plasma \cite{Shen-etal}, as the ponderomotive force due to the 
intense laser pulse gives rise to plasma channels \cite{Yu-etal}. Under such 
extreme circumstances, the effects of pair creation will be pronounced.
Electron-positron plasmas are also produced by interactions of matter with 
powerful multi-terawatt and petawatt laser pulses \cite{liang98,gahn00}. 
The concept of trident pair-production, as described within the framework
of perturbation theory, could give  
a means for creating electron--positron pairs by intense laser pulses
in vacuum \cite{ber92}.  
Moreover, the future x-ray free electron laser systems 
\cite{pat02,Ringwald,Ringwald1,Ringwald2}
could result in methods 
for creating pair plasmas in the laboratory \cite{alk01}. The possible
field strength output could reach $E \approx 0.1E_{\text{crit}}$
\cite{alk01}. Even on an experimental level, pair-production due to collisions of electron backscattered photons with 
the original photon beam has been observed \cite{bur97}. Thus, there are ample evidence that 
the investigation of nonlinear interactions of pair-plasmas and high intensity
electromagnetic fields deserves attention \cite{kozlov79,farina01}. 

\paragraph{The effect of relativistic nonlinearities}

Here we follow \citet{Shukla-Eliasson-Marklund1}. 
The propagation of a circularly polarized intense laser pulse 
in an unmagnetized plasma is governed by \cite{Yu-etal}
 $(\partial_t^2 -c^2 \nabla^2 ){\bf A} 
+ (\omega_p^2 N/\gamma) {\bf A}=0$, 
where ${\bf A}$ is the vector potential of the laser pulse, $\omega_p
=(n_0e^2/\epsilon_0m_e)^{1/2}$ is the unperturbed electron plasma frequency, $\gamma
=\sqrt{1+ e^2 |{\bf A}|^2/m^2c^2}$ is the relativistic gamma factor including
the electron mass variation in intense laser fields, and $N=n_e/n_0$ is the 
ratio of the electron number density to the background plasma number 
density $n_0$.

\begin{figure}[ht]
\includegraphics[width=.8\columnwidth]{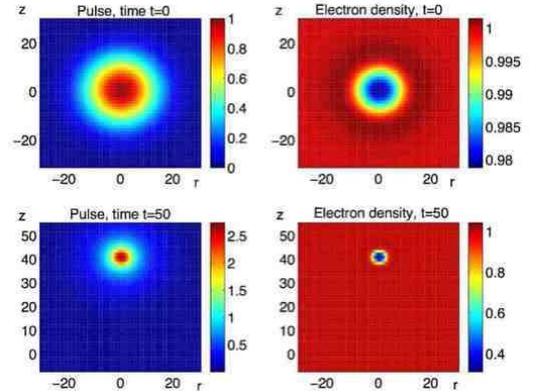}
\caption{The variation of $I$, as given by Eq.\ (\ref{eq:envelope}), 
and $N$ against $r$ (the radial coordinate)
and $z$ for an initial laser pulse which initially has a Gaussian shape. 
We observe the pulse compression and the formation of a  light bullet. 
(Reprinted with permission from \citet{Shukla-Eliasson-Marklund1}.)}
\label{fig:relnlin}
\end{figure}

At intensities beyond $10^{18}$ W/cm$^2$, the electron quiver speed 
$v_{osc}=6 \times 10^{-10}c\lambda \sqrt{I}$ exceeds the speed of light, and
hence nonlinear effects in  plasmas cannot be ignored. Here $I$ is the intensity 
in W/cm$^2$ and $\lambda$ is the laser wavelength in microns.  Thus, the relativistic 
ponderomotive force \cite{Yu-etal,Shukla-etal} $ {\bf F} = -m_ec^2 \nabla \gamma$\, of intense laser 
pulses will separate charges and thereby would create a huge ambipolar electric 
potential $\phi$ in the plasma.  At equilibrium, the balance between the relativistic 
ponderomotive force and a slow electric force $e \nabla \phi$ will yield 
$\phi =(m_ec^2/e)(\gamma-1)$, which, when substituted into Poisson's equation, gives 
$N = 1+  \lambda_e^2 \nabla^2 \gamma$.  Here $\lambda_e =c/\omega_p$ is 
the electron skin depth, and the ions are assumed to be immobile. The electron density  
will be locally evacuated by the relativistic ponderomotive force of 
ultra-intense nonuniform laser fields. The laser pulse localization and 
compression would then occur due to nonlinearities associated with
relativistic  laser ponderomotive force created electron density evacuation 
and relativistic electron mass increase in the laser fields. This phenomena can 
be studied by means of the equation 
\begin{equation}
\frac{\partial^2  \bm{\mathcal{A}}}{\partial t^2 }
-\nabla^ 2\bm{\mathcal{ A}} + \frac{\bm{\mathcal{ A}}}{\sqrt{1+|\bm{
      \mathcal{ A}}|^2}}  
\left(1+ \nabla^2 \sqrt{1+|\bm{\mathcal{ A}}|^2}\right)=0,
\label{eq:compression}
\end{equation}
where $\bm{\mathcal{A}} = e{\bf A}/m_ec$, and the time and space
variables are in units of  $\omega_p^{-1}$ and $\lambda_e$, 
respectively.  For the propagation of a modulated laser pulse along 
the $z$ axis, we obtain from Eq.~(\ref{eq:compression}) after invoking the
slowing varying envelope approximation, 
\begin{eqnarray}
2i\omega \left(\frac{\partial I}{\partial t}+v_g\frac{\partial I}{\partial z}\right)
+ \nabla^2 I + I-\frac{I}{P}(1+\nabla^2 P)=0,
\label{eq:envelope}
\end{eqnarray}
where we have set $\bm{\mathcal{A}} = (1/2) I(r, z, t) (\hat {\bf x}
+ i \hat {\bf y}) \exp(-i \omega t+i k z)+$ complex conjugate, and denoted 
$P=(1+I^2)^{1/2}$.  Here the normalized laser frequency and the
normalized laser group  velocity are denoted by $\omega =(1+k^2)^{1/2}$ 
and $v_g=k/\omega$, respectively.   In the one-dimensional case [viz. set 
$\nabla^2=\partial^2/\partial z^2$ in Eq. (\ref{eq:envelope})],  we have the localization 
of intense electromagnetic waves in the form of a large  amplitude 
one-dimensional bright soliton \cite{Yu-etal}. We have numerically solved Eq.~(\ref{eq:envelope})  
in order to study the evolution of a cylindrically symmetric modulated
laser pulse. The results are displayed in Fig.~\ref{fig:relnlin}.

Initially, the pulse is assumed to have a Gaussian shape, 
$I = I_0 \exp[-(z^2+r^2)/200]$, and we used the normalized wavenumber
$k=2$ and  initial pulse amplitude $I_0=1$.  We notice that the compression of
the pulse envelope  (left panels), which is correlated with the excavation of the
normalized electron  density (right panels). Our numerical results reveal that
self-compression of the pulse is more rapid when one accounts for the relativistic light
ponderomotive force induced  electron density depletion, contrary to the constant density 
case (viz. $N=1$).  Physically, the enhanced compression and self-focusing of an intense laser
pulse occur due to the localization of light in a self-created electron density cavity. 

The fact that evacuation of the plasma takes place as the light
intensification due to self-compression occurs means that the
situation discussed in \citet{Shen-etal}, where the quantum
electrodynamical effect of photon--photon scattering at high
intensities takes place, could be realized in the next generation
laser-plasma systems. Moreover, as the intensities in the evacuated
region increase, the concept of vacuum catastrophic collapse, at
which the pulse due to quantum vacuum nonlinearities self-compresses,
may ensue \cite{r1}. The intensities that can be reached at this
stage, in principle, surpass the Schwinger field, but then the process
of pair creation has to be investigated and removed, since this would 
otherwise quickly dissipate the electromagnetic energy into Fermionic 
degrees of freedom. The problem of self-consistent analysis 
of pair production in a plasma environment has been approached by 
\citet{Bulanov-etal2005} where a model for incorporating a particle 
source term was given (see also Eq.\ (\ref{eq:paircreation})).

\paragraph{Self-interaction in electron--positron plasmas}

Consider the propagation of intense light in
an electron-positron plasma. 
By averaging the inertialess equations of motion for 
electrons and positrons over one electromagnetic wave period, 
the expressions for the electron and positron number densities $n_e$ and 
$n_p$, respectively, 
in the presence of relativistic ponderomotive force \cite{Shukla-etal} of an
arbitrary large amplitude laser pulse takes the form
\begin{eqnarray}
  n_{e,p} = n_0\exp\left[ -\frac{m_ec^2}{k_BT_e}\left(\gamma-1\right) -
  \frac{q_{e,p}}{k_BT_{e,p}}\phi\right] , 
\end{eqnarray}
where $q_e = -e$ and $q_p = e$ are the electron and positron charges, 
$k_B$ is Boltzmann's constant, $T_e (T_p)$ is the electron (positron) temperature, and  
$\gamma =\sqrt{1+ e^2 |\mathbf{A}|^2/m_e^2 c^2}$ for 
circularly polarized light. The
ambipolar potential $\phi$  
associated with the plasma slow motion is found from  Poisson's equation
\begin{equation}
  \nabla^2\phi = \frac{e}{\epsilon_0}(n_e - n_p).
  \label{eq:poisson}
\end{equation}
From the particle momentum conservation equations, the electron and positron 
velocities are given by 
  $\mathbf{v}_{e,p} = -q_{e,p}\mathbf{A}/m_e\gamma_{e,p} $,
where $\gamma_{e,p}= (1- v_{e,p}^2/c^2)^{-1/2} \equiv \gamma$. 

The dynamics of the intense light is obtained from Maxwell's equations and reads
\begin{equation}
  \left( \frac{\partial^2}{\partial t^2} - c^2\nabla^2\right)\mathbf{A} +
  \frac{\omega_p^2}{\gamma n_0}(n_e + n_p)\mathbf{A} = 0 , 
\label{eq:waveA}
\end{equation}
where ${\bf A}$ is the vector potential. 

For a circularly polarized electromagnetic wave
$\mathbf{A}=(1/2){\widetilde A} (\mathbf{r}, t)
(\hat{\mathbf{x}}+\mathrm{i}\hat{\mathbf{y}}) 
\exp(\mathrm{i}\mathbf{k}\cdot\mathbf{r}-\mathrm{i}\omega_0t)$, 
using the scalings $n_{e,p} = n_0N_{e,p}$, $t = \tau\omega_0/\omega_p^2$, $\mathbf{r} = c(\mathbf{\xi} - 
\mathbf{u}_g\tau)/\omega_p$, $\mathbf{u}_g = (\omega_0/\omega_p)\mathbf{v}_g/c$,
$\widetilde{A} = (m_ec/e)\mathcal{A}$, and $\phi = (m_ec^2/e)\Phi$, Eqs.\ (\ref{eq:poisson}) and 
(\ref{eq:waveA}) can be written in the dimensionless form as 
\begin{equation}
\mathrm{i}\frac{\partial\mathcal{A}}{\partial \tau}
+\frac{1}{2}\nabla^2\mathcal{A} +\left(1-\frac{N_e+N_p}{2\sqrt{1+|a|^2}}\right)\mathcal{A}
=0,
\label{eq:nlse-A}
\end{equation}
and
  $\nabla^2\Phi=N_e-N_p$,
respectively, where  
$N_e=\exp[\beta_e(1-\sqrt{1+|\mathcal{A}|^2}+\Phi)]$,
$N_p=\exp[\beta_p(1-\sqrt{1+|\mathcal{A}|^2}-\Phi)]$,
$\beta_{e,p}=c^2/v_{Te,p}^2$, and
$v_{Te,p}=(k_BT_{e,p}/m_e)^{1/2}$. 
Here the dispersion relation 
$\omega_0^2=c^2k^2+\omega_p^2(n_{e0}+n_{p0})/n_0$ has been used, and 
$\mathbf{v}_g = (c^2/\omega_0)\mathbf{k}$ is the group velocity. 
In the quasi-neutral limit
$N_e=N_p$, we have  
$\Phi=(1-\sqrt{1+|\mathcal{A}|^2})(\beta_p-\beta_e)/(\beta_p+\beta_e)$. 
Equation (\ref{eq:nlse-A}) then becomes
\begin{equation}
\mathrm{i}\frac{\partial\mathcal{A}}{\partial \tau}
+\frac{1}{2}\nabla^2\mathcal{A} +\left[1-\frac{\exp\left[
\beta\left(1-\sqrt{1+|\mathcal{A}|^2}\right)
\right]}{\sqrt{1+|\mathcal{A}|^2}}\right]\mathcal{A} = 0,
\label{eq:nlse-A2}
\end{equation}
where $\beta=2\beta_e\beta_p/(\beta_e+\beta_p)$ is the temperature
parameter.

\begin{figure}[ht]
      \includegraphics[width=.8\columnwidth]{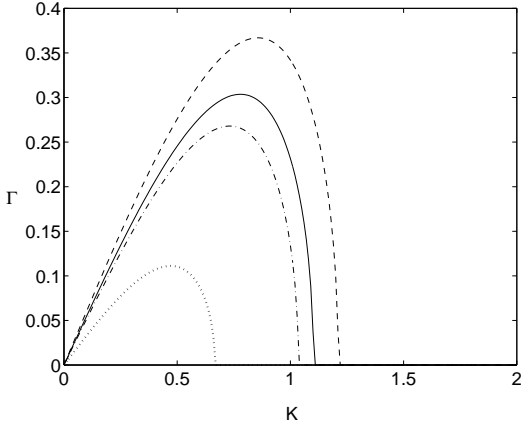}
    \caption{The modulational instability growth rate ${\mng}$ as given by 
    Eq.\ (\ref{eq:nonlin2}) versus the 
    wavenumber $K$, for $\beta=200$ and $a_0=0.1$ (dashed lines), 
    $\beta=100$ and $a_0=0.1$ (solid line), $\beta=100$ and $a_0=0.2$
    (dash-dotted lines),  
    and for $\beta=100$ and $a_0=0.05$ (dotted line). 
(Reprinted from \citet{Shukla-Marklund-Eliasson}, Copyright (2004), with permission from Elsevier.)}
    \label{fig:instability}
\end{figure}

The dispersion relation for the modulational and filamentational instabilities 
for an arbitrary large amplitude electromagnetic pump can be derived
from (\ref{eq:nlse-A2}) following standard techniques \cite{shu87,shu88}. From the ansatz
$\mathcal{A} = (a_0+a_1)\exp(i\delta\tau)$,  where $a_0$ is real, $a_0\gg |a_1|$ and 
$\delta$ is a constant nonlinear frequency shift, the lowest order 
solution gives  
  $\delta = 1 - \exp\left[\beta\left(1 - \sqrt{1 +
  a_0^2}\right)\right]/\sqrt{1 + a_0^2}$.  
Linearizing Eq.\ (\ref{eq:nlse-A2}) with respect to $a_1$, with the ansatz 
$a_1 = (X + \mathrm{i}Y)\exp(\mathrm{i}\mathbf{K}\cdot\mathbf{\xi} -
\mathrm{i}{\mno}\tau)$, where $X$ and $Y$ are real constants and 
${\mno}$ ($\mathbf{K}$) is the frequency (wavevector) of the low-frequency
(in comparison with the light frequency) modulations, the nonlinear 
dispersion relation reads 
\begin{equation}
  {\mno}^2 = \frac{K^4}{4} -\frac{K^2}{2}\frac{a_0^2\left(1 + \beta\sqrt{1 +
  a_0^2}\right)}{(1 + a_0^2)^{3/2}} 
  e^{\beta(1 - \sqrt{1 + a_0^2}\,)} , 
  \label{eq:nonlin1}
\end{equation}
which gives the modulational instability growth rate
${\mng} =-\mathrm{i}{\mno}$ according to  
\begin{equation}
  {\mng} = \frac{K}{\sqrt{2}}\left[\frac{a_0^2\left(1 + 
  \beta\sqrt{1 + a_0^2}\right)}{(1 + a_0^2)^{3/2}}
  e^{\beta(1-\sqrt{1+a_0^2}\,)} -
  \frac{K^2}{2}\right]^{1/2}.
  \label{eq:nonlin2}
\end{equation}
The growth rate increases
with the larger $\beta$ values (i.e. for lower temperature), while
for the intensity field, we do not necessarily obtain higher
growth rates for higher intensities, see Fig. \ref{fig:instability}. 
This is attributed to an interplay between the 
relativistic particle mass variation and the relativistic light
ponderomotive driven density responses. From the expression  
(\ref{eq:nonlin2}) one observes a decrease in the growth rate for 
large enough $\beta$. However, this result should be 
interpreted with caution
since for large-amplitude fields in a low-temperature plasma, electron
inertia effects will become important and may dominate
over the thermal effects. Then, in this case the assumption that the
electrons (and positrons) obey a modified Boltzmann distribution may 
no longer be valid. 

\begin{figure}[ht]
    \includegraphics[width=.8\columnwidth]{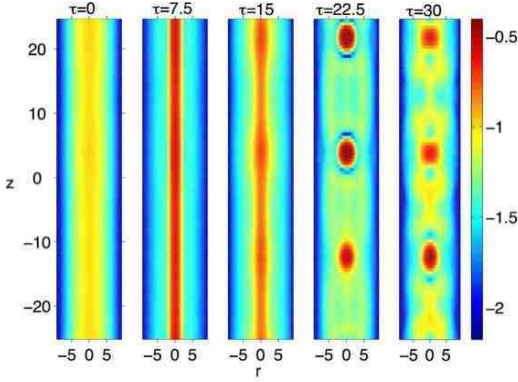}
    \caption{The light amplitude $\mathcal{A}$ in $\log_{10}$-scale
    at five different times $\tau$ (see Eq.\ (\ref{eq:nlse-A2})). The horizontal
    axis represents the distribution along the radial ($r$)
    coordinate, and the axial ($z$) distribution is on the 
    vertical axis.
(Reprinted from \citet{Shukla-Marklund-Eliasson}, Copyright (2004), with permission from Elsevier.)}
    \label{fig:intensity}
\end{figure}

\begin{figure}[ht]
      \includegraphics[width=.8\columnwidth]{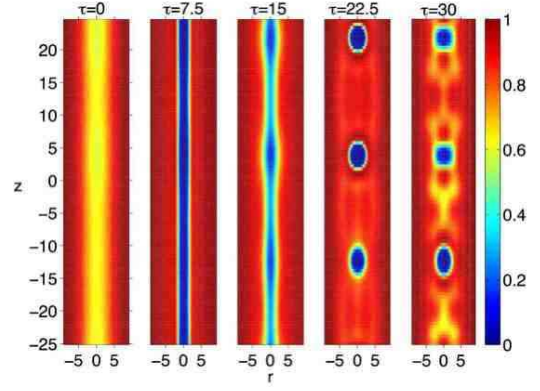}
    \caption{The normalized electron number density
    $N_e = \exp[\beta(1 - \sqrt{1+|\mathcal{A}|^2})]$ 
    as a function of $r$ and
    $z$ at five different times $\tau$ ($\mathcal{A}$ is given by Eq.\
    (\ref{eq:nlse-A2})). Note the
efficient expulsion of plasma particles in the later panels, and their correlation to 
the intensity peaks in Fig. \ref{fig:intensity}.
(Reprinted from \citet{Shukla-Marklund-Eliasson}, Copyright (2004), with permission from Elsevier.)}
    \label{fig:density}
\end{figure}

Since $N_e = N_p = \exp[\beta(1 - \sqrt{1 + |\mathcal{A}|^2})]$, the increase of
the pulse intensity will cause an almost complete expulsion of the  
electrons and positrons from that region, see Figs. \ref{fig:intensity} and \ref{fig:density}. 
The simulations show the evolution of an initially weakly modulated
beam $\mathcal{A} = 10^{-3}[1 + 0.02\sin(z/8) + 0.02\cos(z/4) 
  + 0.02\cos(3z/8)]\exp(-r^2/32)$, where $r^2 = x^2 + y^2$, while the electron
perturbation is zero initially. Thus, as can be seen in Figs.\ \ref{fig:intensity} 
and \ref{fig:density}, the modulated beam self-compresses and breaks up into 
localized filaments. This gives rise to electron and positron holes, and as
the pulse intensity grows, the conditions for pure elastic photon--photon 
scattering improve within these holes.  

\paragraph{Thin-foil amplification}\label{sec:thinfoil}

As noted in the previous two sections, trapping and amplification of
laser pulses can take place given the right plasma environment.
This could be an important tool for stepping up the available electromagnetic
intensities, and could therefore be important for investigations into
photon--photon scattering. 
Here we will describe a method which could yield high intensity pulses.

Laser-foil interactions have been used as a method for proton acceleration on table
top scales \cite{Zepf-etal,Silva-etal,McKenna-etal}. By letting a high intensity 
laser pulse impinge normally on a thin metal foil the foil material is ionized, creating a plasma in which
protons are accelerated up to MeV-energies. The exact mechanism(s) behind the proton 
acceleration is still not completely clear although there exists a number of plausible 
suggestions (see \citet{Zepf-etal} for a discussion). It was suggested by \citet{Shen-MeyerterVehn2}
that this could be used to create confined high density relativistic electron plasmas. Letting 
two counterpropagating laser beams illuminate a thin foil normally, a spatially confined 
high density plasma could be created, and be used for, e.g. harmonic generation, pair production,
and $\gamma$ photon generation \cite{Shen-MeyerterVehn}. 

Building on the work of \citet{Shen-MeyerterVehn2},   
\citet{she02} suggested to let two oppositely directed laser beams interact
via two closely placed thin foils. As above, when the high intensity lasers impinges normally on the
thin foils, the foil material will be ionized and a plasma will be produced. 
\citet{she02} showed that this may lead to electromagnetic trapping.

\citet{she02} started with the trapping of a circularly polarized electromagnetic pulse propagating
in the $z$-direction in a
positive electron density profile, using the stationary equations of
\citet{Shen-MeyerterVehn2} and \citet{Shen-MeyerterVehn} 
\begin{subequations}
\begin{eqnarray}
  &&\!\!\!\!\!\!\!\!\!\!\!\!\!\!\!\!\!\!\!\!\! \!\!\!\!\!
  M = (\gamma^2 - 1){c}\theta'/\omega , \\
  &&\!\!\!\!\!\!\!\!\!\!\!\!\!\!\!\!\!\!\!\!\! \!\!\!\!\!
  W =  \tfrac{1}{2}(\gamma^2 - 1)^{-1}\left[ \left({c\gamma'}/{\omega}\right)^2 
    + M^2 \right] + \tfrac{1}{2}{\gamma}(\gamma - 2N_i) ,
\end{eqnarray}
\label{eq:constantsofmotion}
\end{subequations}
where $M$ and $W$ are two constants motion, $\gamma = \sqrt{1+ e^2 |A|^2/m_e^2 c^2}$ is the
relativistic gamma-factor, $\omega$ is the laser frequency, $A$ is the vector potential 
$\propto \exp[i\omega t + i\theta(z)]$, $N_i = {n_i}/{n_c}$,  $n_i$ is the constant ion density,  
$n_c = 1.1\times 10^{21}\,(\lambda/\mu\mathrm{m})^2\,\mathrm{cm}^{-3}$ is the 
critical electron density, $\lambda$ is the laser wavelength, and the prime denotes 
differentiation with respect to $z$. 
It is possible to find solitary solutions of Eq.\ (\ref{eq:constantsofmotion}) 
representing trapped electromagnetic pulses between parallel high density 
plasma regions \cite{Kim-etal,Esirkepov-etal3}.
These analytical soliton solutions suggest the possibility to trap laser light 
between foils. \citet{she02} performed particle-in-cell (PIC) simulations of the system 
(\ref{eq:constantsofmotion}). Using the foil spacing $\Delta = 0.46\lambda$
they showed that such configurations would yield a 100-fold amplification of
the initial laser pulse intensities, over a trapping time of $26$ laser cycles,
which is in the fs range for $\mu\mathrm{m}$ lasers. 

In multidimensional environments, true analytical soliton solutions
are not known, but it is a well-established fact, due to approximate and 
numerical investigations \cite{Kivshar-Agrawal}, that solitary like solutions
exists in dimensions $\geq 2$. However, these solutions are unstable, and will
suffer either attenuation or self-compression, depending on intensity and pulse
width \cite{Desaix-Anderson-Lisak} (see Fig. \ref{fig:2D-collapse}). Thus,
in a two-dimensional thin-foil environment, light intensification could take 
place. The production of high-intensity pulses by these thin-foil amplification
also has the valuable property of being realizable in a relatively small scale 
setting.

\paragraph{Laser-plasmas and relativistic flying parabolic mirrors}

As we have seen above, the propagation of intense electromagnetic pulses
in plasmas yields interesting nonlinear dynamics, and effects such as 
pulse self-compression can occur. These nonlinear effects act as a very 
promising tool for, e.g. producing intense ion beams \cite{Esirkepov-etal,%
Bulanov-etal2,Esirkepov-etal2}, which is of importance in laboratory 
astrophysics \cite{Chen}. With regards to the nonlinear
quantum vacuum, an interesting proposal has been put forward by 
\citet{Bulanov-etal}. The self-compression of laser pulses, towards intensities  
close to the Schwinger limit, can take place by using relativistic flying 
parabolic mirrors. 

\citet{Bulanov-etal} consider a plasma wakefield in the
wave-breaking regime. They let a short intense laser pulse create a 
wakefield in a plasma, such that the wakefield phase velocity equals the laser pulse 
group velocity (which is close to $c$) in an underdense plasma 
\cite{Tajima-Dawson}.\footnote{We note that the formation of sub-cycle intense 
  solitary waves could penetrate into a highly overdense plasma 
  \cite{Shen-Yu-Li}, transferring energy between low- and high-density
  regions of the plasma, which could be of importance in, e.g. laser fusion.}
Due to the nonlinearity, the resulting wake field will experience wave 
steepening entering the wavebreaking regime, together with a local electron 
density spike approaching infinity. With such a set-up, a  
sufficiently weak counter-propagating laser pulse will be partially 
reflected from the electron density maximum. The relativistic dependence of the 
Langmuir ``mirror'' on the driving laser pulse intensity will cause bending
of the surfaces of constant phase, thus creating a parabolic plasma mirror
\cite{Bulanov-Sakharov}. 
This curvature of the plasma mirror will focus
the counterpropagating (weak) pulse to a spot size $\lambda/4\gamma^2$ along 
the mirror paraboloid axis in the 
laboratory frame, where $\lambda$ is the wavelength of the source of the reflected pulse 
and $\gamma$ is the relativistic 
gamma factor of the wakefield. Similarly, the
focal spot width is $\lambda/2\gamma$ in the transverse direction. With this,
\citet{Bulanov-etal} showed that the intensity gain will be roughly
$64(D/\lambda)^2\gamma^3$, where $D$ is the width of the pulse effectively 
reflected by the mirror. The focal spot intensity of the reflected pulse
can then be estimated to
\begin{equation}\label{eq:focusintensity}
  I_{\text{focal}} \approx 8\left(\frac{\omega_d}{\omega}\right)^2\left(\frac{D}{\lambda}\right)^2\gamma^3 I,
\end{equation}
where $\omega_d$ is the frequency of the pulse driving the Langmuir wave, $\omega$
is the frequency of the source of the reflected pulse, and $I$ is the intensity of the source of
the reflected pulse. \citet{Bulanov-etal} gave the following example of light intensification through
Eq.\ (\ref{eq:focusintensity}). A $1\,\mu\mathrm{m}$ pulse generate the Langmuir mirror
in a plasma where $n_e \sim 10^{17}\,\mathrm{cm}^{-3}$. The estimate 
$\gamma \approx \omega_d/\omega_p$ then gives $\gamma \sim 100$. The 
pulse to be reflected is assumed to have $I \sim 10^{17}\,\mathrm{W/cm^2}$, and
$D = 400\,\mu\mathrm{m}$. Then $I_{\text{focal}} \sim 10^{29}\,\mathrm{W/cm^2}$,
to be compared with the critical intensity $I_c = c\epsilon_0E_{\mathrm{crit}}^2 \approx 
3\times 10^{29}\,\mathrm{W/cm^2}$. The estimated focal intensity thus
seems to reach the Schwinger limit. However, with the given value on $I$ it is likely 
that the backreaction on the Langmuir wave has to be taken into account, thus altering the
estimate. 

\begin{figure}[ht]
  \includegraphics[width=0.7\columnwidth]{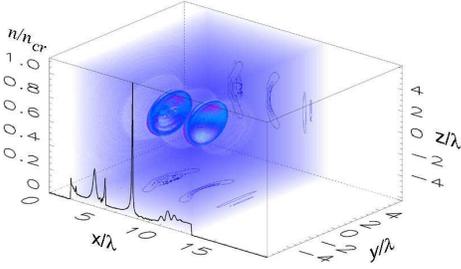}
  \caption{Typical electron density profile in the wake field,  
  where the isosurfaces represents densities $n = 0.15 n_{\text{crit}}$,
  and $n_{\text{crit}}$ denotes the density at which the plasma
  goes from underdense to overdense.
(Reprinted with permission from \citet{Bulanov-etal}.)}
\label{fig:bul1}
\end{figure}

\begin{figure}[ht]
  \includegraphics[width=0.7\columnwidth]{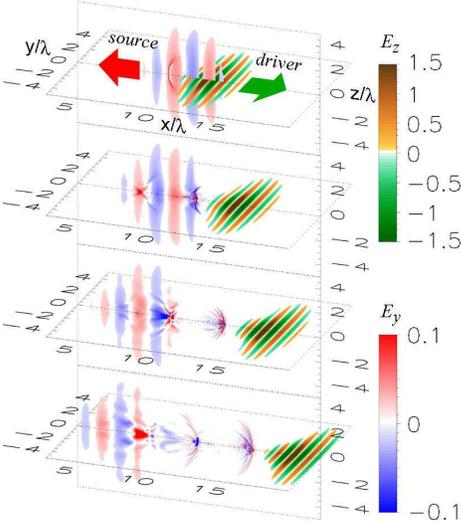}
  \caption{The electric field components at different times.
(Reprinted with permission from \citet{Bulanov-etal}.)}
\label{fig:bul2}
\end{figure}

\citet{Bulanov-etal} also presented numerical results using a fully relativistic code,
see Figs. \ref{fig:bul1} and \ref{fig:bul2}. Using a three-dimensional PIC code, an intense laser pulse
drives the Langmuir wave along the $x$-axis, while the counterpropagating source pulse
for the reflected wave has an intensity $\sim 10^{15}\,\mathrm{W/cm^2}$ in the $\mu\mathrm{m}$
wavelength range. The value of the 
of the source pulse intensity is chosen in order to avoid degradation of the Langmuir
mirror. Figure \ref{fig:bul1} shows the electron density profile of the Langmuir wave.
Using a Langmuir laser driver with intensity $4\times 10^{18}\,\mathrm{W/cm^2}$ and
$\mu\mathrm{m}$ wavelength, the plasma waves move with with a phase velocity $0.87c$,
and the gamma-factor is $2$. The density profile is shown along the $x$-axis, and a steep 
gradient can be seen. Figure \ref{fig:bul2} shows the electric field components of the 
source pulse and its reflection ($y = 0$ plane) and the Langmuir driver ($z = 0$ plane). The focusing
of the reflected  can be seen. The intensity increase in the focal spot is $256$ times the source intensity,
i.e. $I_{\text{focal}} \sim 10^{17} - 10^{18}\,\mathrm{W/cm^2}$ for a $\mu\mathrm{m}$
source laser. This is similar to the intensification obtained for the thin foil setup in Sec. 
\ref{sec:thinfoil}.

\paragraph{Electromagnetic wave localization}

The nonlinear interaction of high-intensity ultrashort electromagnetic 
waves with hot plasmas is of primary interest for the fast ignitor concept 
of inertial confinement fusion and for the development of high
power sources of hard EM radiation, as well as for laser-plasma particle 
and photon accelerators, and compact astrophysical objects containing 
intense electromagnetic bursts.  Recent progress in the development of 
super strong electromagnetic pulses with intensities $I \sim 10^{21}$--$10^{23}$ 
W/cm$^2$ has also made it possible to create relativistic plasmas in the laboratory 
by a number of experimental techniques. At the focus of an ultraintense 
short electromagnetic pulse, the electrons can acquire velocities close 
to the speed of light, opening the possibility of simulating in  
laboratory conditions, by using dimensionless simulation parameters, 
phenomena that belong to the 
astrophysical realm.  In the past, several authors presented theoretical 
\cite{kozlov79b,Kaw,Esirkepov-etal3,Farina} and particle-in-cell 
simulation \cite{bulanov99,naumova} studies of intense electromagnetic envelope 
solitons in a cold plasma, where the slow plasma response to the EM waves is modeled by 
the electron continuity and relativistic momentum equations, supplemented by 
Poisson's equation. Assuming beam-like particle distribution functions, relativistic 
electromagnetic solitons in a warm quasi-neutral electron-ion plasma have been 
investigated \cite{lontano}. Experimental observations \cite{borghesi} 
show bubble-like structures in proton images of laser-produced plasmas, 
which are interpreted as remnants of electromagnetic envelope solitons. 

\citet{Shukla-Eliasson2} presented fully relativistic nonlinear theory and computer simulations 
for nonlinearly coupled intense localized circularly polarized EM waves and relativistic electron hole (REH) structures \cite{Eliasson-Shukla2006} in a relativistically 
hot electron plasma, by adopting the Maxwell-Poisson-relativistic Vlasov system that  
accounts for relativistic electron mass increase in the electromagnetic fields and 
relativistic radiation ponderomotive force \cite{Shukla-etal,bob2}, in addition to trapped 
electrons which support the driven REHs. Such a scenario of coupled intense EM waves 
and REHs is absent in any fluid treatment \cite{kozlov79b,Kaw,Esirkepov-etal3,Farina} of relativistic 
electromagnetic solitons in a plasma. Electromagnetic wave localization is a topic of significant interest in photonics \cite{Mendonca}, as well as in compact astrophysical objects, e.g. gamma-ray bursts \cite{Piran}. 

The electromagnetic wave equation accounting for the relativistic
electron mass increase and the electron density modification due to the radiation 
relativistic ponderomotive force \cite{Mendonca} $F=-m_e c^2 \partial \gamma/\partial z$, 
where  
$  \gamma= (1+ p_z^2/m_e^2 c^2 +e^2 |\mathbf{A}|^2/m_e^2c^2)^{1/2}$ is the relativistic 
gamma factor, are included.  Here, $p_z$ is the $z$ component of the electron momentum, $\mathbf{A}$ is 
the perpendicular (to $\hat {\bf z}$, where $\hat {\bf z}$ is the unit vector along 
the $z$ axis) component of the vector potential of the 
circularly polarized EM waves. The dynamics 
of nonlinearly coupled EM waves and REHs is governed by
\begin{equation}\label{eq:vecpotential-localization}
  \frac{\partial^2 \mathbf{A}}{\partial t^2}-
  \frac{1}{\alpha^2}\frac{\partial^2 \mathbf{A}}{\partial z^2}+
  \int_{-\infty}^{\infty} \frac{f}{\gamma}\,dp_z\,\mathbf{A}=0,
\end{equation}
\begin{equation}\label{eq:vlasov-localization}
  \frac{\partial f}{\partial t}+\frac{p_z}{\gamma}\frac{\partial f}{\partial z}
  +\frac{\partial(\phi-\gamma/\alpha^2)}{\partial z}\frac{\partial f}{\partial p_z}=0,
\end{equation}
and
\begin{equation}\label{eq:poisson-localization}
  \frac{\partial^2\phi}{\partial z^2}=\int_{-\infty}^{\infty} f\,dp_z -1,
\end{equation}
where $\mathbf{A}$ is normalized by $m_e c/e$, $\phi$ by $k_BT_e/e$, $p_z$ by $m_e V_{Te}$ and
$z$ by $r_D$. Here $\gamma=(1+\alpha^2 p_z^2+|{\bf A}|^2)$, $V_{Te}=(k_BT_e/m_e)^2$,
$\alpha=V_{te}/c$, and $r_D=V_{Te}/\omega_p$.
In Eq. (\ref{eq:vecpotential-localization}), we used the Coulomb gauge 
$\nabla \cdot \mathbf{A} =0$ and 
excluded the longitudinal ($z$-) component $\partial^2\phi/\partial t\partial z=j_z$, 
where $j_z$ is the parallel current density, by noticing that this component is 
equivalent to Poisson's equation (\ref{eq:poisson-localization}) \cite{Shukla-Eliasson2}. 

\begin{figure}[floatfix]
\includegraphics[width=.8\columnwidth]{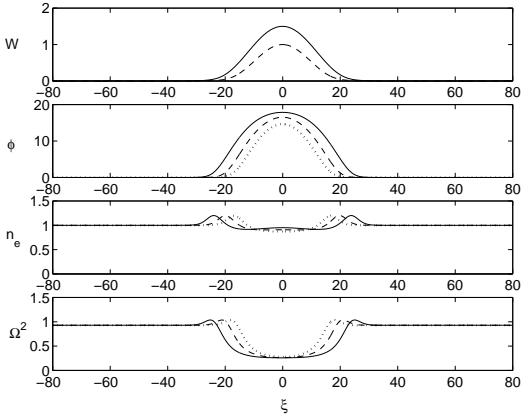}
\bigskip
\caption{
Large-amplitude trapped EM wave envelope (upper panel), 
the potential (second panel), the electron number density (third panel),
and the square of the local electron plasma 
frequency (lower panel) for large amplitude EMWwaves with a 
maximum amplitude of $W_{\text{max}}=1.5$ (solid lines) and 
$W_{\text{max}}=1.0$ (dashed lines), and as a comparison a REH
with small-amplitude EM waves which have $W_{\text{max}}\ll 1$ (dotted lines).
The parameters are: the normalized speed $v_0=0.7$, $\alpha=0.4$ and the trapping parameter $\beta=-0.5$ (corresponding to a vortex distribution presented by \citet{bajurbarua} and \citet{schamel} involving an equilibrium Synge--J\"uttner distribution function 
\cite{deGroot}).
The selected value of $\beta$ are related to the maximum REH potential 
according to a specific relation similar to one in \citet{bajurbarua}
and \citet{schamel}. 
(Reprinted with permission from \citet{Shukla-Eliasson2}.)}
\label{fig:loc1}
\end{figure}

Shukla and Eliasson have discussed the stationary as well as time dependent solutions of Eqs.\ (\ref{eq:vecpotential-localization})--(\ref{eq:poisson-localization}) in the form of REH which traps localized electromagnetic wave envelopes. Typical profiles for the amplitude of the localized EM vector potential $W$ and potential and density of the REH, as well as the local electron plasma frequency squared ($\Omega^2$) including the relativistic electron mass increase, are depicted in Fig.\  \ref{fig:loc1}. We observe that for large electromagnetic field the REH potential becomes larger and the REH wider, admitting larger eigenvalues $\lambda$ that are associated with the nonlinear frequency shift. This is due to  
the relativistic ponderomotive force of localized EM waves pushes the electrons away from the 
center of the REH, leading to an increase of the electrostatic potential and 
a widening of the REH.  We see that the depletion of the electron density in the REH is only 
minimal, while the local electron plasma frequency $\Omega$ is strongly reduced 
owing to the  increased mass of the electrons that are accelerated by the 
REH potential; the maximum potential $\phi_{\text{max}} \approx 15$ in Fig. \ref{fig:loc1} corresponds
in physical units to a potential 
$\alpha^2\phi_{\text{max}}\times0.5\times 10^6\approx 1.2\times 10^6\,\mathrm{V}$, 
accelerating the electrons to gamma factors of $\approx 6$.

\begin{figure}[floatfix]
\includegraphics[width=.8\columnwidth]{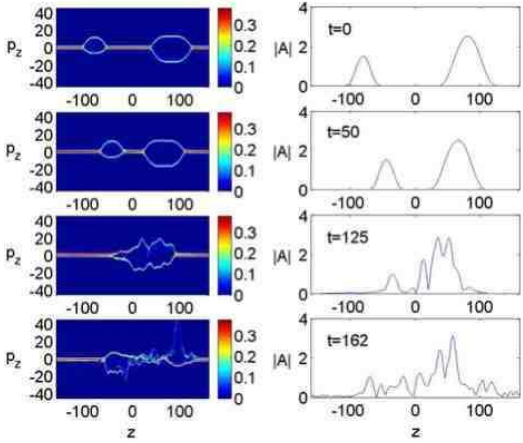}
\bigskip
\caption{Phase space plots of the electron distribution function (left panels) and
the modulus of the electromagnetic field (right panels) for $t=0$, $t=50$,
$t=125$ and $t=162$. (Reprinted with permission from \citet{Shukla-Eliasson2}.)
}
\label{fig:loc3}
\end{figure}

\begin{figure}[floatfix]
\includegraphics[width=.75\columnwidth]{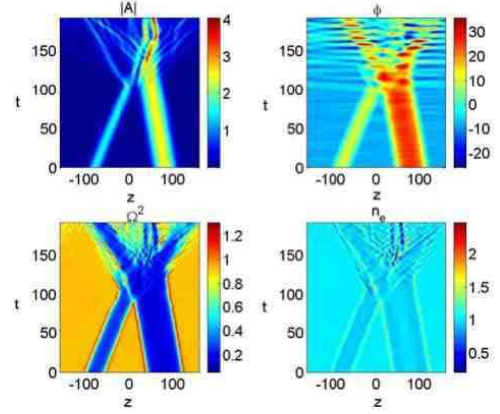}
\bigskip
\caption{The electromagnetic field (upper left panel), potential (upper right panel),
squared local plasma frequency (lower left panel) and electron density (lower right panel)
for two colliding REHs. (Reprinted with permission from \citet{Shukla-Eliasson2}.)
}
\label{fig:loc4}
\end{figure}

In order to study the dynamics of interacting solitary structures composed of localized 
REHs loaded with trapped EM waves, \citet{Shukla-Eliasson2} numerically solved the 
time-dependent, relativistic Vlasov 
equation (\ref{eq:vlasov-localization}) together with the nonlinear Schr\"odinger equation, that is deduced from Eq.\ (\ref{eq:vecpotential-localization}) in the slowly varying envelope approximation. 
The results are displayed in 
Figs. \ref{fig:loc3} and \ref{fig:loc4}. 
As an initial condition to the simulations, Shukla and Eliasson used solutions to the 
quasi-stationary equations described above, where the left 
REH initially has the speed $v_0=0.7$ (normalized by $c$) and is loaded with EM waves with $W_{\text{max}}=1.5$, while the right 
REH has the speed $v_0=-0.3$, and is loaded with EM waves with $W_{\text{max}}=2.5$.  Furthermore, 
\citet{Shukla-Eliasson2} used $k_0=v_g=0$
in the initial condition for $A$ and in the solution of the nonlinear Schr\"odinger equation \cite{Shukla-Eliasson2}.  
Figure \ref{fig:loc3} displays the 
phase space distribution of the electrons and the electromagnetic field amplitude 
at different times.  We see that the REHs loaded with trapped EM waves collide, 
merge and then split into two REHs, while there are two strongly peaked EM wave
envelopes at $z\approx 30$ and $z\approx 70$ remaining after the splitting of the REH.
A population of electrons has also been accelerated to large energies, seen at $z=100$
in the lower left panel of Fig. \ref{fig:loc3}.  The time development 
of the EM wave amplitudes, REH potential, the squared local plasma frequency and the electron 
number density is shown in Fig. \ref{fig:loc4}. Collision and splitting of the REHs can be observed, as well as creation of the two localized EM envelopes at $z\approx 70$; clearly visible in the left two panels at $t>150$. 

\subsubsection{Photon--photon scattering within plasmas}

\paragraph{Charged particle effects and Cherenkov radiation}

As presented by \citet{Dremin} and \citet{NJP}, similar to a charged particle moving in 
a isotropic dielectric, a charged particle can suffer Cherenkov losses
when propagating through a intense gas of photons. The main 
difference compared to the case of a regular medium is the 
frequency spectrum of the emitted radiation. Since the natural 
cut-off in the quantum vacuum is given by the Compton frequency,
$\gamma$-rays may be emitted by such a
particle. 

In 1934, Cherenkov observed the type of radiation now bearing his name
\cite{Cerenkov}. His experimental result was explained by \citet{Tamm-Frank}. In an isotropic 
dielectric medium, a charged particle in rectilinear motion  
satisfying the so called Cherenkov condition, i.e.\  its
velocity exceeds the (parallel) phase speed in the medium in which it 
moves, will radiate \cite{Chefranov}. The radiation shock-front, called the Cherenkov cone, is
analogous to the Mach cone formed as objects move with supersonic 
speeds through air. 
In quantum mechanical terms, the Cherenkov condition corresponds to energy
and momentum conservation. 
Cherenkov radiation has technological uses, e.g.\  in
determining particle velocities. 

The dispersion relation for electromagnetic waves in an isotropic
and homogeneous photon gas with refractive index $n$ 
is $\omega = kc/n$, where $n^2 = 1 + \delta$  
and $\delta = 4\lambda\mathscr{E}/3$ \cite{Bialynicka-Birula,Marklund-Brodin-Stenflo}
(see Eq. (\ref{eq:weakgas})).
Thus, the refractive index 
in this case is always larger than one, and a particle may
therefore have a speed $u$ exceeding the phase velocity in the
medium. The Cherenkov condition $u \geq c/n$ for emission of  
radiation can thus be satisfied. This condition can also be expressed in terms of the 
relativistic gamma factor $\gamma = (1 - u^2/c^2)^{-1/2}$, namely  
$\delta\gamma^2 \geq 1 $.  
We will here assume that a particle with charge $Ze$,  
satisfying the Cherenkov condition, moves through an equilibrium 
radiation gas.  
The energy loss at the frequency $\omega$ per unit length of the path of the 
charged particle is then 
\begin{equation}
  \frac{dU_{\omega}}{ds}d\omega = \frac{Z^2\alpha}{c}\frac{(\delta\gamma^2 - 1)}{(\gamma^2 - 1)}
  \hbar\omega \,d\omega ,
\end{equation}
and the number of quanta $N$ emitted per unit length along the particles path 
is 
\begin{equation}
  \frac{dN}{ds} d\omega= \frac{Z^2\alpha}{c} \frac{(\delta\gamma^2 - 1)}{(\gamma^2 - 1)} \, d\omega .
\end{equation} 
Since $\delta$ is normally much less than one, we need a large gamma factor to satisfy the 
Cherenkov condition. Subsequently, for $\delta \gamma^2 =1$,
we have 
\begin{equation}
  U = N\hbar\omega_e, \quad \text{ and } \quad N = Z^2L\alpha\delta/\lambda_e ,
\end{equation}
respectively, where we have used the Compton frequency as a cut-off. 
Here $L$ is the distance traveled by the charge. 

At the present time, the cosmic microwave background has an energy
density of the order $\mathscr{E} \sim 10^{-15}\,\mathrm{J}/\mathrm{m}^3$, i.e.\
$\delta \sim 10^{-42}$, i.e. the gamma factor has to be 
$\gamma \geq 10^{21}$ for the Cherenkov condition to be satisfied.
Thus Cherenkov radiation is not likely to occur in todays radiation
background. In fact, it is well known that the cosmic rays contain non-thermal hadrons, of which some 
are protons, that can reach gamma factors $10^{11}$, but larger values are
improbable due to the GZK cut-off \cite{Greisen,Zatsepin-Kuzmin}. 
As a comparison, we may consider the situation at the time of
matter--radiation decoupling. 
Since $\mathscr{E}_{\text{emitted}} = \mathscr{E}_{\text{received}}(T/2.7)^4$, 
where the temperature $T$ is given in Kelvin, we have $\mathscr{E} \sim 10^{-2} \, \mathrm{J/m^3}$ 
at the time of decoupling ($T \approx 8000\, \mathrm{K}$), implying $\delta \sim 10^{-28}$. 
Thus, the limiting value on the gamma factor 
for the Cherenkov condition to be satisfied is $\gamma \geq 10^{14} - 10^{15}$,
still out of reach for high energy cosmic rays.
However, as we demonstrate below, the situation changes
drastically for earlier processes at even higher $T$. In particular, we will focus on the era with
$10^9\,\mathrm{K} \leq T \leq 10^{11}\,\mathrm{K}$ when the required $\gamma$-factors range from
$\gamma \sim 10^4$ to $\gamma >3$.

The effect presented above is naturally compared with inverse Compton
scattering. Setting $Z = 1$, the cross-section for this scattering is $\sigma \approx \pi r_e^2m_e^2/M^2\gamma$, 
where $r_e$ the classical electron radius and $M$ is the charged particle mass. We thus 
obtain a collision frequency $\nu = c\mathscr{N}\sigma$, where $\mathscr{N}$ is 
the number density of the photons. Comparing this frequency with 
the frequency $\nu_{\mathrm{ch}} = (\gamma Mc)^{-1}dU/dt$, we note that fast 
particles are mainly scattered due to the Cherenkov effect when $\nu < \nu_{\mathrm{ch}}$, i.e.
\begin{equation}\label{eq:cherenkov}
  1 <  \frac{\delta}{\alpha\pi(m_e/M)\mathcal{N}\lambda_e^3} =  \frac{M}{m_e}\frac{T}{T_{\mathrm{ch}}}.
\end{equation}
Here $T$ is the temperature of the photon gas,  
$\mathcal{N} = [30\zeta(3)a/k_B\pi^4]T^3$, $\mathscr{E} = aT^4$, 
$k_B$ is the Boltzmann constant,  
$a = \pi^2k_B^4/15\hbar^3c^3 \approx 7.6\times 10^{-16}\,\mathrm{J/m^3K^4}$ 
and $T_{\mathrm{ch}}= (2025\zeta(3)/44\pi^3\alpha)m_ec^2/k_B \approx 10^{12}\,\mathrm{K}$
using the polarization averaged effective action charge $\bar{\lambda} = (8\kappa + 14\kappa)/2 = 11\kappa$.
Thus, for a single fast proton to be scattered mainly due to the Cherenkov effect,  
we need $T > T_{\mathrm{ch}}\times 10^{-3} \sim 10^9 \, \mathrm{K}$, well 
within the limit of validity of the theory for photon--photon scattering.    
We note that at radiation gas temperatures around $10^{12}\,\mathrm{K}$ the 
quantum vacuum becomes truly nonlinear, and higher order QED effects must 
be taken into account. 

For the early universe considered above,  
a moderately relativistic plasma is also present, which means that collective 
charged particle interactions can play a role. We take  
these plasma effects into account by introducing the plasma frequency $\omega_p$. 
The photon dispersion relation is $\omega
^{2}\approx k^{2}c^{2}(1-\delta )+\omega _{p}^{2}$. Thus, the
Cherenkov condition is satisfied for charged particles with relativistic factors $\gamma \geq 1/%
\sqrt{\delta -\omega _{p}^{2}/k^{2}c^{2}}$. For the temperatures where the Cherenkov
radiation starts to dominate over inverse Compton scattering, $T\sim
10^{9}-10^{10}\,\mathrm{K}$, we have  $\omega _{p}\sim 10^{15-16}\,\mathrm{rad/s}$,
and thus Cherenkov radiation is emitted in a broad band starting in the 
UV range, $\omega\sim 10^{17}\, \mathrm{rad/s}$, and continuing 
up to the Compton frequency $\sim 8\times 10^{20} \,\mathrm{rad/s}$. 

The Cherenkov radiation emitted during the era when $T \sim 10^9\,\mathrm{K}$
will be redshifted due to the cosmological expansion. Thus, the present value of the 
cut-off frequency will be approximately $2\times 10^{12}\,\mathrm{rad/s}$, i.e. in
the short wavelength range of the microwave spectrum. However, we do not expect 
direct detection of this radiation in the present universe, since the process is only 
expected to be of importance long before the time of radiation decoupling. Still, there are 
possible important observational implications due to the Cherenkov mechanism
presented here. As shown by the inequality (\ref{eq:cherenkov}), the effect will be more pronounced
for massive particles with a given gamma factor, and protons are therefore expected to 
be more constrained than electrons by the QED Cherenkov emission. In particular,
(\ref{eq:cherenkov}) puts stronger limits than Compton scattering for  
supra-thermal protons observed today to be relics of the early universe.
In fact, it seems rather unlikely, given the inequality (\ref{eq:cherenkov}), that
such protons could survive during the $T = 10^9 - 10^{10}\,\mathrm{K}$ era.

\paragraph{Unmagnetized plasmas}

Pair production and pair plasmas play an
important role in the dynamics of the environments surrounding pulsars (see, e.g.
\citet{Beskin-etal,Arendt-Eilek,Asseo}). Charged particles will
attain relativistic energies close to the pulsar magnetic poles and
radiate $\gamma$-ray photons. This, together with the super-strong 
magnetic field present around these objects \cite{Beskin-etal},
is believed to produce a pair plasma \cite{Tsai-Erber}.
Thus, nonlinear QED effects are already known to be an important
ingredient for pulsar physics. Since the pair plasma gives rise to
radio wave emissions, and because of the large energy scales involved,  
pulsar atmospheres are likely to host other QED effects as well, such
as vacuum nonlinearities in the form of photon--photon scattering. 

As presented by \citet{Stenflo-etal}, 
for circularly polarized electromagnetic waves propagating  in a cold
multicomponent plasma rather than in vacuum, the wave operator on the
left-hand sides of Eqs.\ (\ref{WaveE}) and (\ref{WaveB}) is replaced
by 
\begin{equation}
\square \rightarrow \frac{1}{c^{2}}\frac{\partial ^{2}}{\partial t^{2}}
-\nabla ^{2}+\frac{\omega _{p}^{2}}{c^{2}}\rightarrow \frac{-\omega
^{2} + \sum_j\omega _{pj}^{2}/\gamma_j}{c^{2}}+k^{2},  \label{substitution}
\end{equation}
where the sum is over particle species $j$, 
we have assumed that the EM-fields vary as $\exp(ikz-i\omega t)$, 
the relativistic factor of each particle species is  
$\gamma_j = (1 + q_j^2 E_0^2/m_j^2 c^2 \omega^2)^{1/2}$,
where $E_0$ denotes the absolute value of the electric field amplitude 
\cite{Stenflo,Stenflo-Tsintsadze}.
Due to the symmetry of the circularly polarized EM waves,
most plasma nonlinearities cancel, and the above substitution 
holds for arbitrary wave amplitudes. 
Here $\omega_{pj} = (n_{0j}q_{j}^{2}/\epsilon_{0}m_{j})^{1/2}$ is
the plasma frequency of particle species $j$ and $n_{0j}$ denotes the particle
density in the laboratory frame.

Next, we investigate the regime $\omega^2 \ll k^2 c^2$. From Faraday's
law and the above inequality we note that the dominating QED 
contribution to Eq. (\ref{WaveB}) comes from the term proportional to $B^{2}
{\bf B}$. Combining  Eqs. (\ref{WaveB}) and (\ref{eq:magnetization}), noting 
that $B^2 = B_0^2 = k^2 E_0^2/\omega^2 $\,\, is constant for
circularly polarized EM waves, and using $\omega^2 \ll k^2 c^2$,
i.e. ${\bf M} \approx 4\kappa\epsilon_0^2c^4B^{2}{\bf B}$ and
$|{\bf M}| \gg \omega |{\bf P}|/k$, we 
obtain from (\ref{WaveB}) the nonlinear dispersion relation 
\begin{equation}
\omega^2 = \frac{2\alpha}{45\pi}\left(
  \frac{E_0}{E_{\text{crit}}} \right)^2 \frac{k^4 c^4}{\sum_j\omega_{pj}^2/\gamma_j + k^2
  c^2} .
  \label{Dispersion-relation} 
\end{equation}
This low-frequency mode makes the particle motion ultra-relativistic
even for rather modest wave amplitudes. For electrons and positrons in
ultra-relativistic motion ($\gamma _{j}\gg 1$) with equal densities $n_{0}$
and elementary charge $\pm e$, we thus use the approximation
$\sum_j\omega_{pj}^2/\gamma_j \approx 2en_0 c\omega /\epsilon_0E_0 = 2\omega_{p}^2
(\omega/\omega_e) (E_{\text{crit}}/E_0)$ (see (\ref{eq:criticalfield}) and (\ref{eq:constraint1})), 
where $\omega_{p} = (e^2 n_0/\epsilon_0 m_e)^{1/2}$. 
The dispersion relation (\ref{Dispersion-relation}) then reduces to 
\begin{equation}
\omega^3 = \frac{\alpha}{45\pi}\left( \frac{\omega_e}{
  \omega_{p}} \right) \left(
  \frac{E_0}{E_{\text{crit}}} \right)^3\frac{k^4 c^4}{\omega_{p} + 
  (E_0/E_{\text{crit}})(kc\omega_e /2\omega\omega_{p})kc} .
\label{Relativistic-DR}
\end{equation}
We note that the ratio $\omega_e/\omega_{p}$ is much larger than unity for 
virtually all plasmas, i.e.  for electron densities up to $\sim 10^{38}$ 
m$^{-3}$.  In some applications, such as in pulsar astrophysics, it is
convenient  
to re-express the dispersion relation in terms of the
relativistic gamma factor using $E_0/E_{\text{crit}} \approx
(\omega/\omega_e)\gamma$. Thus, we obtain
\begin{equation}
  \lambda  = \gamma\lambda_e\left( \frac{4\alpha}{45\pi} \right)^{1/2}\left[
    1 + \sqrt{1 + \frac{16\alpha}{45\pi}\left(
  \frac{\omega_{p}}{\omega_e} \right)^2\gamma } \right]^{-1/2} 
\label{wavelength}
\end{equation}
from (\ref{Relativistic-DR}) for the wavelength $\lambda = 2\pi/k$.

\paragraph{Magnetized plasmas}

Following \citet{Marklund-Shukla-Stenflo-Brodin-Servin} 
(see also \citet{Marklund-Shukla-Brodin-StenfloC}), 
for a circularly polarized wave $\mathbf{E}_0 = E_{0} 
(\hat{\mathbf{x}} \pm i\hat{\mathbf{y}})\exp(ikz
  - i\omega t)$
propagating along a constant magnetic field ${\bf B}_0 = 
B_{0}\hat{\mathbf{z}}$, the electromagnetic invariants satisfy
\begin{equation} 
  F_{cd}F^{cd} = -2E_0^2\left( 1 - \frac{k^2c^2}{\omega^2}\right) 
    + 2c^2B_0^2 
  \, \text{ and } \,
  F_{cd}\widehat{F}^{cd} = 0 .
\end{equation}
Thus, Eq.\ (\ref{eq:maxwell}) can be written as
\begin{equation}
  \Box A^a = -4\epsilon_0\kappa\left[ E_0^2\left( 1 - \frac{k^2c^2}{\omega^2}\right) 
    - c^2B_0^2 \right]\Box A^a - \mu_0 j^a 
  \label{eq:wave-A}
\end{equation}
in the Lorentz gauge, and $\Box = \partial_a\partial^a$. 
For circularly polarized electromagnetic waves 
propagating in a magnetized cold multicomponent plasma, the four current can 
be `absorbed' in the wave operator on the left-hand side by the replacement
(as in the previous section)
  $\Box \rightarrow -D(\omega, k) $,
where $D$ is the plasma dispersion function, given by 
(see, e.g. \citet{Stenflo,Stenflo-Tsintsadze})
\begin{equation}
  D(\omega,k) = k^2c^2 - \omega^2 + 
  \sum_j\frac{\omega\omega_{pj}^2}{\omega\gamma_j \pm \omega_{cj}} .
\label{eq:dispersionfunction}   
\end{equation}
Here the sum is over the plasma particle species $j$,  
  $\omega_{cj} = {q_jB_0}/{m_{j}}$ and 
  $\omega_{pj} = ({n_{0j} q_j^2}/{\epsilon_0 m_{j}})^{1/2} $
is the gyrofrequency and plasma frequency, respectively, and 
  $\gamma_j = (1 + \nu^{2}_j)^{1/2}$ 
is the gamma factor of species $j$, with $\nu_j$ satisfying
\begin{equation}
  \nu^{2}_j = \left(
  \frac{eE_0}{cm_{j}} \right)^2\frac{1 + \nu^{2}_j}{[\omega(1 +
  \nu^{2}_j)^{1/2} \pm \omega_{cj}]^2} .
  \label{eq:nu}
\end{equation}
Here $n_{0j}$ denotes particle
density in the laboratory frame and $m_{j}$ particle rest mass. 

The dispersion relation, obtained from 
Eq.\ (\ref{eq:wave-A}), reads
\begin{equation}
  D = \frac{4\alpha}{45\pi}(\omega^2 - k^2c^2)
  \left[\left( \frac{E_0}{E_{\text{crit}}}
  \right)^2\frac{\omega^2 - k^2c^2}{\omega^2} - \left(\frac{cB_0}{E_{\text{crit}}}\right)^2 
  \right] .
\label{eq:qeddisp}
\end{equation}
We note that as the 
plasma density goes to zero, the 
effect due to photon--photon scattering, as given by the right-hand 
side of Eq.\ (\ref{eq:qeddisp}), vanishes, since then 
$\omega^2 - k^2c^2 = 0$

Next, we focus on low-frequency ($\omega \ll kc$) mode propagation in an ultra-relativistic 
electron--positron plasma ($\gamma_{e} \gg 1$), 
where the two species have the same number 
density $n_{0}$. Then, Eq.\ (\ref{eq:qeddisp}) gives
\begin{equation}
   \frac{k^{2} c^{2} }{\omega^{2} } \approx \frac{4\alpha}{45\pi}\left[\left( 
    \frac{E_0}{E_{\text{crit}}}  \right)^2 \frac{k^{2} c^{2} }{\omega^{2} } 
    + \left(\frac{cB_0}{E_{\text{crit}}}\right)^2 \right]\frac{k^2c^2}{\omega^2}  \mp 
    \frac{\omega_{p}^{2}}{\omega\omega_e}\frac{E_{\text{crit}}}{E_0}  .  
\label{eq:transverse3}
\end{equation}

For background magnetic field strengths $B_0$ in the pulsar range 
$\sim 10^6 - 10^{10}\,\mathrm{T}$, $cB_{0}\ll E_{\text{crit}}$, and we
therefore drop the term proportional to $B_{0}^{2}$ in Eq.\ (\ref{eq:transverse3}). 
Next, using the normalized quantities 
$\Omega = \omega \omega _{e}/\omega _{p}^{2}$, $K=(4\alpha /45\pi
)^{-1/2}kc\omega _{e}/\omega _{p}^{2}$ and $\tilde{E}=(4\alpha /45\pi
)E_{0}/E_{\text{crit}}$, 
the dispersion relation (\ref{eq:transverse3}) reads 
\begin{equation} \label{eq:norm}
  \Omega^{2} = \tilde{E}^{2}K^{2}\mp \frac{\Omega^{3}}{\tilde{E}K^{2}} .
\end{equation}
The dispersion relation (\ref{eq:norm}) describe three different modes, two with 
$+$ polarization and one with $-$ polarization. 
We note that for $K \ll 1$, the dispersion relation (\ref{eq:norm}) agrees with
that of \citet{Stenflo-Tsintsadze}, whereas in the opposite limit $K \gg 1$, 
the QED term in (\ref{eq:norm}) dominates. 
For the given density, the latter regime applies, except for
extremely long wavelengths ($>10^{8}\,\mathrm{m}$), and thus we note 
that QED effects are highly relevant for the propagation of these modes in 
the pulsar environment. For small $K$ there is
only one mode, but 
two new modes appear for $K \gtrsim 2.6$. Thus for large $K$, applicable in the pulsar
environment, there are three low-frequency modes ($\omega \ll kc$) that
depend on nonlinear QED effects for their existence. 

The effects of the quantum vacuum on electromagnetic wave dispersion
also allows for nonlinear effects, such as wave steepening, shock front
formation, and soliton propagation \cite{Marklund-etal2005}.

\paragraph{Magnetohydrodynamic plasmas}

When analysing low frequency magnetised plasma phenomena, 
magnetohydrodynamics (MHD)
gives an accurate and computationally economical description.
Specifically, a simple plasma model is obtained if the
characteristic MHD time scale is much longer than both the plasma
oscillation and plasma particle collision time scales, and the
characteristic MHD length scale is much longer than the plasma
Debye length and the gyroradius. These assumptions will make it
possible to describe a two-component plasma in terms of a
one-fluid description. 
The one-fluid description means a tremendous computational
simplification, especially for complicated geometries. Moreover,
if the mean fluid velocity, the mean particle velocity, and the
Alfv\'en speed are much smaller than the speed of light in
vacuum, the description becomes non-relativistic and simplifies
further.

\citet{Heyl-Hernquist3} considered the propagation of MHD modes,
including the effects of photon--photon scattering and an axion field.
Following \citet{Thompson-Blaes}, Heyl and Hernquist start with 
the Lagrangian [see Eq. (\ref{eq:lagrangian2})]
\begin{equation}
  \mathscr{L} = \mathscr{L}_{\text{QED}} + \tfrac{1}{2}\alpha\epsilon_0\theta\mathscr{G}
  = \mathscr{L}_0 + \mathscr{L}_c 
  + \tfrac{1}{2}\alpha\epsilon_0\theta\mathscr{G},
  \label{eq:mhd-lagrangian}  
\end{equation}
where the field invariant $\mathscr{G}$ is defined by (\ref{eq:invariants}),
and $\theta$ is the axion field, that acts as a Lagrange multiplier for 
the MHD condition $\mathscr{G} = 0$. 
The modified Maxwell's equations can be derived from Eq. (\ref{eq:mhd-lagrangian}). They become 
[cf. Eq. (\ref{eq:exact-evol1})]
\begin{eqnarray}
  && \partial_aF^{ab} 
 = -4\left( \frac{\partial\mathscr{L}_{\text{QED}}}{\partial\mathscr{F}} \right)^{-1}
 \Bigg[ \widehat{F}^{cb}\partial_c\left( 2\alpha\epsilon_0\theta + 
 8\mathscr{G}\frac{\partial\mathscr{L}_{c}}{\partial\mathscr{G}^2} \right) 
\nonumber \\ &&\quad
 + 4F^{cb}\partial_c\left( \frac{\partial\mathscr{L}_{c}}{\partial\mathscr{F}} 
 \right)
 \Bigg] .
\end{eqnarray}
Given a background magnetic field, these equations allow for both fast 
and Alfv\'en modes. The fast modes will suffer the same type of shock wave
formation as presented by \citet{Heyl-Hernquist2}, in the absence of the MHD effects.
A single Alfv\'en mode will not experience the effects of photon--photon scattering 
due to the absence of self-interactions. This is not true for the case of counter-propagating 
Alfv\'en modes for which photon--photon scattering introduces higher order
corrections to their propagation.

\section{Applications}

\subsection{Measuring photon--photon scattering}

Classically, electromagnetic waves only interact indirectly, via scattering, 
by passing through a suitable medium such as a nonlinear optical
fibre \cite{Kivshar-Agrawal,Hasegawa}. To some extent, this is still
true in QED. One may view the quantum vacuum as a 
medium through which photons scatter off virtual charged particles,
predominantly electron--positron pairs, producing nonlinear effects
similar to the ones found in nonlinear optics. However, since the 
nonlinear effects enter the effective Lagrangian through the 
Lorentz invariants, a plane wave will not self-interact, and 
more sophisticated techniques are needed in order to 
excite the nonlinear quantum vacuum. In this section, such means
will be reviewed with the aim of establishing methods for 
direct detection of low energy elastic real photon--photon scattering.

The concept of elastic photon--photon scattering is theoretically well-established.
Furthermore, the scattering of virtual photons is routinely observed in 
particle accelerator environments, and is thus well confirmed in experiments. 
Moreover, inelastic photon--photon scattering is also experimentally well-confirmed,
but this is not the case for elastic photon--photon scattering 
(although experiments have been made where it in principle would have been 
possible to make modification such that a direct measurement of elastic 
photon--photon scattering could have been made \cite{Bamber-etal}). 
Thus, as a fundamental test of QED and its predictions
about the properties of the quantum vacuum, an experiment on the latter type of 
scattering may be considered an important issue. 

\begin{figure}
\subfigure[]{\includegraphics[width=0.4\columnwidth]{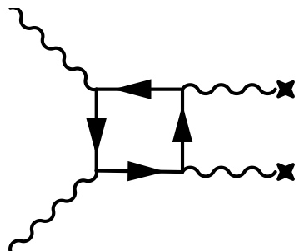}}
\subfigure[]{\includegraphics[width=0.4\columnwidth]{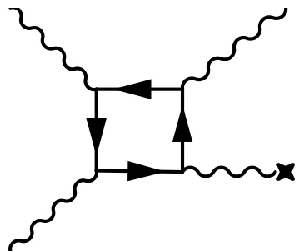}}
\caption{Feynman box diagrams for
  (a) Delbr\"uck scattering and (b) photon splitting, respectively.
  Here each cross denotes external field legs, e.g. an atomic 
  Coulomb field or a strong background magnetic field.}
  \label{fig:feynman2}
\end{figure}

Closely related to photon--photon scattering is Delbr\"uck scattering \cite{Delbruck} 
and photon splitting \cite{Adler-etal,Adler,Chistyakov-etal}, see Fig. \ref{fig:feynman2} 
(cf. Fig. \ref{fig:weak} for a comparison with
photon--photon scattering). Delbr\"uck scattering is the elastic scattering 
of photons in a Coloumb field, e.g. an atomic nucleus, mediated by virtual 
electron--positron pairs, while photon splitting is the down conversion of a  
photon into two photons of lower frequency through an external field, e.g.
a strong magnetic field. These processes contain
external fields mediating the interaction between the photons, making the 
cross section larger than for pure photon--photon scattering. In fact, 
using high-$Z$ atomic targets, Delbr\"uck 
scattering for high energy photons has been detected \cite{Jarlskog-etal},
and photon splitting, although not detected in a laboratory environment,
is assumed to be prominent component of many astrophysical 
environments, such as magnetars and soft $\gamma$-ray repeaters 
 \cite{Adler-Shubert,Baring-Harding2,Baring-Harding,Harding-etal}. In fact, the
 splitting of photons in the atomic Coulomb field has been reported by 
 \citet{Akhmadaliev-etal}, where good agreement with the calculated 
 exact Coulomb field cross-section was obtained.

The effects of photon scattering also manifest themself in the anomalous
magnetic moments of the electron and the muon \cite{Berestetskii-etal,Calmet-etal,%
Bailey-etal,Rodionov}. Even so, the detection of
direct light-by-light scattering of real photons remains elusive, even though
considerable efforts have been made in this area. The possibility to detect 
low energy photon--photon scattering would open up for new tests of QED, since 
fermion loop diagrams could give gauge invariant tests of the fermion
propagator, as well discerning between QED and other theories predicting
or postulating photon properties. 
Thus, photon--photon scattering can both produce  
interesting effects, as described in previous sections, as well as 
produce important tests for fundamental physical theories.

\subsubsection{Pair production in external fields}

\begin{figure}
  \includegraphics[width=0.5\columnwidth]{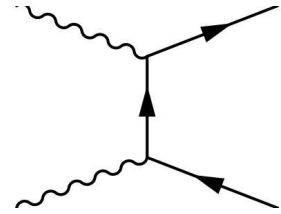}
  \caption{Feynman diagram for the pair creation process.}
  \label{fig:paircreation}
\end{figure}

The case of inelastic photon--photon scattering deserves some attention
in this context. Here the aim is, to some extent, anti-matter production 
on a large scale (see Fig.\ \ref{fig:paircreation}). 
There are a number of ways, both experimentally confirmed
as well as schemes suggested on numerical or theoretical grounds, to produce 
and store \cite{Oshima-etal} positronium and antimatter, e.g. 
laser generated relativistic superthermal electrons interacting with high-$Z$ materials
\cite{liang98}, the trident 
process in conjunction with ultra-intense short
laser pulses in plasmas \cite{ber92}, pair production by circularly polarized 
waves in plasmas \cite{Bulanov1}, laser-thin foil interactions 
\cite{Shen-MeyerterVehn,she02}, 
using Bose--Einstein condensation 
traps \cite{Surko-etal,Greaves-etal} (see \citet{Surko-Greaves} for an overview), 
and
using fullerenes  \cite{Oohara-Hatakeyama}. The formation of anti-plasmas and 
long lifetime 
trapping of antimatter is currently intensely studied, and could shed light on 
the fundamental laws of nature, e.g. giving new CPT and Lorentz invariance 
tests \cite{Bluhm-etal,Bluhm}, 
or producing an annihilation laser \cite{Mills}. Since electron-positron pairs 
also constitute
a unique type of plasma, prominent in e.g. the pulsar magnetosphere, 
the formation of large collections of pairs in the laboratory will further enable the 
study
of astrophysical conditions \cite{Greaves-Surko2}, which we so far have only 
been able to observe over astronomical 
distances, and without control over the physical parameter range. The anti-matter 
production
in most laboratory applications rely on the plasma being cold. However, as laser 
powers approach
the Schwinger critical regime, we will see an increased interest in using these for 
producing 
high temperature pair plasmas as well, and for exciting the quantum vacuum. 

After the laser was introduced, it was realized that future laser systems
could be tools for fundamental physics research, and the pair creation process was reconsidered by
\citet{Reiss}, \citet{Nikishov-Ritus1,Nikishov-Ritus2,Nikishov-Ritus3,Nikishov-Ritus4},
and \citet{Narozhny-etal}. Thus, the mechanism behind the
production of electron--positron pairs from electromagnetic fields and photons
is well-known, and was first directly observed by \citet{bur97} at the SLAC 
facility. Since this observation, schemes making use of the next generation 
laser systems has therefore been in the focus of research efforts by using 
the immense laser intensities for producing, not necessarily cold, pair plasmas 
in the laboratory. Indeed, as reported by \citet{gahn00}, femtosecond table-top
lasers can indirectly create positrons due to electron acceleration in plasma
channels. 

Since the pair production from the nonlinear quantum vacuum formally 
depends crucially on the invariant $|\mathbf{E}|^2 - c^2|\mathbf{B}|^2$ being positive 
\cite{Schwinger},
schemes with strong pure electric fields have been also attracted interest 
\cite{Sauter,Schwinger,Narozhny-Nikishov,Marinov-Popov,Brezin-Itzykson,%
Casher-etal,Kluger-etal,Popov1,Popov2,Popov3,Popov4,%
Popov-Marinov,Mostepanenko-Frolov,Grib-etal,Ringwald} 
(by the same argument, strong static magnetic fields does not 
excite the quantum vacuum, unless perturbed). The pair production rate per
unit volume at the one-loop level is given by \cite{Schwinger}
\begin{equation}
  w = \frac{\omega_e^4}{(2\pi c)^3}\left(\frac{|\mathbf{E}|}{E_{\text{crit}}}\right)^2%
  \sum_{n = 1}^{\infty}\frac{1}{n^2}%
  \exp\left( -n\pi\frac{E_{\text{crit}}}{|\mathbf{E}|} \right)
\label{eq:paircreation}
\end{equation}
for a uniform electric field $\mathbf{E}$. Here the sum is over the real 
poles in the imaginary part of the integral (\ref{eq:lagrangian2}). 
Thus, the pair creation rate is vanishingly 
small in most circumstances. The electron--nucleus electric field (although not uniform) 
requires a nucleus
charge of the order $\alpha^{-1}$ for vacuum breakdown, and such nuclei are unlikely
to exist in any other state than a transient one \cite{Reinhardt-Greiner,Greiner,Milonni}. However, the 
situation may be different for laser fields, where ultra-short high intensity fields are
available. \citet{Brezin-Itzykson} derived the pair creation rate for varying fields and
generalized the pair creation rate (\ref{eq:paircreation}). In the 
low frequency limit (i.e.\ $\omega \ll \omega_e$, where $\omega_e$ is the Compton 
frequency), their expression coincides with Eq. (\ref{eq:paircreation}), taking 
into account only the first term in the sum. Thus, Eq.\ (\ref{eq:paircreation}) can be 
used, with good accuracy, to predict the pair production efficiency of different processes, 
even if the fields are alternating. 

In all the cases above, the derivations of the pair creation rate rely on 
the assumption of an electric field dominating over the magnetic field. 
In plasmas, the phase velocity $v$ can exceed the velocity of light. This 
was used by \citet{Bulanov1} to analyze pair production
in the field of a circularly polarized electromagnetic wave in an 
underdense plasma.
Since for a circularly polarized wave $c|\mathbf{B}| = (kc/\omega)|\mathbf{E}| 
= (c/v)|\mathbf{E}|$, we see that $|\mathbf{E}|^2 - c^2|\mathbf{B}|^2 > 0$. 
Thus, the condition for pair creation according to \citet{Schwinger} is
satisfied, and positrons are therefore predicted to be produced in a 
laser-plasma environment. Moreover, 
\citet{Avetisyan-etal} solved the Dirac equation perturbatively to find the production
of electron--positron pairs by inelastic multi-photon scattering in a plasma. They
found the probability distribution for transverse electromagnetic perturbations in 
the plasma, and used this \cite{Avetissian-etal} to investigate pair
production due to nonlinear photon--photon scattering from oppositely 
directed laser beams. Analytical results for the number of particles created
on short interaction time scales were found. \citet{Fried-etal} investigated the 
possibility for pair production via crossing laser beams, and concluded that 
laser intensities has to reach $10^{29}\,\mathrm{W/m^2}$ before this
could be used as a means for electron--positron generation. 

Pair production may possibly also be achieved without the intervention
of a plasma or other dispersive media. According to \citet{Narozhny-etal1,Narozhny-etal2}, focused
and/or counter-propagating laser pulses can interact via the
nonlinear quantum vacuum as to produce real electron--positron pairs.
The prediction of \citet{Narozhny-etal2} is that pair creation for colliding 
pulses is expected for intensities of the order $10^{26}\,\mathrm{W/cm}^2$,
which is two orders of magnitude lower than for single pulse generation.
Moreover, \citet{Narozhny-etal1} claim that the effect of pair creation 
puts an upper theoretical limit on laser focusing, since the electromagnetic
energy will be dissipated into fermionic degrees of freedom for high enough
intensities. 

As intense fields create electron--positron pairs, the particle density increases. If 
intense photon beams can be sustained for long enough times, this will create
a pair plasma. In this case, the effects of this plasma on the electromagnetic field
need to be taken into account.  
The back-reaction of pair creation on the electromagnetic field was considered by 
\citet{Kluger-etal} in $1+1$ dimensions. Starting from a semi-classical approximation, a kinetic model 
taking pair production into account using an emissive term in the electron equation of 
motion was presented. From a numerical analysis of the governing equations it was found that 
high enough field intensities will induce plasma oscillations. Due to the realization 
that the right conditions for pair creation by lasers could soon be at our disposal, the problem 
of back-reaction and the dynamics
of the interaction of the electron--positron plasma on the photons has produced an increasing 
number of publications over the years.                                                                                                                                                                                                                                                                                                                                                                                                                                                                                                                                                                                                                                                                                                                                                                                                                                                                                                                                                                                                                                                                                                                                                                                                                                                                                                                                                                                                                                                                                                                                                                                                                                                                                                                                                                                                                                                                                                                                                  \citet{alk01}, \citet{proz00}, and \citet{Roberts-etal} have similarly developed 
self-consistent schemes where a collisionless
plasma is coupled to the time-dependent electric field, via Maxwell's equations 
and the pair creation source term. In their application to the X-ray free electron laser, they 
arrived at plasma behaviour reminiscent of a modulational instability, 
and suggested necessary and sufficient conditions to generates a pair plasma
using the XFEL. Collisions in the plasmas created
due to intense electromagnetic fields may also be taken into account using a 
quantum kinetic  description with a pair creation source term \cite{Bloch-etal2,Bloch-etal}.

A somewhat different scheme using intense lasers was suggest by \citet{liang98}.
Letting two intense laser pulses impinge on the surface of a thin foil made of
a suitable material, e.g.\ gold, plasma formation takes place. The jitter energy
for a large fraction [$\sim$ 50\% \cite{Wilks-etal}] of the produced plasma electrons 
is suggested to exceed the pair creation threshold $2m_ec^2$. Thus, in this scheme
the pair creation is a result of the thermal plasma, instead of direct laser interaction
with the quantum vacuum. Similarly, \citet{Helander-Ward} suggested that runaway
electrons in tokamak plasmas could have the same effect. Since electrons with
sufficient energy experience a decreasing plasma friction force as the energy 
increases, such particles will in effect be accelerated to very high energies until 
direct collisions with plasma particles occur. The typical runaway electron energy 
is $\gtrsim 3m_ec^2$, and these collisions could therefore trigger positron production,
as the electrons loose their energy via brehmsstrahlung in the Coulomb field,
the so called Bethe--Heitler process. 
The number of positrons in a facility such as JET was estimated to $\sim 10^{13} 
- 10^{14}$, a very large number compared to other laboratory positron production 
methods. 

The predicted pair production rates normally assume spatially uniform 
electromagnetic fields, which is often in good agreement with experimental 
parameters. However, recently oriented crystals have become an important tool
in studying effects of quantum electrodynamics in strong fields, such as spin effects
in electron energy loss and crystal assisted pair production (see \citet{Kirsebom-etal} 
and references therein). In these experiments, the fields may not be considered 
uniform, and the models described above can therefore only partially account for the
observed effects. \citet{Nitta-etal} remedied this shortcoming by using the trial
trajectory method \cite{Khokonov-Nitta}, based on the method developed by
\citet{Baier-Katkov}. Previous attempts to analyze the experimental results
were based on numerical schemes, but \citet{Nitta-etal} found an analytical 
expression for the pair creation rate in a inhomogeneous field, in good
agreement with the observed pair creation rate. Furthermore, the inhomogeneous
case also displayed pair creation for low amplitude fields, where the uniform field
treatment effectively gives a zero pair creation rate. This result could be of interest
in the case of strongly magnetized stars, which have a characteristic dipole 
behavior.      

Since waves in vacuum are described in terms of their behavior
along the null coordinates $u = z - ct$ and $v = z + ct$, it is of interest to 
generalize the Schwinger results of pair creation to the case of
fields depending on $u$ and/or $v$. This was done by \citet{Tomaras-etal},
and later generalized by \citet{Avan-etal} to a more complicated coordinate dependence.
Furthermore, the momentum spectrum of the produced pairs was derived for arbitrary 
time dependent gauge fields by \citet{Dietrich}, via the exact solution of the
equation of motion for the Dirac Green's function. 
 
\subsubsection{Laser induced pair creation}

The production of anti-matter is of great importance for a variety of experimental tests of 
fundamental issues in physics, e.g.\ Lorentz invariance tests, as well as being of interest 
in its own right. Moreover, there can also be a test of the nonlinear properties of QED, 
since high energy photons may create matter and anti-matter out of the quantum vacuum. 
This process is well established as a model for pair production in the vicinity of neutron 
stars, and corresponds to the imaginary part of the full Heisenberg--Euler Lagrangian, 
and can thus be interpreted as energy being dissipated from bosonic to fermionic degrees 
of freedom.  

The implications of pair creation was understood very early on in the history of QED, but 
the direct creation of electron--positron plasmas from photons has long escaped experimental 
efforts. Thus, an important piece in our view of the quantum vacuum had long eluded the 
attempts of detection. However, with the rapid advances in laser intensity, the prospects 
for performing a successful experiment in pair creation using laser sources took a turn for the better. 
As described below, inelastic photon--photon scattering, where two real photons gives
rise to a real electron--positron pair, has now been experimentally 
confirmed \cite{bur97,Bamber-etal}, and holds the promise of further elucidating 
our picture of the nonlinear quantum vacuum (see also \citet{Meyerhofer}).

\begin{figure}
\includegraphics[width=.8\columnwidth]{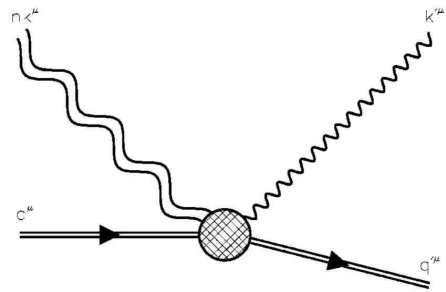}
\caption{Nonlinear Compton scattering, as given in (\ref{eq:nlin-compton}).  
  (Reprinted with permission from \citet{Bamber-etal}.)}
\label{fig:nlincompton}
\end{figure}

In nonlinear Compton scattering, multi-photon absorption by an electron
results in the emission of a single high-energy photon according to (see Fig. 
\ref{fig:nlincompton})
\begin{equation}\label{eq:nlin-compton}
  e + n\omega \rightarrow e' + \gamma .
\end{equation}
The effect (\ref{eq:nlin-compton}) was first measured by 
\citet{Bula-etal}, using a GeV electron beam and a terawatt laser source, 
obtained by chirped-pulse amplification.  In the experiment, up to four laser photons 
interacted with a single electron. The high-energy photons produced by nonlinear Compton 
scattering can be used in the laser assisted production of a pair plasma. The usage 
of laser produced photons for the electron--positron pair production was suggested long before 
lasers reached the necessary 
intensities \cite{Reiss,Nikishov-Ritus1,Nikishov-Ritus2,Nikishov-Ritus3,Nikishov-Ritus4,Narozhny-etal}
(the direct production of pairs
by photons requires $\hbar\omega \gtrsim 2m_ec^2$ in the center-of-mass system).
By re-colliding the high-frequency photons with the original laser photons, according 
to the Breit--Wheeler\footnote{\citet{Breit-Wheeler} considered the single photon scattering 
$\omega_1 + \omega_2 \rightarrow e^+e^-$, thus somewhat different from the multi-photon process discussed here.}
process \cite{Breit-Wheeler,Bethe-Heitler}
\begin{equation}\label{eq:multiphoton}
  \gamma + n\omega \rightarrow e^+ e^- ,
\end{equation}
the production of electron--positron pairs can be achieved in a laboratory environment. This 
can be compared to the trident process
\begin{equation}\label{eq:trident}
  e + n\omega \rightarrow e' e^+ e^- .
\end{equation}
While the multi-photon process (\ref{eq:multiphoton}) requires $n \geq 4$ with experimental 
values used by \citet{bur97}, the trident process requires $n \geq 5$ with the same
 experimental data.
The two-step process (\ref{eq:nlin-compton}) and (\ref{eq:multiphoton}) 
was used by \citet{bur97} in the first reported laser production 
of electron--positron pairs.

\citet{Bula-etal} reported on the observation of the effect of nonlinear Compton
scattering (\ref{eq:nlin-compton}), where the scattered electrons were detected using
a $46.6\,\mathrm{GeV}$ electron beam in conjunction with a $1054\,\mathrm{nm}$ and
a $527\,\mathrm{nm}$ laser with focal intensity $\sim 10^{18}\,\mathrm{W/cm^2}$.
This process can also be understood in terms of a plane wave interaction with
an electron. For a weak electromagnetic field with amplitude $E$, the maximum speed attained
by an electron (initially at rest) due to the passing of a plane wave is
\begin{equation}
  v_{\text{max}} = \frac{eE}{m_e\omega},
\end{equation}
where $m_e$ is the rest mass of the electron and $\omega$ is the frequency 
of the plane wave. As the field strength increases, higher order radiation effects 
becomes important as $v_{\text{max}} \rightarrow c$, which in terms of light quanta
can be interpreted as multi-photon absorption by the electron, with the release
of a single distinguishable light quanta as a result, i.e. the process 
(\ref{eq:nlin-compton}). In this sense, nonlinear Compton scattering 
becomes important as the parameter 
\begin{equation}\label{eq:eta}
  \eta = \frac{v_{\text{max}}}{c} = \frac{eE}{m_ec\omega} = \frac{e|A_bA^b|^{1/2}}{m_ec^2}
\end{equation}
approaches unity. Here, the four-vector potential $A^b$ satisfies the Lorentz gauge. 

\begin{figure}
\includegraphics[width=.6\columnwidth]{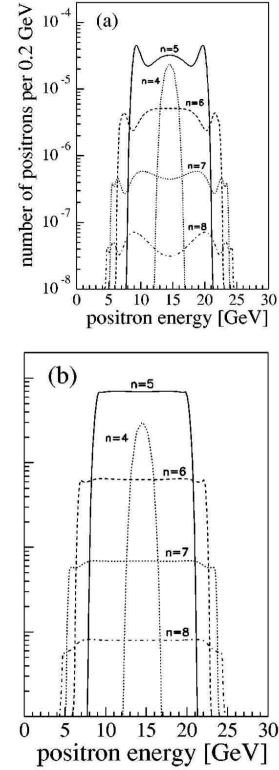}
\caption{Calculated positron energy spectra for a $30\,\mathrm{GeV}$ photon
  interacting with a $527\,\mathrm{nm}$ laser beam. In panel (a) the polarization 
  is parallel while in panel (b) the polarization is perpendicular. $n$ gives the number
  of photons involved in the interaction. (Reprinted with permission from \citet{Bamber-etal}.)}
\label{fig:nopositrons}
\end{figure}

\begin{figure}
\includegraphics[width=.7\columnwidth]{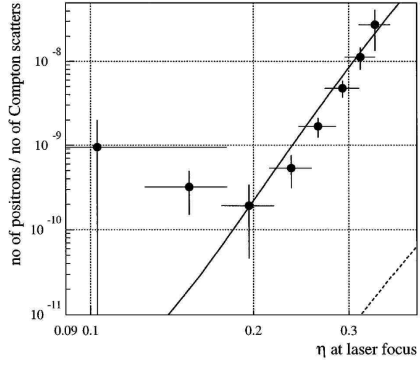}
\caption{Positron production rate per Compton scatterer as a function
  of the intensity parameter $\eta$, as given by (\ref{eq:eta}). 
  The solid line is the numerical estimate from the 
  two-step process (\ref{eq:nlin-compton}) and (\ref{eq:multiphoton}), while the
  dashed line represents the trident process (\ref{eq:trident}). The measurements performed
  by \citet{bur97} are given by the dots in the plot. (Reprinted with permission from 
  \citet{bur97}.)}
\label{fig:positronrate}
\end{figure}

For an electron with initial energy $\mathcal{E}_0$, the absorption of $n$ photons of 
the frequency $\omega$ at an angle $\theta$ between the electron and laser beam,
results in the minimum electron energy
\begin{equation}
  \mathcal{E}_{\text{min}} = \frac{\mathcal{E}_0}{1 + ns/m_{\text{eff}}^2c^4} ,
\end{equation} 
where $s = 2\mathcal{E}_0\omega(1 + \cos\theta)$ is the scattering parameter and 
$m_{\text{eff}} = m(1 + \eta^2)^{1/2}$ gives the effective mass. With the experimental 
parameters used by \citet{Bula-etal}, the intensity parameter becomes $\eta \approx 0.6$. 
Linear Compton scattering ($\eta \ll 1$, $n = 1$) would then result in $\mathcal{E}_{\text{min}} 
\approx 25.6 \,\mathrm{GeV}$ at $\theta = 17^{\circ}$. Since the spectrum of multi-photon 
Compton scattered electrons extends below $25.6\,\mathrm{GeV}$, it was possible to identify 
the nonlinear effects \cite{Bula-etal}.

\begin{figure}
\includegraphics[width=.7\columnwidth]{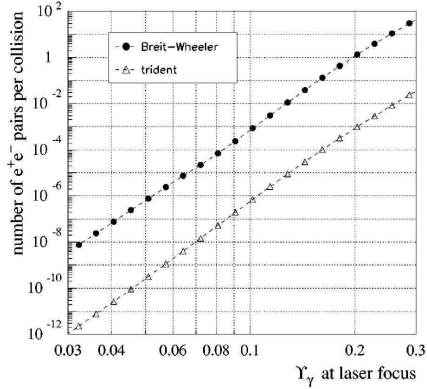}
\caption{Pair production rate, as compared between the multi-photon Breit--Wheeler
  process (\ref{eq:multiphoton}) and the trident process (\ref{eq:trident}), as a function of $\Upsilon$ given by (\ref{eq:upsilon}).
  (Reprinted with permission from \citet{Bamber-etal}.)}
\label{fig:pairrate}
\end{figure}

In the same way, as the intensity parameter $\eta$ approaching unity signifies the onset of 
the nonlinear Compton effect, the parameter \cite{bur97,Bamber-etal} 
\begin{equation}\label{eq:upsilon}
  \Upsilon = \frac{|F_{ab}p^b|}{m_ec^2E_{\text{crit}}}
\end{equation}
characterizes the strength of the vacuum polarization, as it contains both information of 
the photon frequency as well as the background field strength, the two important parameters 
for vacuum breakdown. Here, $F_{ab}$ is the Maxwell 
tensor of the background electromagnetic field, and $p_a$ is the four-momentum of 
the probe photon. As $\Upsilon$ approaches unity, the pair production rate according 
to the process (\ref{eq:multiphoton}) becomes 
significant \cite{Nikishov-Ritus1,Nikishov-Ritus2,Nikishov-Ritus3,Nikishov-Ritus4,Narozhny-etal,bur97}. 
For the case of single particle ($n = 1$) Breit--Wheeler scattering, laser wavelengths of $527\,\mathrm{nm}$ 
would require single photon energies of $111\,\mathrm{GeV}$ in order for significant pair production
to occur, while for the multi-photon Breit--Wheeler process the  photons of the same 
wavelength colliding with backscattered photons with energies $29\,\mathrm{GeV}$ gives
$\Upsilon \approx 0.5\eta$ \cite{bur97}. 
Thus, for large enough $\eta$, the pair production rate would 
yield a detectable level of electrons and positrons (see Figs. \ref{fig:positronrate} 
and \ref{fig:pairrate}), with a well-defined energy spectrum (Fig. \ref{fig:nopositrons}). 

\citet{bur97} for the first time presented the results of a successful measurement along the lines 
presented above. The signal consisted of $\sim 100$
positrons above the background value using a $46.6\,\mathrm{GeV}$ electron beam 
and a $527\,\mathrm{nm}$ Nd:glass laser with focal intensity 
$\sim 10^{18}\,\mathrm{W/cm^2}$ \cite{Meyerhofer}.

\subsubsection{Other mechanisms for pair production}

\citet{Narozhny-etal1} considered pair production in an electromagnetic field created by two counter propagating laser pulses, and showed that pair production can be experimentally observed when the intensity of each beam is similar to $10^{26}\,\mathrm{W/cm^2}$, three orders of magnitude lower than that of a single pulse. However, the cross-section for the Schwinger process at optical frequencies (or below) is so small at any laser intensity that this effect is insignificant \cite{Mittelman}. 

Production of pairs is also possible in the Coulomb field of a nucleus via virtual photons ("tridents"), which is a dominant energy loss mechanism at high energies. In a trident Bahba process, high energy electrons, with kinetic energies exceeding the pair production threshold $2m_ec^2$, can produce electron--positron pairs by scattering in the Coulomb potential of the nucleus. In the past, some authors \cite{Bunkin1970,Scharer1973} had presented a preliminary discussion about pair production by relativistic electrons accelerated by intense laser, while others \cite{ber92} presented a detailed investigation of pair production due to scattering of high energy electrons produced in strong wake fields driven by intense short laser pulses. This was found to be an efficient mechanism for a "pair factory". Recently, \citet{ber2005} carried out computer simulations of laser plasma dynamics in overdense plasmas and showed that an intensive production of pairs by the drive motion of plasma electron takes place due to the trident process. 
Furthermore, \citet{Bulanov-etal2005} have shown that electromagnetic waves could be damped due to electron--positron pair production (see also \citet{Mikheev-Chistyakov} for a discussion on the process in a strong magnetic field).

\subsubsection{Laser experiments on photon--photon scattering}

The evolution of laser intensity is truly astounding \cite{mou98,Perry-Mourou,taj02,mou05}
(see Fig. \ref{fig:laserevol}), and with the event of the X-ray free electron laser, 
a new domain in experimental physics will open up. There have been 
an interesting set of both suggested and performed experiments using lasers
of previous and current intensities. Note that one of the major obstacles in these investigations have been residual gas components in the vacuum environment. However, depending on the problem of study, the means for inhibiting the residual gas to have a detrimental effect on the measurement varies. In high intensity laser experiments on elastic photon--photon scattering, the electron expulsion at the leading edge of the laser pulses will in fact make the generation of background radiation weaker (at a vacuum of $10^{-9}$\,torr), and particle effects would therefore have a negligible effect in these experiments \cite{Lundstrom-etal}. This is contrast to weak field experiment, such as cavity environments, where the effects due to residual gas may be significant. However, it is possible to design the mode interaction such as to produce a unique signature of photon--photon interaction, thus making it possible, in principle, to detect the scattering by the proper filtering techniques \cite{Eriksson-etal}. 

\paragraph{Vacuum birefringence}
The concept of vacuum birefringence is well known and has been 
theoretically explored in many publications \cite{Klein-Nigam1,Klein-Nigam2,%
Erber,Adler,Adler-Shubert,Heyl-Hernquist}. The birefringence of the vacuum
manifests itself as the difference in the refractive index between the propagating 
ordinary and extraordinary modes \cite{Rikken-Rizzo1}. Thus, 
although a very difficult high precision experiment, this difference may in principle become
measurable in strong enough background magnetic (or electric) fields. This idea 
has been exploited in the PVLAS set-up \cite{Bakalov-etal2,Melissinos}, for which the 
difference \cite{Bakalov-etal}
\begin{equation}
  \Delta n = n_{\|} - n_{\perp} = 3\kappa\epsilon_0c^2|\mathbf{B}_0|^2 
     \approx 4\times10^{-24} |\mathbf{B}_0|^2 , 
\end{equation}
is to be measured. Here $|\mathbf{B}_0|$ is given in Tesla.   

A linearly polarized laser beam is sent through the static field $\mathbf{B}_0$, 
with $\mathbf{E}\cdot\mathbf{B}_0 = |\mathbf{E}||\mathbf{B}_0|\cos\theta$. Due to the 
birefringence of the magnetized vacuum, as given by  $\Delta n$, the beam will 
attain an ellipticity  
\begin{equation}
  \Psi = \frac{\pi L}{\lambda}\Delta n\sin(2\theta)
\end{equation} 
over a propagation distance $L$, where $\lambda$ is the wave length of
the radiation. The change in ellipticity is proposed as a measurement of the birefringence 
of vacuum. \citet{Bakalov-etal} also presented a detailed discussion
of noise sources as well as a rather detailed description of the actual 
experimental setup. Current superconducting magnets can reach 
field strength up to $5 - 25\,\mathrm{T}$, and could in principle yield 
detectable changes in the polarization state of a laser beam traversing 
it. Unfortunately, the strong magnetic fields generate forces within the 
detection equipment which may interfere with the ellipsometric 
measurement. Moreover, magnetic fields cannot be shielded in
any efficient way, and this is therefore a problem that is likely to persist
\cite{Cameron-etal}. 

\begin{figure}
  \includegraphics[width=.75\columnwidth]{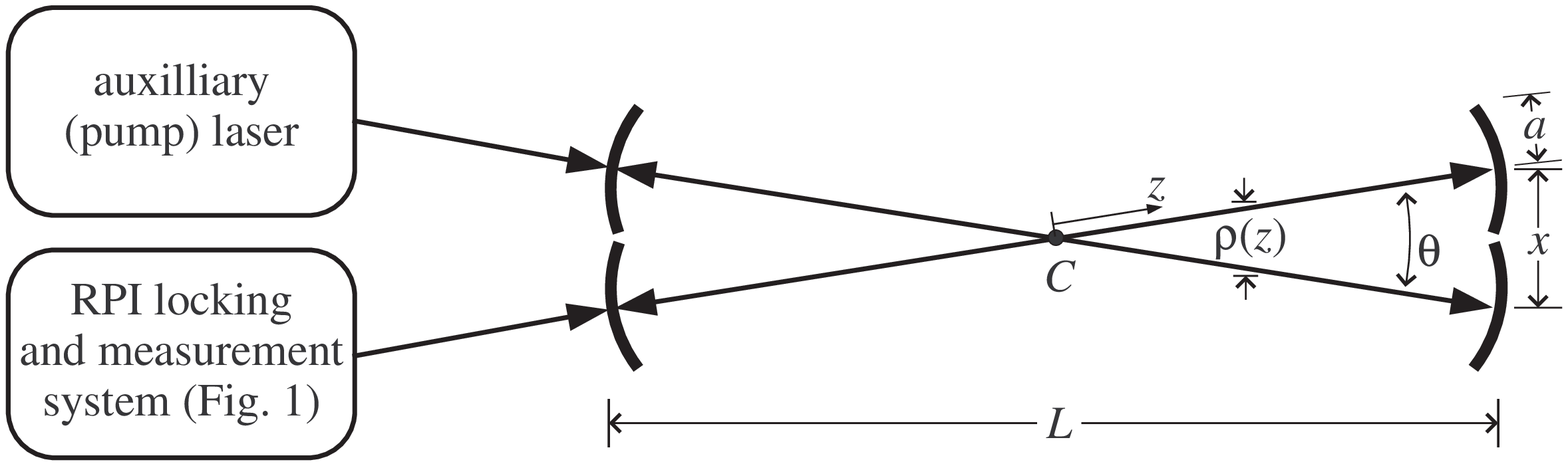}
  \caption{The interferometric set up for detection 
  of vacuum birefringence. The figure reference in the
  above set up is figure 1 in \citet{Luiten-Petersen}.
  (Reprinted with permission from \citet{Luiten-Petersen}.)}
  \label{fig:skew}
\end{figure}

Another approach towards measuring vacuum birefringence is by
using laser interferometry. This technique can reach astonishing 
accuracy and sensitivity, and is currently the most promising method
of choice in gravitational wave detection \cite{Saulson}. Using laser interferometry 
for detecting light-by-light scattering through vacuum birefringence 
rests on the same principle as described above, but replacing 
the strong magnetic field by ultra-short laser pulses \cite{Partovi2,%
Boer-vanHolten,Luiten-Petersen}. Since laser beams,
i.e. laser light with typical pulse length much larger than its wavelength,
has a very low energy density compared to the strongest 
laboratory magnetic fields, one has instead to resort to ultra-short
highly focused laser pulses. Such configurations could indeed
result in magnetic field components of the order $10^5\,\mathrm{T}$,
orders of magnitude larger than quasi-stationary magnetic fields 
produced by superconducting coils \cite{Lee-Fairbanks}. Due to the 
degree of focusing of the pulse, interaction of the strong field with 
the detector can be almost eliminated. On the other hand, 
the ultra-short time- and length scales require a very high 
resolution in the detection. An experimental suggestion along
these lines was put forward by \citet{Luiten-Petersen2,Luiten-Petersen}, using 
a high precision birefringence measuring technique 
\cite{Hall-etal}. \citet{Luiten-Petersen} argue that this technique 
may be used to construct a table-top detector of vacuum polarization
using current state-of-the-art optical techniques. The set up 
consists of two concentric resonant cavities with an interaction 
cross-section. One of the cavities acts as the vacuum
polarizer, while the other cavity supplies the test photons
for which the ellipticity is to be detected (see Fig. \ref{fig:skew}).
Depending on the Fabry--Perot resonator reflectivity, the integration
time was estimated. With a reflectivity $R=99.97\,\%$ the necessary
operation time of the device would be $2.6$ years, $R=99.994\,\%$
yields $1.7$ days, and $R = 99.997\,\%$ gives $2.5\,\mathrm{h}$,
using a $20\,\mathrm{W}$ $200\,\mathrm{fs}$ laser and a resonator of 
length $3\,\mathrm{m}$. 

\citet{Jeah-Sheng-etal} have built and tested a $3.5\,\mathrm{m}$ 
Fabry--Perot inteferometer with a precision ellipsometer 
for QED tests and axion search, along the lines of the vacuum 
birefringence test presented above. Note that the results 
presented by \citet{Jeah-Sheng-etal} are for a prototype
detector, and, although promising,  
a measurement of the vacuum polarization has not been performed (see
also \citet{Sheng-Jui-etal}). Moreover, the PVLAS collaboration has recently 
claimed \cite{Zavattini-etal} to have measured the dichroism of a magnetized vacuum, an effect possible
due to interaction between light pseudoscalars and photons. 

\paragraph{Harmonic generation}

\citet{Ding} suggested that the nonlinear vacuum
could be given measurable properties by the possible harmonic
generation of radiation in an external field. The work attracted 
lots of attention, and some questions as to whether the result was correct 
or not was raised \cite{Raizen-Rosenstein,Ford-Steel,Ding-Kaplan2}.
While some of the critiques were flawed, the main problem 
in the work of Ding and Kaplan was the assumption of a constant 
background field \cite{Ford-Steel}. It is well-known that a homogeneous and
time-independent background field cannot transfer momentum
to the photons, and such a field is therefore not capable of
driving a frequency upshift as suggested by Ding and Kaplan
(see \citet{McKenna-Platzman}). However, \citet{Ding-Kaplan}
showed that a spatially inhomogeneous background field could indeed
result in higher harmonics. This idea was further developed by
\citet{Kaplan-Ding}, where Maxwell's equations were analyzed 
with a weakly varying background magnetic field. 

Using the slowly varying amplitude approximation \cite{Hasegawa}, 
\citet{Kaplan-Ding} showed that the envelope of the electromagnetic
field satisfies the second harmonic generation equation
\begin{equation}\label{eq:kaplan-ding}
  4ik\frac{\partial \mathbf{a}}{\partial t} + \nabla_{\perp}^2\mathbf{a} 
  = -2ik\kappa\mathbf{F} ,
\end{equation}
where 
\begin{equation}
  \mathbf{F} = \frac{1}{\kappa}\left( \frac{\partial\mathbf{D}^{(2)}}{\partial t}
  + \nabla\times\mathbf{H}^{(2)}
  \right)\exp(2ikz - 2i\omega t)
\end{equation}
is the second harmonic generation background source term. Here
$\mathbf{D}^{(2)}$ and $\mathbf{H}^{(2)}$ are derived from the 
Heisenberg--Euler Lagrangian (\ref{eq:lagrangian1}) from
the nonlinear field combinations giving rise to terms proportional
to $\exp(2i\omega t)$. \citet{Kaplan-Ding} use Eq. (\ref{eq:kaplan-ding})
to study the evolution of 2-D Gaussian beams propagating
in an external non-constant magnetic field, giving the estimated 
output power. Moreover, a discussion of more complicated background
magnetic field geometries, e.g. the magnetic quadrupole case, 
was given. Considering $\mu\mathrm{m}$ lasers with focal intensities
$\sim 10^{22}\,\mathrm{W/cm}^2$ generating a pulse propagating
through the background magnetic field strengths $10^3\,\mathrm{T}$
[which in the paper by \citet{Kaplan-Ding} is suggested to be produced 
by explosive mechanisms], a rough estimate gives a production
of $85$ photons$/$day by second harmonic generation. However,
in this estimate, temporal effects, which may be of importance
in the next generation ultra-short intense lasers, have been omitted,
and could yield alterations in their estimates.

\paragraph{Four-wave interactions}

In the second harmonic generation presented above, the interaction
of photons is mediated by a background magnetic field. However,
crossing electromagnetic waves would similarly interact and
yield new modes of different frequencies. One of the more
prominent modes in such a mechanism is given by the 
four-wave interaction mediated mode satisfying 
resonance condition between the frequencies and wavevectors 
(i.e. photon energy and momentum conservation) \cite{Rozanov93}. 
It is therefore not a surprise, given the evolution of laser powers and frequencies, 
that the search for photon--photon scattering using resonant 
four-wave interactions has caught the attention of researchers
in this area. This approach has
also come furthest in the experimental attempts to detect 
elastic scattering among photons 
\cite{Bernard,Moulin-Bernard,Bernard2,Bernard-etal,Bernard3}.  

\citet{Moulin-etal} presented experiments on light-by-light
scattering performed in the optical regime. With this, they managed to
put new experimental upper limits on the photon--photon scattering
cross-section. Unfortunately, no scattering was detected, but stimulated
the continued research along the lines of four-wave interactions as
an experimental tool for probing the quantum vacuum. 
Using the resonance conditions\footnote{%
  The interaction between modes of different frequencies
  gives rise to several new modes, but the resonance conditions
  and time averaging, mimicking the act of detection
  over certain timescales, yield the desired equations.}
  $\omega_4 = \omega_1 + \omega_2 - \omega_3$
 and
  $\mathbf{k}_4 = \mathbf{k}_1 + \mathbf{k}_2 - \mathbf{k}_3$,
one may, in general, derive a set of wave interaction equations
for slowly varying amplitudes $a_i, \, i = 1, \ldots, 4$, of the form
\cite{Weiland-Wilhelmsson}
\begin{equation}
  \frac{da_i}{dt} = Ca_ja_ka^{\ast}_l ,
\end{equation}
given any type of media through which the waves may 
interact. Here the coupling constants $C$ depend on the 
the interaction in question, as well as on the physical 
parameters of the system around which the waves are
modulated. In the case of a nonlinear quantum vacuum, 
the coupling constant will depend on $\kappa$ of the Lagrangian
(\ref{eq:lagrangian1}).

The coupling constants may be interpreted
in terms of the nonlinear susceptibility of the vacuum.
\citet{Moulin-Bernard} considered the interaction of three
crossing waves, characterized by their respective electric field
vectors $\mathbf{E}_i$, producing a fourth wave $E_4$. 
Starting from Maxwell's equations with the usual 
weak field limit Heisenberg--Euler third order nonlinear
corrections (see the Lagrangian (\ref{eq:lagrangian1})), 
they derive the equation 
\begin{equation}
 i\left( \frac{\partial}{\partial t} + c\frac{\partial}{\partial z} 
 \right)E_4 + \frac{c^2}{2\omega_4}\nabla^2_{\perp}E_4 = 
 -\frac{\omega_4}{2}\chi^{(3)} E_1E_2E_3^{\ast}
\end{equation}
for the driven wave amplitude $E_4$, where the overall 
harmonic time dependence $\exp(-i\omega t)$ has been factored out. 
Here $\chi^{(3)}$ is the third order nonlinear susceptibility given by
\begin{equation}
  \chi^{(3)} 
  = \frac{\alpha}{45\pi}\frac{K}{E_{\text{crit}}^2}
  \approx 3\times 10^{-41}\times K \, \mathrm{m^2/V^2} ,
\end{equation}
where $K$ is a dimensionless form factor of order unity. The value
of $K$ depends on the polarization and propagation directions
of the pump modes, and reaches a maximum of $K = 14$
for degenerate four-wave mixing \cite{Moulin-Bernard}. 
\citet{Moulin-Bernard} furthermore discuss the influence
of a non-perfect vacuum, where the susceptibility of the gas
will introduce a threshold, in terms of a critical gas pressure, 
for the nonlinear QED effect to be detected. \citet{Bernard-etal}
and \citet{Bernard3}  
recently presented experiments on four-wave mixing in vacuum,
improving previous attempts by nine orders of magnitude, although
no direct detection of photon--photon scattering was achieved. 
Experiments along
the same lines as described for four-wave mixing above can 
also be used for a large number of other, non-QED, test, such
as axion\footnote{Axions are bosons which were introduced in order 
to explain the 
absence of CP symmetry breaking in QCD \cite{Peccei-Quinn,%
Weinberg,Wilczek}, 
and the axion is still to be detected.} 
search \cite{Bernard,Dupays-etal,Bradley-etal}. Thus, progress of low-energy 
QED experiments could also prove to be useful for, e.g.
dark matter searches. 

There are more recent proposals for detection of photon--photon
scattering using four-wave interactions. \citet{Lundstrom-etal} has done more detailed
calculations concerning experimental constraints, in particular for the 
Astra Gemini laser (operational in 2007) at the Rutherford Appleton Laboratory \cite{RAL}, 
as well as non-perfect vacuum problems etc., and concluded that it will be feasible to detect elastic scattering among photons
if using a high repetition rate high intensity laser system.

\subsubsection{Cavity experiments}\label{sec:cavityexp}

As mentioned in Sec. \ref{sec:cavity}, the effects of photon--photon
scattering on cavity EM fields is to produce new wave modes. The new modes 
excited in the cavity will (approximately) satisfy the cavity
dispersion relation. Thus, by varying the cavity cross-section,
the pump modes may be filtered out, leaving the new modes for
detection. The treatment of cavity mode interaction in the quantum
vacuum was described by \citet{Brodin-Marklund-Stenflo,%
Brodin-Marklund-Stenflo2} and \citet{Eriksson-etal}.
 
If no damping or dissipation is present, Eqs. (\ref{Evolution2})
yield a linear growth of the vector potential amplitude $A_3$
of mode 3. In order to gain an understanding of the saturation level, 
we make the following modification to Eq. (\ref{Evolution2}). 
Let $d/dt\rightarrow d/dt-(\omega _{3}/2\pi
Q)$, where $Q$ is the cavity quality factor. A steady state amplitude 
\begin{equation}
A_{3}=\frac{\mathrm{i}\pi QK_{\text{cyl}}}{4}\frac{\alpha}{90\pi}\frac{\omega _{3}^{2}A_{1}^{2}%
}{E_{\text{crit}}^{2}}A_{2}^{\ast } . 
\label{Saturated-amp}
\end{equation}
is thus obtained. Here $\omega_3$ is the frequency of the mode generated
by the third order QED nonlinearities.  
The number of excited photons in the cavity mode can be described by $%
N\approx (\epsilon _{0}\int |E_{3}|^2\, d^{3}r)/\hbar \omega _{3}$. 
Using the saturation value (\ref{Saturated-amp}) for the vector
potential of mode 3, the number of photons generated  
by the nonlinear interaction of two cavity modes is given by the
expression 
\begin{equation}
  N_{\text{QED}}=\frac{\epsilon_0\alpha^2 
  V Q^2\omega_3^5 K_{\text{cyl}}^2 J_0^2(\beta_3) |A_1|^4 |A_2|^2 }{129600\,\hbar E_{\text{crit}}^4} ,
\label{photon-numb}
\end{equation}
where the coupling constant $K_{\text{cyl}}$ can be found in 
\citet{Eriksson-etal}, $V = \pi a^2z_0$, $a$ is the cylindrical cavity radius, 
$z_0$ is the cavity length, and $\beta_3$ is a zero for the Bessel function 
$J_1$ corresponding the the generated mode satisfying the resonance
condition (\ref{frequency-matching}).
We note that the number of generated photons depends on a large
number of parameters, and one thus needs to specify the 
cavity geometry etc. in order to obtain an estimate of the magnitude
of the effects. \citet{Eriksson-etal} choose the following 
wave mode numbers: 
$(\ell_{1},\ell_{2},\ell_{3})=(3,15,21)$
(fulfilling $\ell_{3}=2\ell_{1}+\ell_{2}$), $\beta _{2}=$ $\beta _{3}=3.83$,
corresponding to the first zero of $J_{1}$, and $\beta _{1}=7.01$
corresponding to the second zero. This gives $z_{0}/a=9.53$ through the
frequency matching condition (\ref{frequency-matching}) and determines the
frequency relations to $\omega _{3}/\omega _{2}=1.26$ and $\omega
_{3}/\omega _{1}=1.12$. Substituting these values gives 
$K_{\text{cyl}}=3.39$ (see \citet{Eriksson-etal}). The
remaining key parameters are the quality factor and the pump field strength.  
\citet{Q-factor2} has shown that it is possible to reach intense 
cavity surface fields, of the order $|A_1|, |A_2| \sim 0.01 - 0.03$ $\mathrm{Vs/m}$, 
 with quality factors as high as $Q=4\times
10^{10}$ at temperatures of $1\,\mathrm{K}$. Thus, in this case
Eq. (\ref{photon-numb}) gives
\begin{equation}
  N_{\text{QED}} \approx 18 . \label{Photon-numb2}
\end{equation}
For a cavity wall temperature of $0.5\,\mathrm{K}$,  
the number of thermal photons is $N_{\text{thermal}}=1/[\exp (\hbar \omega
_{3}/k_BT)-1]\approx 7$, where $k_B$ is the Boltzmann constant, which 
is thus lower than $N_{\text{QED}}$. In order to reach further
accuracy in the measurement, a cavity filtering system can be set up,
so that the pump modes may be reduced or eliminated. It can furthermore
be shown that the nonlinearities of the cavity walls will not generate
modes swamping the QED photons \cite{Eriksson-etal}.

\subsection{Laser-plasma systems and the X-ray free electron laser}

The X-ray free electron laser promises new and exciting applications for 
a coherent electromagnetic source. The applications range from 
probing astrophysical conditions in the laboratory to new possibilities to 
do molecular biology. Could the XFEL also provide insight into quantum 
electrodynamical nonlinearities, such as photon--photon scattering? 
If affirmative, this would enhance our understanding of
the quantum vacuum, as well as providing new prospects of testing
fundamental properties of physical laws, such a Lorentz invariance and 
symmetry breaking. Indeed, 
it has been stated that the facilities at DESY and SLAC would be able
to produce electron--positron pairs directly from the vacuum 
\cite{Ringwald,Ringwald1,Ringwald2}, due to the estimated focal intensities
at these sources. If this scenario is demonstrated, it is likely that the effects of 
elastic photon--photon scattering would come into play at an even
earlier stage. Due to the possible effects of scattering amongst photons, 
such as photonic self-compression and collapse, it is therefore of interest 
to include such effects into the analytical and numerical models used in 
predicting the behaviour of these systems. Furthermore, the creation
of a pair plasma in the laboratory could be affected by new low-frequency 
modes from nonlinear quantum vacuum effects, thus altering the 
properties of energy transfer within such plasmas, as well as
providing indirect tests for QED. XFEL will also gives the opportunity
to do laboratory astrophysics in a new parameter regime, making the quantum vacuum more accessible.

However, it is not necessary to enter the new regime of XFEL 
in order to facilitate tests of QED and Lorentz invariance, as
well as doing laboratory astrophysics. Such effects as Unruh
radiation \cite{Unruh} and the related Hawking effect \cite{Hawking}
can hopefully be investigated using the next generation 
laser-plasma systems \cite{bob}, such as the high repetition-rate Astra Gemini laser (to be operational 2007) \cite{RAL}. In such regimes, it will also be
of interest to investigate QED effects, such as photon--photon
scattering. As seen in the previous sections, the introduction
of plasma dispersion allows for new electromagnetic wave modes
in both unmagnetized and magnetized plasmas 
\cite{Marklund-Shukla-Brodin-StenfloC,Marklund-Shukla-Stenflo-Brodin-Servin}, when 
nonlinear quantum vacuum effects are included. 
For example, at the laboratory level, current high laser powers are able to
accelerate particles to highly relativistic speeds. 
Furthermore, pulse self-compression in laser-plasma systems 
may play an important role in attaining power levels well above
current laser limits (see, e.g. \citet{Bulanov-etal,puk03}).
As the intensities approach the Schwinger limit in future laser-plasma
setups, effects of pair-creation and photon--photon scattering have to
be taken into account \cite{Bulanov-etal,Bulanov1}.
Laser-plasma systems can have electron densities of the order $10^{26}$
m$^{-3}$, and laser intensities can be close to
$10^{23}-10^{25}$ W/cm$^2$ \cite{mou98,bob}. Moreover, as stated by
\citet{Bulanov-etal}, laser self-focusing in plasmas could come close to 
$E_{\text{crit}}$, at which pair creation is likely to follow. In
fact, it has been estimated that the National Ignition Facility
would be able to produce pairs by direct irradiation of a deuterium
pellet \cite{nif}.  On the other hand, the creation of laboratory electron--positron
plasmas is already a feasible task \cite{Surko-etal,Greaves-etal}, as
is the usage of these plasmas for making pair plasma experiments
\cite{Greaves-Surko}. Thus, the possibility to study laser-plasma 
interactions in pair plasmas could be a reality in the nearby future.  
The currently available positron densities in the laboratories are
well below those of regular laser plasma systems, but according to
\citet{nif} there is a possibility of reaching densities of order
$10^{27}$ m$^{-3}$. 
Using $n_0 \sim 10^{26}$ m$^{-3}$, and the
field intensity $10^{16}$ V/m (due to laser self-compression
\cite{Bulanov-etal}) at the wavelength $0.3\times 10^{-6}$ m, we find from
Eq.\ (\ref{Relativistic-DR}) that $\omega \approx 7.8 \times 10^5$
rad/s, i.e. the frequency is in the LF band, as an example of QED effects in 
laboratory plasmas. 

In combination with plasma particle expulsion due to 
electromagnetic wave trapping, the possibility of catastrophic
collapse due to photon--photon collisions arises \cite{Bulanov-etal,r1}.
This scenario it highly interesting, since it would make 3-dimensional 
ultra-intense solitonic structures possible in vacuum bounded by a 
plasma or wave-guide structure, a truly exciting prospects. This may
even prove a valuable tool for intense pulse storage, if a
successful cavity nonlinear QED experiment is performed. 

\subsection{Astrophysical importance}

The implications of the QED vacuum is well-known within astrophysics,
and the pair plasma in pulsar surroundings is partly dependent on
mechanisms which has no classical counterpart \cite{Asseo}. Furthermore,
photons splitting \cite{Bialynicka-Birula,Adler-etal,Adler} supports the notion
that strongly magnetized objects, such as neutron stars and magnetars, could
be used as probes of nonlinear QED effects. 

However, most of the effects discussed within astrophysical applications
concerning QED deals with single-photon effects, and thus do not take collective 
effects into account. It is well-known from plasma physics that 
collective effects alter the charged particle behaviour in non-trivial 
and important ways. In fact, it would not be possible to understand
most plasma effects without resorting to a collective  description.
The analogy between the quantum and a plasma system has been
stated before \cite{Dittrich-Gies}, and is both useful and imaginative. 
Thus, in line with this, it is likely that collective quantum vacuum
effects could yield crucial information about astrophysical systems,
where extreme energy levels are reached. Even kinetic effects,
such as Landau damping, could play a role in the dynamics
of photons in the vicinity of strongly magnetized objects. This
could prove a new realm of photon kinetics, and the applications
to astrophysical sources, such as magnetar quakes \cite{Kondratyev},
are of interest for future research directions.  

Especially strong magnetic field effects due to the quantum vacuum
is of great interest in astrophysical applications. Since the earth-based
magnetic field strengths are very limited, and are likely to remain so
for the foreseeable future, magnetars and similar objects offers a
unique perspective on the quantum vacuum \cite{Erber,Baring-Harding}. 
Pulsar magnetospheres exhibit extreme field strengths in a highly energetic
pair plasma. Ordinary neutron stars have surface magnetic field strengths of 
the order of $10^{6}-10^{9}\,\mathrm{T}$, while magnetars can reach 
$10^{10}-10^{11}\,\mathrm{T}$ \cite{magnetar}, coming close to,
or even surpassing, energy densities $\epsilon_0E_{\text{crit}}^2$ 
corresponding to the Schwinger limit. Such strong fields will make the  
vacuum fully nonlinear, due to the excitation of virtual pairs. 
Photon splitting can therefore play a significant role in these extreme 
systems \cite{Harding,Baring-Harding}.

 Neutron
stars have surface magnetic field strengths of the order of
$10^{6}-10^{9}\,\mathrm{T}$, while magnetars can reach
$10^{10}-10^{11}\,\mathrm{T}$ \cite{magnetar}, coming close to energy
densities corresponding the Schwinger critical limit $\epsilon_0E_{\text{crit}}$; 
here, the quantum 
vacuum becomes fully nonlinear. Single-particle QED effects, such as
photon splitting can play a significant role in the understanding and
interpretation of observations from these extreme systems
\cite{Baring-Harding,Harding}.  
In fact, the pair plasma creation in pulsar environments itself rests on
nonlinear QED vacuum effects. The emission of short wavelength photons
due to the acceleration of plasma particles close to the polar caps
results in a production of electrons and positrons as the
photons propagate through the pulsar intense magnetic field 
\cite{Beskin-etal}. The precise density of the pair plasma created
in this fashion is difficult the estimate, and the answer is model
dependent. However, given the Goldreich--Julian density $n_{GJ} = 7\times 10^{15} (0.1\,\mathrm{s}/P)(B/10^8\,\mathrm{T})$ m$^{-3}$,
where $P$ is the pulsar period and $B$ the pulsar magnetic field, the
pair plasma density is expected to satisfy $n_0 = M n_{GJ}$, $M$ being
the multiplicity \cite{Beskin-etal,Luo-etal}. The multiplicity is
determined by the model through which the pair plasma is assumed to be
created, but a moderate estimate is $M = 10$ \cite{Luo-etal}. Thus,
with these pre-requisites, the density in a hot dense pair plasma is
of the order $10^{18}$ m$^{-3}$, and the pair plasma experiences a
relativistic factor $\sim 10^2 - 10^3$ \cite{Asseo}. We may use these
estimates to obtain estimates for particular QED processes in plasmas. For example,
inserting the above values in (\ref{wavelength}), we obtain $\lambda \sim
10^{-12} - 10^{-11}$\,m. 
On the other hand, the primary beam
will have $n_0 \sim n_{GJ}$ and $\gamma \sim 10^6 -10^7$ \cite{Asseo},
at which (\ref{wavelength}) yields $\lambda \sim 10^{-8} - 10^{-7}$\,m.
Thus, in this case we obtain short wavelength effects. 

The field of laboratory astrophysics ties the experimental domain of laser--plasma systems to areas of research where we so far have been restricted to observations \cite{HEDLA}. Interesting studies, such as shock front formation relevant to supernova explosions, could in principle be achieved in facilities such as NIF. However, the scales of the astrophysical event and the laboratory setup differs by orders of magnitude. Thus, it is reasonable to ask if it is possible to apply laboratory findings to astrophysical events. \citet{Ryutov-etal} consider the prospects of investigating MHD phenomena of relevance for supernova hydrodynamics. From self-similarity in the governing system of equations and boundary conditions, as well as from dimensionless variables (such as the magnetic Reynolds number $Re_M$) they argued that the laboratory results could be translated to astrophysical settings (however, $Re_M$ in the laboratory cannot reach the extreme values of supernova ejecta but can reach values much larger than $1$). Similarly, \citet{Budil-etal} discussed the applicability of peta-watt lasers to radiative-hydrodynamics relevant to, e.g.\ supernova remnant evolution. The testbed experimental results presented by \citet{Budil-etal} indicated that the results could be useful in calibrate models of radiation hydrodynamics in supernova remnants (see also \citet{Shigemori-etal}). Thus, the use of high intensity lasers for probing astrophysical phenomena, in particular as tool for testing and calibrating simulations of certain events, has undergone rapid development over the last decade. For testing QED effects within astrophysical systems the relevant dimensionless parameters are the frequency compared to the Compton frequency, the field strength over the Schwinger critical field strength (\ref{eq:criticalfield}), as well as the sign of the relativistic invariant $c^2{\bf B}^2 - {\bf E}^2$. As can be seen by the second of these requirements, the laboratory experiments of today will at most be weakly nonlinear, whereas the effects in astrophysical systems, such as magnetars, can be strongly nonlinear. However, the combined effect of laser--plasma dynamics and vacuum nonlinearities would yield unique signatures, and could be probes of more exotic phenomena in astrophysical plasmas. One such example in the testing of the Unruh effect \cite{Unruh,Chen-Tajima} as a means of understanding the Hakwing effect \cite{Hawking}.

\section{Conclusion and outlook}

The possibility of simulating astrophysical events in a laboratory
environment has, during the last decade, progressed \cite{Remington,Chen}. 
Apart from the astrophysical tests,
laser-plasma systems also provide an opportunity to test certain aspects of fundamental
physics, e.g. the properties if the quantum vacuum, via strong fields. 
Strong ($\sim 10-100\, \mathrm{MV/m}$) 
coherent electromagnetic fields can nowadays be produced in
superconducting cavities \cite{Graber}, and fields within plasmas could come close to the Schwinger limit (\ref{eq:criticalfield}). Moreover, QED effects are part of many
astrophysical phenomena, such as pair cascading, and thus 
laboratory astrophysics has a natural connection to investigations of the quantum vacuum. 

Here, we have reviewed the implications of QED corrections to 
classical electrodynamics and the propagation of electromagnetic waves and pulses.
In particular, QED corrections on photon--plasma interactions were described. The modifications introduced by the nonlinear quantum vacuum were
considered for, e.g.\ coherent and incoherent pulse propagation. Analytical, perturbative, and numerical ways of analyzing the governing equations were presented. Moreover, the properties of nonlinear collective effects were 
presented, such as three-dimensional pulses collapse and the formation of light
bullets. 

The application of the results can be seen both in an astrophysical
context as well as in a laboratory setting. For example, in magnetar environments 
\cite{magnetar} photon splitting \cite{Adler} is important 
and it is believed to give a plausible 
explanation for the radio silence of magnetars \cite{Harding}. 
On the other hand, collective effects, such as the ones 
presented here, could give valuable insight of QED phenomena in 
astrophysical environments. 
In the laboratory, the
formation of ultra-high intensity pulse trains, due to
self-compression and pulse splitting, is a truly exiting prospect. 
The fact that
such configurations are within laboratory reach, using the next
generation laser-plasma facilities, makes the predicted effects and
their connection to astrophysical events even more interesting and may  
open up new possibilities for basic and applied research in
the future.    

\section*{Acknowledgments}

The authors are grateful to G. Brodin, L. Stenflo,
B. Eliasson, J.T. Mendon\c{c}a, R. Bingham, D.D. Tskhakaya, J. Collier, and
P.A.\ Norreys for valuable comments, collaboration, 
and stimulating discussions. We are indebted to D.D. Meyerhofer for going through our manuscript and offering valuable suggestions for its improvements.  

This research was supported by the Swedish Research Council Contract 
No. 621-2004-3217.

\input{bibliographyR}

\end{document}

%% file: bibliographyR.tex






%% file: review_final_twocol.bbl
\begin{thebibliography}{42}

  \bibitem[Adler et al.(1970)]{Adler-etal}
  Adler, S.L. , Bahcall, J.N., Callan, C.G., and Rosenbluth, M.N., 
  Phys. Rev. Lett. \textbf{25} 1061 (1970).
  
  \bibitem[Adler(1971)]{Adler} Adler, S.L., Ann.\ Phys.-NY  \textbf{67} 599 (1971). 

  \bibitem[Adler and Shubert(1996)]{Adler-Shubert}
  Adler, S.L., and Shubert, C., Phys. Rev. Lett. \textbf{77} 1695 (1996).

  \bibitem[Agrawal(2001)]{Agrawal} 
  Agrawal, G., \textit{Nonlinear Fiber Optics} 
    (Academic Press, 2001). 
    
  \bibitem[Akhmadaliev et al.(2002)]{Akhmadaliev-etal}
  Akhmadaliev, Sh. Zh., Kezerashvili, G. Ya., Klimenko, S. G., Lee, R. N., et al.,
  Phys. Rev. Lett. \textbf{89}, 061802 (2002).
  
  \bibitem[Alexandrov et al.(1985)]{Alexandrov-etal}
   Alexandrov, E.B., Anselm, A.A., and Moskalev, A.N.,
  {Zh.\ Eksp.\ Teor.\ Fiz.}\ {\bf 89} 1181 (1985)
         [{Sov.\ Phys.\--JETP} {\bf 62} 680 (1985))]. 
  
  \bibitem[Alkhofer et al.(2001)]{alk01} 
  Alkhofer, R., et al., Phys.\ Rev.\ Lett.\ \textbf{87} 193902
  (2001). 
  
  \bibitem[Anderson et al.(1999a)]{Anderson-Cattani-Lisak}  
  Anderson, D., Cattani F., and Lisak, 
  M.,Phys.\ Scripta {\bf T82} 32 (1999). 
  
  \bibitem[Anderson et al.(1999b)]{dan99} 
  Anderson, D., Fedele, R., Vaccaro,V., et al., Phys.\ Lett. A \
  \textbf{258} 244 (1999b).
  
  \bibitem[Arendt and Eilek(2002)]{Arendt-Eilek} Arendt, P.N. Jr., and Eilek, J.A., 
    Astrophys. J. \textbf{581} 451 (2002).

  \bibitem[Asseo(2003)]{Asseo} Asseo, E., Plasma Phys.\ Control.\ Fusion
    \textbf{45} 853 (2003).  

  \bibitem[Avan et al.(2003)]{Avan-etal}
  Avan, J., Fried, H.M., and Gabellini, Y., Phys. Rev. D
  \textbf{67} 016003 (2003).
 
  \bibitem[Avetissian et al.(2002)]{Avetissian-etal} 
  Avetissian, H.K., et al., Phys. Rev. E \textbf{66} 016502 (2002).

  \bibitem[Avetisyan et al.(1991)]{Avetisyan-etal} 
  Avetisyan, G.K., Avetisyan, A.K., and Serdrakyan, Kh.V., JETP \textbf{72}
  26 (1991).
  
  \bibitem[Bahk et al.(2004)]{Bahk-etal}
  Bahk, S.-W., Rousseau, P., Planchon, T.A., et al., 
  Opt. Lett. \textbf{29}, 2837 (2004).
  
  \bibitem[Ba{\u\i}er and Katkov(1968)]{Baier-Katkov}
  Ba{\u\i}er,V.N., and Katkov, V.M., Sov. Phys. JETP \textbf{26} 854 (1968).

  \bibitem[Baier et al.1996()]{Baier-Milstein-Shaisultanov}
    Baier, V.N., Milstein, A.I., and Shaisultanov, R.Zh., Phys. Rev. Lett. 
    \textbf{77} 1691 (1996).
  
  \bibitem[Bailey et al.(1979)]{Bailey-etal}
  Bailey, J., et al., Nucl. Phys. B \textbf{150} 1 (1979).
  
  \bibitem[Bakalov et al.(1994)]{Bakalov-etal2}
  Bakalov, D., et al., Nucl. Phys. B \textbf{35} 180 (1994).
    
  \bibitem[Bakalov et al.(1998)]{Bakalov-etal}
    Bakalov, D., et al., Quantum Semiclass. Opt. \textbf{10} 239 (1998).

  \bibitem[Bamber et al.(1999)]{Bamber-etal}
  Bamber, C., et al., Phys. Rev. D \textbf{60} 092004 (1999).
  
  \bibitem[Baring and Harding(1997)]{Baring-Harding2}
  Baring, M.G., and Harding, A.K., Astrophys. J. \textbf{482} 372 (1997).

  \bibitem[Baring and Harding(2001)]{Baring-Harding}
  Baring, M.G., and Harding, A.K., Astrophys. J. \textbf{547} 929 (2001).

  \bibitem[Barton(1990)]{Barton}
  Barton, G., Phys. Lett. B \textbf{237} 559 (1990).
  
  \bibitem[Barton and Scharnhorst(1993)]{Barton-Scharnhorst}
  Barton, G., and Scharnhorst, K., J. Phys. A \textbf{26} 2037 (1993).
  
  \bibitem[Berestetskii et al.(1982)]{Berestetskii-etal} 
  Berestetskii, V.B., Lifshitz, E.M., and Pitaevskii, L.P.,
  \textit{Quantum Electrodynamics} (Pergamon Press, Oxford, 1982).
  
  \bibitem[Berezhiani et al.(1992)]{ber92} 
  Berezhiani, V.I., Tskhakaya, D.D., and Shukla, P.K., 
  Phys.\ Rev.\ A \textbf{46} 6608 (1992).
  
  \bibitem[Berezhiani et al.(2005)]{ber2005} 
  Berezhiani, V.I., Garochava, D. P., Mikhladze, S. V., Sigua, K. I., Tsintsadze, N. L., Mahajan, S. M., Kishimoto, Y., and Nishikawa, K.,  
  Phys.\ Plasmas \textbf{12}, 062308 (2005).
  
  \bibitem[Berge et al.(1995)]{r4} Berge, L., Kuznetsov, E.A., and Rasmussen, J.J., 
    Phys.\ Rev.\ E, \textbf{53}, R1340 (1995).

  \bibitem[Berge and Rasmussen(1996)]{r5} 
    Berge, L.,  and Rasmussen, J.J., Phys.\ Plasmas,
    \textbf{3}, 324 (1996).

  \bibitem[Berge et al.(1998)]{r6} Berge, L., Rasmussen, J.J., and Schmidt, M.R., 
    Physica Scripta, \textbf{T75}, 18 (1998). 
  
  \bibitem[Bernard(1998)]{Bernard2} Bernard, D., Experiments on Photon--Photon 
  Scattering, in \textit{Frontier Tests of QED and Physics of the Vacuum}, eds.
  E. Zavattini, D. Bakalov, and C. Rizzo (Heron Press, Sofia, Hungary, 1998).

  \bibitem[Bernard(1999)]{Bernard} Bernard, D., Nucl. Phys. B (Proc. Suppl.)
  \textbf{72} 201 (1999).
  
  \bibitem[Bernard(2000)]{Bernard3} Bernard, D., Nucl. Phys. B (Proc. Suppl.)
  \textbf{82} 439 (2000). 
  
  \bibitem[Bernard et al.(2000)]{Bernard-etal}
  Bernard, D., et al., Eur. Phys. J. D \textbf{10} 141 (2000).

  \bibitem[Beskin et al.(1993)]{Beskin-etal} 
    Beskin, V.I., Gurevich, A.V., and Istomin, Ya.N., 
    \textit{Physics of the Pulsar Magnetosphere}
    (Cambridge University Press, Cambridge, 1993).

  \bibitem[Bethe and Heitler(1934)]{Bethe-Heitler}
   Bethe, H.A., and Heitler, W., Proc. R. Soc. London A \textbf{146}, 83 (1934).
      
  \bibitem[Bialynicka--Birula and Bialynicki--Birula(1970)]{Bialynicka-Birula} 
    Bialynicka--Birula, Z., and
    Bialynicki--Birula, I., Phys.\ Rev.\ D \textbf{2} 2341 (1970).

  \bibitem[Bingham(2003)]{bob} Bingham, R., Nature \textbf{424} 258 (2003).

  \bibitem[Bingham et al.(2004)]{bob2} 
  Bingham, R., Mendon\c{c}a, J.T., and Shukla, P.K., 
  Plasma Phys.\ Controll.\ Fusion
    \textbf{46}, R1 (2004).

  \bibitem[Bloch et al.(1999)]{Bloch-etal2}
  Bloch, J.C.R., et al., Phys. Rev. D
  \textbf{60} 116011 (1999).

  \bibitem[Bloch et al.(2000)]{Bloch-etal}
  Bloch, J.C.R., Roberts, C.D., and Schmidt, S.M., Phys. Rev. D \textbf{61} 117502 (2000).
  
  \bibitem[Bloembergen(1996)]{Bloembergen}
  Bloembergen, N., \textit{Nonlinear Optics} (World Scientific, 1996).

  \bibitem[Bluhm et al.(1999)]{Bluhm-etal} Bluhm, R., Kosteleck\'y, V.A., 
  and Russel, N., Phys. Rev. Lett. \textbf{82} 2254 (1999).

  \bibitem[Bluhm(2004)]{Bluhm} Bluhm, R., Nucl. Instrum. Meth. B \textbf{221}
  6 (2004).
  
  \bibitem[Boer and van Holten(2002)]{Boer-vanHolten}
  Boer, D., and van Holten, J.-W., Exploring the QED vacuum with
  laser interferometry, hep-ph/0204207 (2002).

  \bibitem[Bordag et al.(2001)]{Bordag-Mohideen-Mostepanenko} 
  Bordag, M., Mohideen, U., and Mostepanenko, V.M., 
  Phys. Rep. \textbf{353} 1 (2001).
  
  \bibitem[Borghesi et al.(2002)]{borghesi} 
  Borghesi, M., et al., Phys. Rev. Lett. {\bf 88}, 135002 (2002). 
  
  \bibitem[Bradley et al.(2003)]{Bradley-etal}
  Bradley, R., Clarke, J., Kinion, D., Rosenberg, L. J., et al., 
  Rev. Mod. Phys. \textbf{75}, 777 (2003).

  \bibitem[Breit and Wheeler(1934)]{Breit-Wheeler}
    Breit, G., and Wheeler, J.A., Phys. Rev. \textbf{46}. 1087 (1934).
    
  \bibitem[Bressi et al.(2002)]{Bressi-etal}
  Bressi, G., Carugno, G., Onofrio, R., and Ruoso, G.,
  Phys. Rev. Lett. \textbf{88} 041804 (2002).
  
  \bibitem[Brezin and Itzykson(1970)]{Brezin-Itzykson}
  Brezin, E., and Itzykson, C., Phys. Rev. D \textbf{2} 1191 (1970).

  \bibitem[Brodin et al.(2001)]{Brodin-Marklund-Stenflo} 
    Brodin, B., Marklund, M.,  and 
    Stenflo, L., Phys.\ Rev.\ Lett.~\textbf{87} 171801 (2001).
    
  \bibitem[Brodin et al.(2002)]{Brodin-Marklund-Stenflo2}
  Brodin, G., Marklund, M., and Stenflo, L.,
  Physica Scr. \textbf{T98} 127 (2002).
 
  \bibitem[Brodin et al.(2003)]{Brodin-etal} 
   Brodin, G., Stenflo, L., Anderson, D., 
   Lisak, M., Marklund, M., and Johannisson, P., Phys.\ Lett. A
  \textbf{306} 206 (2003).
  
  \bibitem[Budil et al.(2000)]{Budil-etal}
  Budil, K. S., Gold, D. M., Estabrook, K. G., Remington, B. A., et al., 
  Astrophys. J. Suppl. \textbf{127}, 261 (2000).

  \bibitem[Bujarbarua and Schamel(1981)]{bajurbarua} 
  Bujarbarua, S., and Schamel, H., J. Plasma Phys. {\bf 25}, 515 (1981).
    
  \bibitem[Bula et al.(1996)]{Bula-etal} 
    Bula, C., et al., Phys. Rev. Lett. \textbf{76}, 3116 (1996).

  \bibitem[Bulanov and Sakharov(1991)]{Bulanov-Sakharov}
    Bulanov, S.V., and Sakharov, A.S., JETP Lett. \textbf{54} 203 (1991).
    
  \bibitem[Bulanov et al.(1999)]{bulanov99} 
  Bulanov, S.V.,  et al., Phys. Rev. Lett. {\bf 82}, 3440 (1999).
  
  \bibitem[Bulanov et al.(2000)]{Bulanov-etal3} 
  Bulanov, S.V., et al., JETP Lett. \textbf{71} 407 (2000).
  
  \bibitem[Bulanov et al.(2003)]{Bulanov-etal} 
    Bulanov, S.V., Esirkepov, T.,  and Tajima, T., 
    Phys.\ Rev.\ Lett.\ {\bf 91}, 085001 (2003); Erratum, ibid. \textbf{91}
    085001 (2003). 
  

  \bibitem[Bulanov(2004)]{Bulanov1} Bulanov, S.S., Phys. \ Rev.\  E~\textbf{69},
  036408 (2004). 
 
  \bibitem[Bulanov et al.(2004)]{Bulanov-etal2} 
  Bulanov, S.V., et al., Plasma Phys. Rep. \textbf{30} 196 (2004).
  
  \bibitem[Bulanov et al.(2005)]{Bulanov-etal2005}
  Bulanov, S.S., Fedotev, A.M., and Pegoraro, F., 
  Phys. Rev. E \textbf{71}, 016404 (2005).  

  \bibitem[Bunkin and Kasakov(1970)]{Bunkin1970}
  Bunkin, F. V., and Kazakov, A. E., Dokl. Akad. Nauk. SSR \textbf{93}, 1274 (1970). 
   
  \bibitem[Burke et al.(1997)]{bur97} Burke, D.L., {et al.}, Phys.\ Rev.\ Lett.\
  \textbf{79} 1626 (1997).

  \bibitem[Cairns et al.(2004)]{cai04} 
    Cairns, R.A., Reitsma, A., and Bingham, R., Phys.\
    Plasmas \textbf{11} 766 (2004).
  
  \bibitem[Calmet et al.(1977)]{Calmet-etal}
  Calmet, J., et al., Rev. Mod. Phys. \textbf{49} 21 (1977).
  
  \bibitem[Cameron et al.(1993)]{Cameron-etal}
  Cameron, R., et al., Phys. Rev. D \textbf{47} 3707 (1993).
  
  \bibitem[Casher et al.(1979)]{Casher-etal}
  Casher, A., Neuberger, H., and Nussinov, S., 
  Phys. Rev. D \textbf{20} 179 (1979).

  \bibitem[Casimir and Polder(1948)]{Casimir1} 
  Casimir, H.B.G., and Polder, D., Phys.\ Rev.\ \textbf{73}
  1948.
  
  \bibitem[Casimir(1948)]{Casimir2} 
  Casimir, H.B.G., {Proc.\ Kon.\ Ned.\ Akad.\
  Wetenschap.}, ser.\ B \textbf{52} 793 (1948). 

  \bibitem[\v{C}erenkov(1934)]{Cerenkov} 
    \v{C}erenkov, P.A., Doklady Akad.\ Nauk SSSR \textbf{2} 451 (1934).
    
  \bibitem[Chefranov(2004)]{Chefranov} 
  Chefranov, S.G., Phys. Rev. Lett. \textbf{93} 254801 (2004).

  \bibitem[Chen(2003)]{Chen} Chen, P., AAPPS Bull. \textbf{13} 3 (2003).

  \bibitem[Chen and Tajima(1999)]{Chen-Tajima}
  Chen, P., and Tajima, T., Phys. Rev. Lett. \textbf{83} 256 (1999). 

  \bibitem[Chernev and Petrov(1992))]{Chernev-Petrov} 
    Chernev, P.,  and Petrov, V., Opt.\ Lett.\
    \textbf{17} 172 (1992). 
    
  \bibitem[Chistyakov et al.(1998)]{Chistyakov-etal}
  Chistyakov, M. V., Kuznetsov, A. V., and Mikheev, N. V., 
  Phys. Lett. B \textbf{434}, 67 (1998). 
    
  \bibitem[Colloday and Kosteleck\'y(1998)]{Colloday-Kostelecky}
  Colloday, D., and Kosteleck\'y, V.A., Phys. Rev. D \textbf{58} 116002 (1998).

  \bibitem[Curtis(1982)]{CurtisMichel} 
  Curtis, M.F., Rev. Mod. Phys. \textbf{54} 1 (1982).
  
  \bibitem[de Groot et al.(1980)]{deGroot} 
  de Groot, S.R., van Leeuwen, W.A., and van Weert, C.G., {\it Relativistic Kinetic Theory: 
  Principles and Applications} (North-Holland, Amsterdam, 1980).
  
  \bibitem[Delbr\"uck(1933)]{Delbruck}
  Delbr\"uck, M., Z. Phys. \textbf{84} 144 (1933).
    
  \bibitem[De Lorenci et al.(2000)]{DeLorenci-etal}   
    De Lorenci, V.A., Klippert, R., 
    Novello, M., and Salim, J.M., Phys.\ Lett.\ B \textbf{482} 134
    (2000).
  
  \bibitem[Desaix et al.(1991)]{Desaix-Anderson-Lisak}  
  Desaix, M., Anderson, D., and Lisak M.,
  J.\ Opt.\ Soc.\ Am.\ B {\bf 8} 2082 (1991). 
  
  \bibitem[DESY XFEL(2005)]{DESY}
  DESY X-Ray Free Electron Laser, \url{http://xfel.desy.de/} (2005).

  \bibitem[Dewar(1974)]{Dewar}  Dewar, R.L., {Phys.\ Rev.\ A} {\bf 10} 2017
  (1974).

  \bibitem[Dicus et al.(1998)]{Dicus-etal}
  Dicus, D.A., Kao, C., and Repko, W.W., 
    {Phys.\ Rev.\ D} \textbf{57} 2443 (1998).
  
  \bibitem[Dietrich(2003)]{Dietrich}
  Dietrich, D.D., Phys. Rev. D \textbf{68} 105005 (2003).

  \bibitem[Ding and Kaplan(1989)]{Ding} 
   Ding, Y.J.,  and Kaplan, A.E., 
  {Phys.\ Rev.\ Lett.}~\textbf{63} 2725 (1989).

  \bibitem[Ding and Kaplan(1990)]{Ding-Kaplan2}
  Ding, Y.J., and Kaplan, A.E., Phys. Rev. Lett. \textbf{65}, 2746 (1990).
   
  \bibitem[Ding and Kaplan(1992)]{Ding-Kaplan} 
  Ding, Y.J., and Kaplan, A.E., {J.\ Nonlinear Opt.\ Phys.\
  Mater.}\ {\bf 1} 51 (1992).
  
  \bibitem[Dittrich(1979)]{Dittrich} 
  Dittrich, W., Phys. Rev. D \textbf{19} 2385 (1979).

  \bibitem[Dittrich and Gies(1998)]{Dittrich-Gies2} 
    Dittrich, W., and Gies, H., Phys.\ Rev.\ D
    \textbf{58} 025004 (1998). 
  
  \bibitem[Dittrich and Gies(2000)]{Dittrich-Gies} Dittrich W., and Gies, H., 
    \textit{Probing the Quantum Vacuum} (Springer-Verlag, Berlin, 2000).
    
  \bibitem[Dremin(2002)]{Dremin} 
  Dremin, I.M., JETP Lett.\ \textbf{75} 167 (2002).

  \bibitem[Duncan(2002)]{Duncan} Duncan, R.C., in 
  \textit{Fifth Huntsville $\gamma$-Ray Burst Symposium}, astro-ph/0002442 
  (2002).

  \bibitem[Dunne(2004)]{Dunne} Dunne, G.V., Heisenberg--Euler Effective Lagrangians: 
  Basics and Extensions, hep-ph/0406216, to appear in \textit{From Fields to 
  Strings: Circumnavigating 
  Theoretical Physics}, eds. M. Shifman, A. Vainshtein, and J. Weather 
  (World Scientific, 2004).
  
  \bibitem[Dupays et al.(2005)]{Dupays-etal}
  Dupays, A., Rizzo, C., Roncadelli, M., and Bignami, G. F., 
  Phys. Rev. Lett. \textbf{95}, 211302 (2005).
  
  \bibitem[Eliasson and Shukla(2006)]{Eliasson-Shukla2006}
  Eliasson, B., and Shukla, P. K., Phys. Rep. \textbf{422}, 225 (2006).
  
  \bibitem[Eliezer(2002)]{Eliezer}
  Eliezer, S., \textit{The Interaction of High-Power Lasers with Plasmas} 
  (Institute of Physics, Bristol, 2002),
  
  \bibitem[Elmfors and Skagerstam(1995)]{Elmfors-Skagerstam}
  Elmfors, P., and Skagerstam, B.-S., Phys. Lett. B \textbf{348} 141 (1995);
  \textbf{348} 141(E) (1995).

  \bibitem[Erber(1966)]{Erber}
  Erber, T.,Rev. Mod. Phys. \textbf{38} 626 (1966).
  
  \bibitem[Eriksson et al.(2004)]{Eriksson-etal}
  Eriksson, D., Brodin, G., Marklund, M., and Stenflo, L.,
  Phys. Rev. A \textbf{70} 013808 (2004).
  
  \bibitem[Esirkepov et al.(1998)]{Esirkepov-etal3} 
  Esirkepov, T.Zh., et al., JETP Lett. \textbf{68} 36 (1998).

  \bibitem[Esirkepov et al.(1999)]{Esirkepov-etal} 
  Esirkepov, T.Zh., et al., JETP Lett. \textbf{70} 82 (1999).
  
  \bibitem[Esirkepov et al.(2004)]{Esirkepov-etal2}
  Esirkepov, T., et al., Phys. Rev. Lett. \textbf{92} 175003 (2004).

  \bibitem[Farina and Bulanov(2001a)]{farina01}
  Farina, D., and Bulanov, S.V., Phys. Rev. E {\bf 64} 066401 (2001).
  
  \bibitem[Farina and Bulanov(2001b)]{Farina} 
  Farina, D., and Bulanov, S.V., Phys. Rev. Lett. {\bf 86}, 5289 (2001).

  \bibitem[Ford and Steel(1990)]{Ford-Steel}
  Ford, G.W., and Steel, D.G., Phys. Rev. Lett. \textbf{65} 2745 (1990). 
  
  \bibitem[Fradkin et al.(1991)]{Fradkin-etal}
  Fradkin, E.S., Gitman, D.M., and Shvartsman, Sh.M.,
  \textit{Quantum Electrodynamics with Unstable Vacuum}
  (Springer-Verlag, Berlin, 1991).

  \bibitem[Fried et al.(2001)]{Fried-etal}
  Fried, H.M., Gabellini, Y., McKellar, B.H.J., and Avan, J., 
  Phys. Rev. D \textbf{63} 125001 (2001).

  \bibitem[Gaeta(2003)]{Gaeta} Gaeta, A.L., Science \textbf{301} 54 (2003).
  
  \bibitem[Gahn et al.(2000)]{gahn00} Gahn, C., {et al.}, Appl. Phys. Lett. {\bf 77}
  2662 (2000). 


  \bibitem[Gies(1999a)]{Gies}
  Gies, H., Phys. Rev. D \textbf{60} 105002 (1999a).
  
  \bibitem[Gies(1999b)]{Gies2}
  Gies, H., Phys. Rev. D \textbf{60} 105033 (1999b).
  
  \bibitem[Gies(2000)]{Gies3}
  Gies, H., Phys. Rev. D \textbf{61} 085021 (2000).
  
  \bibitem[Goloviznin and Shep(1999)]{gol99}
  Goloviznin, V.V., and Shep, T.J., JETP Lett.\
  \textbf{70} 450 (1999).
  
  \bibitem[Graber(1993)]{Graber} Graber, J., Ph.D.\ Dissertation (Cornell 
    University, 1993), see also \\ 
  \url{http://w4.lns.cornell.edu/public/CESR/SRF/BasicSRF/SRFBas1.html}  
  
  \bibitem[Greaves et al.(1994)]{Greaves-etal} Greaves, R.G., Tinkle, M.D., and
  Surko, C.M.,  Phys.\ Plasmas \textbf{1} 1439 (1994). 

  \bibitem[Greaves and Surko(1995)]{Greaves-Surko} 
  Greaves, R.G., and Surko, C.M., Phys.\
    Rev.\ Lett.\ \textbf{75} 3846 (1995).

  \bibitem[Greaves and Surko(1997)]{Greaves-Surko2}
  Greaves, R.G., and Surko, C.M., Phys. Plasmas \textbf{4} 1528 (1997).

  \bibitem[Greiner et al.(1985)]{Greiner} 
  Greiner, W., M\"uller, B., and Rafelski, J., 
  \textit{Quantum electrodynamics of strong fields} (Springer, Berlin,
  1985).
  
  \bibitem[Greisen(1966)]{Greisen}  
    Greisen, K., Phys.\ Rev.\ Lett.\ \textbf{16} 748 (1966).

  \bibitem[Grib et al.(1994)]{Grib-etal} 
  Grib, A.A., Mamaev, S.G., and Mostepanenko, V.M., 
  \textit{Vacuum Effects in Strong Fields} (Atomizdat, Moscow, 1988).
  
  \bibitem[Hall et al.(2000)]{Hall-etal}
  Hall, J.L., Ye, J., and Ma, L.-S., Phys. Rev. A \textbf{62} 013815 (2000).
  
  \bibitem[Harber etal.(2005)]{Harber-etal}
  Harber, D. M., Obrecht, J. M., McGuirk, J. M., and Cornell, E. A., 
  Phys. Rev. A \textbf{72}, 033610 (2005).

  \bibitem[Harding(1991)]{Harding}
   Harding, A.K.,Science \textbf{251} 1033 (1991).
   
  \bibitem[Harding et al.(1997)]{Harding-etal}
  Harding, A.K., Baring, M.G., and  Gonthier,ÊP.L., Astrophys. J.
  \textbf{476} 246 (1997).

  \bibitem[Hasegawa(1975)]{Hasegawa} 
  Hasegawa, A., \textit{Plasma Instabilities and
  Nonlinear Effects} (Springer-Verlag, Berlin, 1975).

  \bibitem[Hawking(1974)]{Hawking} Hawking, S.W., Nature \textbf{248} 30 (1974).
  
  \bibitem[Hayata and Koshiba(1993)]{Hayata-Koshiba} 
  Hayata, K., and Koshiba, M., Phys.\ Rev.\ E
    \textbf{48} 2312 (1993). 
  
  \bibitem[HEDLA(2005)]{HEDLA}
  HEDLA-2004, \textit{Proceedings of the 5th International Conference on 
    High Energy Density Laboratory Astrophysics (Tuscon Arizona, March
    10--13 2004)}, Astrophys. Space Sci. \textbf{298}, Issue 1--2 (2005).

  \bibitem[Heisenberg and Euler(1936)]{Heisenberg-Euler} 
  Heisenberg, W., and Euler, H., Z.\
  Phys.~\textbf{98} 714 (1936).

  \bibitem[Helander and Ward(2003)]{Helander-Ward}
  Helander, P., and Ward, D.J., Phys. Rev. Lett. \textbf{90} 135004 (2003).
  
  \bibitem[Heyl and Hernquist(1997a)]{Heyl-Hernquist} 
  Heyl, J.S., and Hernquist, L., J.\ Phys.\ A: Math.\ Gen.~\textbf{30}
  6485 (1997a).
  
  \bibitem[Heyl and Hernquist(1997b)]{Heyl-Hernquist2} 
  Heyl, J.S., and Hernquist, L., Phys. Rev D \textbf{55} 2449 (1997b).
  
  \bibitem[Heyl and Hernquist(1999)]{Heyl-Hernquist3}
  Heyl, J., and Hernquist, L., Phys. Rev. D \textbf{59} 045005 (1999).
  
  \bibitem[Heyl and Shaviv(2002)]{Heyl-Shaviv} 
  Heyl, J.S., and Shaviv, N.J., Phys. Rev. D \textbf{66}
  023002 (2002).
  
  \bibitem[Heyl et al.(2003)]{Heyl-etal}
  Heyl, J.S., Shaviv, N.J., and Lloyd, D., Mon. Not. R. Astron. Soc. 
  \textbf{342} 134 (2003).

  \bibitem[Iacopini and Zavattini(1979)]{Iacopini-Zavattini}
  Iacopini, E., and Zavattini, E., Phys. Lett. B \textbf{85} 151 (1979).
  
  \bibitem[ILE/Osaka(2005)]{ILE}
  Institute for Laser Engineering, Osaka University, \url{http://www.ile.osaka-u.ac.jp/}
  (2005).
  
  \bibitem[Jackiw and Kosteleck\'y(1999)]{Jackiw-Kostelecky}
  Jackiw, R., and Kosteleck\'y, V.A., Phys. Rev. Lett. \textbf{82} 3572 (1999).
  
  \bibitem[Jarlskog et al.(1973)]{Jarlskog-etal}
  Jarlskog, C., et al., Phys. Rev. D \textbf{8} 3813 (1973).
  
  \bibitem[Jeah-Sheng et al.(2004)]{Jeah-Sheng-etal}
  Jeah-Sheng, W., Wei-Tou, N., Sheng-Jui, C., Class. Quantum Grav.
  \textbf{21} S1259 (2004).

  \bibitem[Kaplan and Ding(2000)]{Kaplan-Ding}
   Kaplan, A.E., and Ding, Y.J., 
    {Phys.\ Rev.\ A} \textbf{62} 043805 (2000).
  
  \bibitem[Karpman(1971)]{Karpman} 
  Karpman, V.I., Plasma Phys.\ {\bf 13}, 477 (1971).
  
  \bibitem[Karpman(1998)]{Karpman2} 
  Karpman, V.I.,
    Phys.\ Plasmas {\bf 5}, 932 (1998).  

  \bibitem[Karpman and Washimi(1977)]{Karpman-Washimi} 
  Karpman, V.I., and Washimi, H., J.\ Plasma
    Phys.\ \textbf{18} 173 (1977).

  \bibitem[Kaw et al.(1982))]{r10} 
  Kaw, P.K., Tsintsadze, N.L., and Tskhakaya, D.D., Sov.\
    Phys.\ JETP, \textbf{55, }839 (1982).

  \bibitem[Kaw et al.(1992)]{Kaw} Kaw, P.K., et al., Phys. Rev. Lett. {\bf 68}, 
  3172 (1992).

  \bibitem[Khokonov and Nitta(2002)]{Khokonov-Nitta}
  Khokonov, M.Kh., and Nitta, H., Phys. Rev. Lett. \textbf{89} 094801 (2002).
  
  \bibitem[Kim et al.(2000)]{Kim-etal}
  Kim, A., et al., JETP Lett. \textbf{72} 241 (2000).

  \bibitem[Kirsebom et al.(2001)]{Kirsebom-etal}
  Kirsebom, K., et al., Phys. Rev. Lett. \textbf{87} 054801 (2001).

  \bibitem[Kivshar and Agrawal(2003)]{Kivshar-Agrawal} 
  Kivshar, Y.S., and Agrawal, G.P., 
    \textit{Optical Solitons} (Academic Press, San Diego, 2003). 

 \bibitem[Klein and Nigam(1964a)]{Klein-Nigam1} Klein, J.J., and Nigam, B.P., 
  Phys. Rev. \textbf{136} B1279 (1964a).

  \bibitem[Klein and Nigam(1964b)]{Klein-Nigam2} Klein, J.J., and Nigam, B.P., 
  Phys. Rev. \textbf{136} B1540 (1964b).

  \bibitem[Kluger et al.(1991)]{Kluger-etal} 
  Kluger, Y., et al., Phys. Rev. Lett \textbf{67} 2427 (1991).

  \bibitem[Kondratyev(2002)]{Kondratyev} 
  Kondratyev, V.N., Phys.\ Rev.\
    Lett.~\textbf{88} 221101 (2002).

  \bibitem[Kouveliotou(1998)]{magnetar} 
  Kouveliotou, C., Dieters, S., Strohmayer, T., 
    {et al.}, Nature \textbf{393} 235 (1998). 
 
  \bibitem[Kozlov et al.(1979a)]{kozlov79}
  Kozlov, V.A., {et al.}, Sov. Phys. JETP {\bf 49} 75 (1979).

 \bibitem[Kozlov et al.(1979b)]{kozlov79b} 
 Kozlov, V.A., et al., JETP {\bf 76}, 148 (1979).
  
  \bibitem[Kuznetsov et al.(1995)]{r3} 
  Kuznetsov, E.A., Rasmussen, J.J., Rypdal, K., 
    Turitsyn, S.K., Physica D, \textbf{87} 273 (1995).

   \bibitem[Lamoreaux(1998)]{Lamoreaux}
  Lamoreaux, S.K., Phys. Rev. Lett. \textbf{78} 5 (1997); erratum, ibid.
  \textbf{81} 5475 (1998).
  
  \bibitem[Latorre et al.(1995)]{Latorre-etal}
  Latorre, J.I., Pascual, P., and Tarrach, R., 
    {Nucl.\ Phys.\ B} \textbf{437} 60 (1995).
  
  \bibitem[Lee et al.(1995)]{nif} Lee, R.W., {et al.},
    \textit{Science on High-Energy Lasers} (Lawrence Livermore
    National Laboratory, 1995). 
  
  \bibitem[Lee and Fairbanks(2002)]{Lee-Fairbanks}
  Lee, S.A., and Fairbanks, W.M., in \textit{Laser Physics at the Limits},
  eds. Figger, H., Meschede, D., and Zimmermann, C. (Springer-Verlag,
  Berlin, 2002). 

  \bibitem[Liang et al.(1998)]{liang98} 
  Liang, E.P., Wilks, S.C., and Tabak, M., Phys. Rev. Lett. {\bf 81}
  4487 (1998). 
  
  \bibitem[Liepe(2000)]{Q-factor2}  
  Liepe, M., in \textit{eConf C00082 WE204}, (2000), see also \textit{%
  Pulsed Superconductivity Acceleration}, physics/0009098.
 
  \bibitem[Lipa et al.(2003)]{Lipa-etal} 
  Lipa, J.A., Nissen, J.A., Wang, S., Stricker, D.A., and Avaloff, D.,
  Phys. Rev. Lett. \textbf{90} 060403 (2003).

  \bibitem[Lodenqual et al.(1974)]{Lodenqual-etal}
  Lodenqual, J., et al., Astrophys. J. \textbf{190} 141 (1974).
  
  \bibitem[Lontano et al.(2003)]{lontano} 
  Lontano, M., et al., Phys. Plasmas {\bf 10}, 639 (2003).
  
  \bibitem[Luiten and Petersen(2004a)]{Luiten-Petersen2}
  Luiten, A.N., and Petersen, J.C., Phys. Lett. A \textbf{330} 429 (2004a)
  
  \bibitem[Luiten and Petersen(2004b)]{Luiten-Petersen}
  Luiten, A.N., and Petersen, J.C., Phys. Rev. A \textbf{70}
  033801 (2004b).
  
  \bibitem[Lundstr\"om et al.(2005)]{Lundstrom-etal}
  Lundstr\"om, E., Brodin, G., Lundin, J., et al., hep-ph/0510076 (2005)
  
  \bibitem[Luo et al.(2002)]{Luo-etal} Luo, Q., {et al.}, Phys.\ Rev.\ E {\bf 66}
  026405 (2002).

  \bibitem[Luther et al.(1994)]{r8} 
    Luther, G.G., Newell, A.C., and Moloney, J.V., Physica
    D, \textbf{74, }59 (1994).

  \bibitem[Mamaev et al.(1981)]{Mamaev-etal} 
    Mamaev, S.G., Mostepanenko, V.M., and 
    E\u{\i}des, M.I., Sov.\ J.\ Nucl.\ Phys.\ \textbf{33} 569 (1981).

  \bibitem[Marinov and Popov(1977)]{Marinov-Popov}
  Marinov, M.S., and Popov, V.S., Fortsch. Phys. \textbf{25} 373 (1977). 
    
  \bibitem[Marklund(2004)]{Marklund}
  Marklund, M., Phys. Scr. \textbf{T113} 59 (2004).

  \bibitem[Marklund et al.(2003)]{Marklund-Brodin-Stenflo} 
  Marklund, M., Brodin, G., and 
    Stenflo, L., Phys.\ Rev.\ Lett.\ \textbf{91}  163601 (2003).

  \bibitem[Marklund et al.(2004a)]{r1} 
  Marklund, M., Eliasson, B., and Shukla, P.K., JETP
    Lett. {\bf 79}, 262 (2004a).

  \bibitem[Marklund et al.(2004b)]{Marklund-Shukla-Brodin-StenfloC}
  Marklund, M., Shukla, P.K., Brodin, G., and Stenflo, L.,
  Proceedings of the 12$^{\text{th}}$ International Congress on Plasma Physics,
  Nice, France, 2004, 03--0100; e-Proceedings available at
  \url{http://hal.ccsd.cnrs.fr/ccsd-00003094/en} 
  (2004b).
  
  \bibitem[Marklund et al.(2004c)]{Marklund-Shukla-Brodin-Stenflo2}
  Marklund, M., Shukla, P.K., Brodin, G., and Stenflo, L.,
  New J. Phys. \textbf{6} 172 (2004c).
  
  \bibitem[Marklund et al.(2005a)]{NJP}
  Marklund, M., Brodin, G., Stenflo, L., and Shukla, P.K.,
  New J. Phys. \textbf{7} 70 (2005a).

  \bibitem[Marklund et al.(2005b)]{Marklund-Shukla-Eliasson} 
  Marklund, M., Shukla, P.K., 
  and Eliasson, B., Europhys. Lett.\ \textbf{71}, 327 (2005b).

  \bibitem[Marklund et al.(2005c)]{Marklund-Shukla-Brodin-Stenflo}
  Marklund, M., Shukla, P.K., Brodin, G., and Stenflo, L.,
  J. Plasma Phys.\ \textbf{71}, 527 (2005c).
    
  \bibitem[Marklund et al.(2005d)]{Marklund-Shukla-Stenflo-Brodin-Servin}
  Marklund, M., Shukla, P.K., Stenflo, L., Brodin, G., and Servin, M.,
  Plasma Phys. Control. Fusion \textbf{47} L25 (2005d).
  
  \bibitem[Marklund et al.(2005e)]{Marklund-etal2005}
  Marklund, M, Tskhakaya, D.D, and Shukla, P.K.,
  Europhys. Lett. \textbf{72}, 950 (2005e).
  
  \bibitem[Max et al.(1974)]{Max-etal}
  Max, C.E., Arons, J., and Langdon, J.B., Phys. Rev. Lett. \textbf{33}
  209 (1974).

  \bibitem[McKenna and Platzman(1963)]{McKenna-Platzman}
  McKenna, J.M., and Platzman, P.M., Phys. Rev. \textbf{129} 2354 (1963).
  
  \bibitem[McKenna et al.(2005)]{McKenna-etal}
    McKenna, P., et al., Phys. Rev. Lett. \textbf{94} 084801 (2005).
  
  \bibitem[Melissinos(2002)]{Melissinos}
  Melissinos, A.C., Measuring the Phase Velocity of Light in a 
  Magnetic Field with the PVLAS Detector, hep-ph/0205169 (2002).

  \bibitem[Mendon\c{c}a(2001)]{Mendonca} 
  Mendon\c{c}a, J.T., \textit{Theory of Photon
  Acceleration} (Institute of Physics Publishing, Bristol, 2001).

  \bibitem[Mentzel et al.(1994)]{Mentzel-Berg-Wunner} 
  Mentzel, M., Berg, D., and Wunner, G., 
  Phys. Rev. D \textbf{50} 1125 (1994).
  
  \bibitem[Meyerhofer(1997)]{Meyerhofer}
  Meyerhofer, D.D., IEEE J. Quant. Electronics \textbf{33} 1935 (1997).

  \bibitem[Mikheev and Chistyakov(2001)]{Mikheev-Chistyakov}
  Mikheev, N. V., and Chistyakov, N. V., JETP Lett. \textbf{73}, 642 (2001).
   
  \bibitem[Mills(2002)]{Mills} Mills, A.P. Jr., Nucl. Instrum. Meth. B \textbf{192}
  107 (2002).
  
  \bibitem[Milonni(1994)]{Milonni} 
  Milonni, P.W., \textit{The Quantum Vacuum} (Academic Press, San Diego, 1994).
  
  \bibitem[Mittelman(1987)]{Mittelman}
  Mittelman, M., Phys. Rev. A \textbf{35}, 4624 (1987).
  
  \bibitem[Morse and Feshbach(1953)]{r9} 
  Morse, P.M., and Feshbach, H.,  \textit{Methods of
    Theoretical Physics} (McGraw-Hill, NY, 1953), p.\ 841.
    
  \bibitem[Mostepanenko and Frolov(1974)]{Mostepanenko-Frolov}
  Mostepanenko, V.M., and Frolov, V.M., Sov. J. Nucl. Phys. 
  \textbf{19} 451 (1974)

  \bibitem[Mostepanenko and Trunov(1997)]{Mostepanenko-Trunov}
  Mostepanenko, V.M., and Trunov, N.N., \textit{The Casimir effect
  and its Applications} (Oxford Science Publications, Oxford, 1997).
  
  \bibitem[Moulin et al.(1996)]{Moulin-etal} Moulin, F., Bernard, D.,
  and Amiranoff, F., Z. Phys. C \textbf{72} 607 (1996).
  
  \bibitem[Moulin and Bernard(1999)]{Moulin-Bernard}
  Moulin, F., and Bernard, D., Opt. Comm. \textbf{164} 137 (1999). 
  
  \bibitem[Mourou et al.(1998)]{mou98} 
  Mourou, G.A.,  Barty, C.P.J., and  Perry, M.D., Phys.\
    Today \textbf{51}, 22 (1998).

  \bibitem[Mourou et al.(2005)]{mou05}
  Mourou, G.A., Tajima, T., and Bulanov, S.V., Rev.\ Mod.\ Phys.\ \textbf{77},
  in press (2005).

  \bibitem[Narozhny et al.(1965)]{Narozhny-etal}
    Narozhny, N.B., et al., Sov. Phys. JETP \textbf{20}, 622 (1965).
    
  \bibitem[Narozhny and Nikishov(1970)]{Narozhny-Nikishov}
  Narozhny, N.B., and Nikishov, A.I., Sov. J. Nucl. Phys. \textbf{11} 
  596 (1970).

  \bibitem[Narozhny et al.(2004a)]{Narozhny-etal1}
  Narozhny, N.B., et al., Phys. Lett A \textbf{330} 1 (2004a).

  \bibitem[Narozhny et al.(2004b)]{Narozhny-etal2}
  Narozhny, N.B., et al., JETP Lett. \textbf{80} 382 (2004b).
    
  \bibitem[Naumova et al.(2001)]{naumova} 
  Naumova, N.M., et al., Phys. Rev. Lett. {\bf 87}, 185004 (2001).  
  
  \bibitem[Nibbelink and Pospelov(2005)]{Nibbelink-Pospelov}
  Nibbelink, S.G., and Pospelov, M., Phys. Rev. Lett. \textbf{94} 081601 (2005).
    
  \bibitem[Nikishov and Ritus(1964a)]{Nikishov-Ritus1}
    Nikishov, A.I., and Ritus, V.I., Sov. Phys. JETP \textbf{19}, 529 (1964a).
  
  \bibitem[Nikishov and Ritus(1964b)]{Nikishov-Ritus2}
   Nikishov, A.I., and Ritus, V.I., Sov. Phys. JETP \textbf{19}, 1191 (1964b).

  \bibitem[Nikishov and Ritus(1965)]{Nikishov-Ritus3}
    Nikishov, A.I., and Ritus, V.I., Sov. Phys. JETP \textbf{20}, 757 (1965).
  
  \bibitem[Nikishov and Ritus(1967)]{Nikishov-Ritus4}
    Nikishov, A.I., and Ritus, V.I., Sov. Phys. JETP \textbf{25}, 1135 (1967).
    
  \bibitem[Nitta et al.(2004)]{Nitta-etal}
  Nitta, H., Khokonov, M.Kh., Nagata, Y., and Onuki, S.,
  Phys. Rev. Lett. \textbf{93} 180407 (2004).

  \bibitem[Novello et al.(2001)]{Novello-etal} 
  Novello, M.,  Salim, J.M., De Lorenci, V.A., 
    et al., Phys.\ Rev.\ D \textbf{63} 103516 (2001).
    
  \bibitem[OMEGA EP(2005)]{OMEGA}
  The OMEGA EP Laser Facility, University of Rochester, 
  \url{http://omegaep.lle.rochester.edu/} (2005).

  \bibitem[Oohara and Hatakeyama(2003)]{Oohara-Hatakeyama}
  Oohara, W., and Hatakeyama, R., Phys. Rev. Lett. \textbf{91} 205005 (2003). 
  
  \bibitem[Oshima et al.(2004)]{Oshima-etal}
  Oshima, N. et al., Phys. Rev. Lett \textbf{93} 195001 (2004).
 
  \bibitem[Partovi(1993)]{Partovi2} Partovi, M.H., Detecting the 
  Photon-Photon Interaction by Colliding Laser Beam Interferometry, 
  hep-ph/9308293 (1993).

  \bibitem[Partovi(1994)]{Partovi} 
  Partovi, M.H., Phys.\ Rev.\ D \textbf{50},
    1118 (1994). 

  \bibitem[Patel(2004)]{pat02} Patel, N., Nature \textbf{415} 110 (2002). 
  
  \bibitem[Peacock(1998)]{Peacock}
  Peacock, J.A., \textit{Cosmological Physics} (Cambridge University Press, 1998).
  
  \bibitem[Peccei and Quinn(1977)]{Peccei-Quinn}
  Peccei, R.D., and Quinn, H.R., Phys. Rev. Lett. \textbf{38} 1440 (1977).

  \bibitem[Perry and Mourou(1994)]{Perry-Mourou} 
  Perry, M.D., and Mourou, G., Science \textbf{264} 917 (1994). 
  
  \bibitem[Piran(2004)]{Piran}
  Piran, T., Rev. Mod. Phys. \textbf{76}, 1143 (2004).
  
  \bibitem[Popov(1971)]{Popov1}
  Popov, V.S., JETP Lett. \textbf{13} 185 (1971).
  
  \bibitem[Popov(1972)]{Popov2}
  Popov, V.S., Sov. Phys. JETP \textbf{34} 709 (1972).
  
  \bibitem[Popov(1973)]{Popov3}
  Popov, V.S., JETP Lett. \textbf{18} 255 (1973).
  
  \bibitem[Popov(1974)]{Popov4}
  Popov, V.S., Sov. J. Nucl. Phys. \textbf{19} 81 (1974).
  
  \bibitem[Popov and Marinov(1973)]{Popov-Marinov}
  Popov, V.S., and Marinov, M.S., Sov. J. Nucl. Phys. \textbf{16}
  449 (1973).

  \bibitem[Prozorkevich et al.(2000)]{proz00}
  Prozorkevich, A.V., et al., Pair creation and plasma oscillations, 
  in \textit{Proceedings of Quark Matter in Astro- 
  and Particlephysics, University of Rostock, Germany, 2000}, Eds. D. Blaschke, G. Burau,
  and S.M., Schmidt, nucl-th/0012039 (2000).

  \bibitem[Pukhov(2003)]{Pukhov} Pukhov, A., Rep. Prog. Phys. \textbf{66} 47 (2003).

  \bibitem[Raizen and Rosenstein(1990)]{Raizen-Rosenstein}
  Raizen, M.G., and Rosenstein, B., Phys. Rev. Lett. \textbf{65} 2744 (1990).

  \bibitem[Reinhardt and Greiner(1977)]{Reinhardt-Greiner} 
  Reinhardt,ÊJ., and Greiner,ÊW., Rep. Prog. Phys. \textbf{40} 219 (1977).
  
  \bibitem[Reiss(1962)]{Reiss}
    Reiss, H.R., J. Math. Phys. \textbf{3}, 59 (1962).
  
  \bibitem[RHIC(2005)]{RHIC}
  Relativistic Heavy Ion Collider, Brookhaven National Laboratory,  
  \url{http://www.bnl.gov/rhic/} (2005).
    
  \bibitem[Remington(2005)]{Remington}
  Remington, B.A., Plasma Phys. Control. Fusion \textbf{47} A191 (2005).

  \bibitem[Rikken and Rizzo(2000)]{Rikken-Rizzo1} 
  Rikken, G.L.J.A., and Rizzo, C., Phys. Rev. A \textbf{63} 012107 (2000).

  \bibitem[Rikken and Rizzo(2003)]{Rikken-Rizzo2} 
  Rikken, G.L.J.A., and Rizzo, C., Phys. Rev. A \textbf{67} 015801 (2003).
  
  \bibitem[Ringwald(2001a)]{Ringwald}
  Ringwald, A., Phys. Lett. B \textbf{510} 107 (2001a).

  \bibitem[Ringwald(2001b)]{Ringwald1}  Ringwald, A., 
  Fundamental physics at an X-ray free electron laser, 
  in \textit{Electromagnetic Probes 
  of Fundamental Physics, Erice, Italy}, hep-ph/0112254
  (2001b).
 
  \bibitem[Ringwald(2003)]{Ringwald2} 
  Ringwald, A., 
  Boiling the vacuum with an X-ray free electron laser,  
  in \textit{Quantum Aspects of Beam Physics,
  Hiroshima, Japan}, hep-ph/0304139 (2003).
  
   \bibitem[Ritus(1976)]{Ritus} 
   Ritus, V.I., Sov.\ Phys.\ JETP \textbf{42} 774
    (1976). 
 
  \bibitem[Roberts et al.(2002)]{Roberts-etal}
  Roberts, C.D., Schmidt, S.M., and Vinnik, D.V., Phys. Rev. Lett. 
  \textbf{89} 153901 (2002).
  
  \bibitem[Rodionov(2004)]{Rodionov}
  Rodionov, V.N., JETP \textbf{98} 395 (2004).
  
  \bibitem[Rothenberg(1992)]{Rothenberg} 
  Rothenberg, J.E., Opt.\ Lett.\ \textbf{17} 583
    (1992).  

  \bibitem[Rozanov(1993)]{Rozanov93} 
  Rozanov, N.N., {Zh.\ Eksp.\ Teor.\ Fiz.}\ {\bf
  103} 1996 (1993) [{JETP} {\bf 76} 991 (1993)].

  \bibitem[Rozanov(1998)]{Rozanov}
  Rozanov, N.N., {Zh.\ Eksp.\ Teor.\ Fiz.}\ {\bf 113} 513
  (1998) [{JETP} {\bf 86} 284 (1998)].
  
  \bibitem[CCLRC(2005)]{RAL} Rutherford Appleton Laboratory Central 
  Laser Facility, \url{http://www.clf.rl.ac.uk/} (2005). 
  
  \bibitem[Ryutov et al.(2000)]{Ryutov-etal}
  Ryutov, D. D., Drake, R. P., and Remington, B. A., 
  Astrophys. J. Suppl. \textbf{127}, 465 (2000). 
 
  \bibitem[Saulson(1994)]{Saulson}
  Saulson, P.R., Fundamentals of Interferometric Gravitational
  Wave Detectors (World Scientific, Singapore, 1994).

  \bibitem[Sauter(1931)]{Sauter}
  Sauter, F., Z. Phys. \textbf{69} 742 (1931).
  
  \bibitem[Scharer et al.(1973)]{Scharer1973}
  Scharer, J. W., Garrison, J., Wong, J., and Swain, J. E., Phys, Rev. A \textbf{8}, 1582 (1973).

   \bibitem[Scharnhorst(1990)]{Scharnhorst}
  Scharnhorst, K., Phys. Lett B \textbf{236} 354 (1990).
  
  \bibitem[Scharnhorst(1998)]{Scharnhorst2}
  Scharnhorst, K., Ann. Phys. (Leipzig) \textbf{7} 700 (1998).

  \bibitem[Schwinger(1951)]{Schwinger} Schwinger, J., Phys.\ Rev.~\textbf{82}
    664 (1951).
  
  \bibitem[Schamel(2000)]{schamel}
  Schamel, H., Phys. Plasmas {\bf 7}, 4831 (2000).

  \bibitem[Scott(2003)]{Scott}
  Scott, A., \textit{Nonlinear Science} (Oxford University Press, Oxford, 2003).
    
  \bibitem[Shaviv et al.(1999)]{Shaviv} 
  Shaviv, N.J.,  Heyl, J.S., and Lithwick, Y., {Mon.\
  Not.\ R.\ Soc.} \textbf{306} 333 (1999).
  
  \bibitem[Shen and Meyer-ter-Vehn(2001a)]{Shen-MeyerterVehn2}
  Shen, B., and Meyer-ter-Vehn, J., Phys. Plasmas \textbf{8} 
  1003 (2001).
 
  \bibitem[Shen and Meyer-ter-Vehn(2001b)]{Shen-MeyerterVehn}
  Shen, B., and Meyer-ter-Vehn, J., Phys. Rev. E. \textbf{65} 
  016405 (2001).
  
  \bibitem[Shen and Yu(2002)]{she02} 
  Shen, B., and Yu, M.Y., Phys.\ Rev.\ Lett.\ \textbf{89} 
  275004 (2002).

  \bibitem[Shen and Yu(2003))]{Shen-Yu} 
  Shen, B., and Yu, M.Y., Phys.\ Rev.\ E \textbf{68},
    026501 (2003).  
    
  \bibitem[Shen et al.(2003)]{Shen-etal} Shen, B., Yu, M.Y., and Wang, X., Phys.\ Plasmas
  \textbf{10} 4570 (2003).

  \bibitem[Shen et al.(2004)]{Shen-Yu-Li}
  Shen, B., Yu, M.Y., and Li, R., Phys. Rev. E \textbf{70} 036403 (2004). 
  
  \bibitem[Sheng-Jui et al.(2003)]{Sheng-Jui-etal} Sheng-Jui, C., et al., 
  Improving ellipticity detection sensitivity for the Q \& A 
  vacuum birefringence experiment, hep-ex/0308071 (2003).
  
  \bibitem[Shigemori et al.(2000)]{Shigemori-etal}
  Shigemori, K., Ditmire, T., Remington, B. A., Yanovsky, V., et al., 
  Astrophys. J. Lett. \textbf{533}, L159 (2000). 

  \bibitem[Shukla et al.(1986)]{Shukla-etal} Shukla, P.K., Rao, N.N., Yu, M.Y., 
  and  Tsintsadze, N.L., Phys.\ Reports \textbf{138}, 1 (1986).
  
  \bibitem[Shukla et al.(1987)]{shu87} 
  Shukla, P.K., Bharuthram, R., and Tsintsadze, N.L., 
  Phys. Rev. A {\bf 35} 4889 (1987).

  \bibitem[Shukla et al.(1988)]{shu88} 
  Shukla, P.K., Bharuthram, R., and Tsintsadze, N.L.,  
  Physica Scripta {\bf 38} 578 (1988).
  
  \bibitem[Shukla(1992)]{Shukla} Shukla, P.K., Phys.\ Scr.\ \textbf{45} 618 (1992).
 
  \bibitem[Shukla and Stenflo(1998)]{Shukla-Stenflo-PoP} 
  Shukla, P.K., and Stenflo, L., Phys.\ Plasmas
  \textbf{5} 1554 (1998).
  
  \bibitem[Shukla and Eliasson(2004a)]{Shukla-Eliasson} 
  Shukla, P.K., and Eliasson, B.,  Phys.\
    Rev.\ Lett.\ \textbf{92}, 073601 (2004a).

  \bibitem[Shukla et al.(2004b)]{Shukla-Eliasson-Marklund} 
  Shukla, P.K., Eliasson, B., and
    Marklund, M., Opt.\ Comm. \textbf{235} 373 (2004b). 

  \bibitem[Shukla et al.(2004c)]{Shukla-Marklund-Tskhakaya-Eliasson}
    Shukla, P.K., Marklund, M., Tskhakaya, D.D., and Eliasson, B.,
    Phys. Plasmas \textbf{11} 3767 (2004c).
    
  \bibitem[Shukla et al.(2004d)]{Shukla-Marklund-Brodin-Stenflo}
  Shukla, P.K., Marklund, M., Brodin, G., and Stenflo, L., 
  Phys. Lett. A \textbf{330} 131 (2004d).  
  
  \bibitem[Shukla et al.(2004e)]{Shukla-Marklund-Eliasson}
  Shukla, P.K., Marklund, M., and Eliasson, B.,  
  Phys. Lett. A \textbf{324} 193 (2004e).
  
  \bibitem[Shukla and Eliasson(2005)]{Shukla-Eliasson2}
  Shukla, P.K., and Eliasson, B., Phys. Rev. Lett. \textbf{94} 065002 (2005).
  
  \bibitem[Shukla et al.(2005)]{Shukla-Eliasson-Marklund1} 
  Shukla, P.K., Eliasson, B.,  
  and Marklund, M., J. Plasma Phys.\ \textbf{71}, 213 (2005).

  \bibitem[Shorokhov et al.(2003)]{puk03} 
  Shorokhov, O., Pukhov, P., and Kostyukov, I., Phys.\
    Rev.\ Lett.\ \textbf{91} 265002 (2003). 
  
  \bibitem[Silva et al.(2004)]{Silva-etal}
    Silva, L.O., et al., Phys. Rev. Lett. \textbf{92} 015002 (2004).
  
  \bibitem[SLAC LCLS(2005)]{SLAC}
  SLAC Linac Coherent Light Source, \url{http://www-ssrl.slac.stanford.edu/lcls/}
  (2005).
    
  \bibitem[Solja\v{c}i\'c et al.(1998)]{Soljacic-etal}  
  Solja\v{c}i\'c, M., Sears, S., and Segev, M., Phys. Rev. Lett. 
  \textbf{81} 4851 (1998).
  
  \bibitem[Solja\v{c}i\'c and Segev(2000a)]{Soljacic-Segev2}
  Solja\v{c}i\'c, M., and Segev, M., Phys. Rev. E \textbf{62} 2810 (2000a) 
    
  \bibitem[Solja\v{c}i\'c and Segev(2000b)]{Soljacic-Segev} 
  Solja\v{c}i\'c, M., and Segev, M., 
    Phys.\ Rev.\ A \textbf{62} 043817 (2000b).
    
  \bibitem[Stenflo(1976)]{Stenflo} 
  Stenflo, L., Phys.\ Scripta {\bf 14} 320 (1976).

  \bibitem[Stenflo and Tsintsadze(1979)]{Stenflo-Tsintsadze} 
  Stenflo, L., and Tsintsadze, N.L., 
  Astrophys. Space Sci. \textbf{64} 513 (1979).
    
  \bibitem[Stenflo et al.(2005)]{Stenflo-etal}
  Stenflo, L., Brodin, G., Marklund, M., and Shukla, P.K., J. Plasma Phys.,
  in press, (2005).

  \bibitem[Sukenik et al.(1993)]{Sukenik-etal}
  Sukenik, C.I., Boshier, M.G., Cho, D., Sandoghdar, V., and Hinds, E.A., 
  Phys. Rev. Lett. \textbf{70} 560 (1993).  
  
  \bibitem[Surko et al.(1989)]{Surko-etal} Surko, C.M., Leventhal, M., and Passner, A., 
    Phys.\ Rev.\ Lett.\ \textbf{62} 901 (1989).

  \bibitem[Surko and Greaves(2004)]{Surko-Greaves}
  Surko, C.M., and Greaves, R.G., Phys. Plasmas \textbf{11} 2333 (2004).
  
  \bibitem[Tajima and Dawson(1979)]{Tajima-Dawson} 
  Tajima, T., and Dawson, J., Phys. Rev. Lett. \textbf{43}, 262 (1979).
  
  \bibitem[Tajima and Taniuti(1990)]{Tajima-Taniuti}
  Tajima, T., and Taniuti, T., Phys. Rev. A \textbf{42} 3587 (1990).

  
  \bibitem[Tajima and Mourou(2002)]{taj02} 
  Tajima, T., and Mourou, G., Phys.\ Rev.\ ST Accel.\
    Beams \textbf{5} 031301 (2002).
 
  \bibitem[Tajima(2003)]{taj03} Tajima, T., Plasma\  Phys.\  Rep.
    \textbf{29}, 207 (2003).
    
  \bibitem[Tamm and Frank(1937)]{Tamm-Frank} 
    Tamm, I.E., and Frank, I.M., Doklady Akad.\ Nauk SSSR \textbf{14} 107 (1937).

  \bibitem[Thoma(2000)]{Thoma} Thoma, M.H., Europhys.\ Lett.\ \textbf{52} 498
    (2000).  
    
  \bibitem[Thompson and Blaes(1998)]{Thompson-Blaes}
  Thompson, C., and Blaes, O., Phys. Rev. D \textbf{57} 3219 (1998).

  \bibitem[Tomaras et al.(2000)]{Tomaras-etal}
  Tomaras, T.N., Tsamis, N.C., and Woodard, R.P., Phys. Rev. D
  \textbf{62} 125005 (2000). 
 
  \bibitem[Tsai(1974a)]{Tsai1} Tsai, W.-Y., Phys. Rev. D \textbf{10} 1342 (1974a).
  
  \bibitem[Tsai(1974b)]{Tsai2} Tsai, W.-Y., Phys. Rev. D \textbf{10} 2699 (1974b).
  
  \bibitem[Tsai and Erber(1975)]{Tsai-Erber}
  Tsai, W., and Erber, T., Phys. Rev. D \textbf{12} 1132 (1975).
  
  \bibitem[Tsintsadze and Mendon\c{c}a(1998)]{Nodar} 
  Tsintsadze, N.L, and Mendon\c{c}a, J.T., Phys.\
  Plasmas \textbf{5}, 3609 (1998).   

  \bibitem[Tskhakaya(1982)]{r2} Tskhakaya, D.D., Phys.\ Rev.\ Lett.\ \textbf{48}
  484 (1982). 
  
  \bibitem[Unruh(1976)]{Unruh} Unruh, W., Phys. Rev. D \textbf{14} 870 (1976). 
  
  \bibitem[Valluri et al.(2003)]{Valluri-etal} Valluri, S.R., Jentschura, U.D., 
  and Lamm, D.R., The Study of the Heisenberg--Euler Lagrangian and
  Some of its Applications, hep-ph/0308223 (2003). 

  \bibitem[Ventura(1979)]{Ventura}
  Ventura, J., Phys. Rev. D \textbf{19} 1684 (1979).
  
  \bibitem[Weiland and Wilhelmsson(1977)]{Weiland-Wilhelmsson}
  Weiland, J.C., and Wilhelmsson, H., \textit{Coherent Non-linear 
    Interaction of Waves in Plasmas} (Pergamon Press, Oxford, 1977).
    
  \bibitem[Weinberg(1978)]{Weinberg}
  Weinberg, S., Phys. Rev. Lett. \textbf{40} 223 (1978).

  \bibitem[Weisskopf(1936)]{Weisskopf} Weisskopf, V.S., K.\ Dan.\
  Vidensk.\ Selsk.\ Mat.\ Fy.\ Medd.~\textbf{14} 1 (1936).

  \bibitem[Wilczek(1978)]{Wilczek}
  Wilczek, F., Phys. Rev. Lett. \textbf{40} 279 (1978).
  
  \bibitem[Wilks et al.(1992)]{Wilks-etal}
  Wilks, S.C., et al., Phys. Rev. Lett. \textbf{69} 1383 (1992).
  
  \bibitem[Woolsey et al.(2004)]{Woolsey-etal}
  Woolsey, N.C., Courtois, C., and Dendy, R.O., Plasma Phys.
  Control. Fusion \textbf{46} B397 (2004).

  \bibitem[Yu et al.(1982)]{Yu-etal} Yu, M.Y., Shukla, P.K., and Tsintsadze, N.L.,
    Phys.~Fluids \textbf{25}, 1049 (1982). 
  
  \bibitem[Zatsepin and Kuzmin(1966)]{Zatsepin-Kuzmin}
    Zatsepin, G.T., and Kuzmin, V.A., Sov.\ Phys.\ JETP Lett.\ \textbf{4} 78 (1966).
    
  \bibitem[Zavattini et al.(2005)]{Zavattini-etal}
  Zavattini, E., Zavattini, G., Ruoso, G., Polacco, E., et al.,
  hep-ex/0507107 (2005). 

  \bibitem[Zepf et al.(2003)]{Zepf-etal}
    Zepf, M., et al., Phys. Rev. Lett. \textbf{90} 064801 (2003).

  \bibitem[Zharova et al.(2003)]{Zharova-Litvak-Mironov} 
  Zharova, N.A., Litvak, A.G., and
    Mironov, V.A., JETP \textbf{96} 643 (2003). 

  
  

\end{thebibliography}
